\documentclass[a4paper,fleqn]{cas-dc}

\usepackage{lineno,hyperref}
\usepackage[numbers,longnamesfirst]{natbib}
\usepackage{amssymb}
\usepackage{tablefootnote} %

\usepackage{braket}

\usepackage{booktabs, makecell, tabularx}

\usepackage{siunitx}%
\usepackage{adjustbox}
\usepackage{physics} %Vasile
\usepackage{graphics}
\usepackage{makecell}
\usepackage[numbers]{natbib}
\usepackage{notoccite}

\usepackage{float}
\usepackage[scaled=0.92]{helvet}
\usepackage[T1]{fontenc}
\usepackage{textcomp}
\usepackage{color}
\usepackage[normalem]{ulem}
\usepackage[dvipsnames]{xcolor}
\def\tsc#1{\csdef{#1}{\textsc{\lowercase{#1}}\xspace}}
\tsc{WGM}
\tsc{QE}
\begin{document}
\let\WriteBookmarks\relax
\def\floatpagepagefraction{1}
\def\textpagefraction{.001}

\shorttitle{High-precision measurements of the atomic mass and electron-capture decay $Q$ value of $^{95}$Tc}   

\shortauthors{Z.~Ge, T.~Eronen, V.~A.~Sevestrean et al.} 

\title [mode = title]{High-precision measurements of the atomic mass and electron-capture decay $Q$ value of $^{95}$Tc}

%\tnotemark[1,2]

%\tnotetext[1]{This document is the results of the research   project funded by the National Science Foundation.}

%\tnotetext[2]{The second title footnote which is a longer text  matter to fill through the whole text width and overflow   into another line in the footnotes area of the first page.}
%\author[1]{Z.~Ge}[orcid=0000-0001-8586-6134]
\author[1]{Zhuang~Ge}[orcid=0000-0001-8586-6134]
%[orcid=0000-0001-0000-0000,style=chinese]
%[type=editor,                     auid=000,bioid=1, prefix=Sir, role=Researcher,   orcid=0000-0001-0000-0000]
\cormark[1]
%\ead{z.ge@gsi.de}
%{zhuang.z.ge@jyu.fi; zhuang@ribf.riken.jp}
%\ead[url]{zhuang@ribf.riken.jp}
%\ead{zhuang@ribf.riken.jp}
\ead{zhuang.z.ge@jyu.fi}
%\author[1]{T.~Eronen}[orcid=0000-0003-0003-6022]
\author[1]{Tommi~Eronen}[orcid=0000-0003-0003-6022]
%\author[2,3]{William {J. Hansen}}[% role=Co-ordinator,suffix=Jr, ]
%\ead{tommi.eronen@jyu.fi}
%\cormark[1]

%\thanks{Present address: KU Leuven, Instituut voor Kern- en Stralingsfysica, B-3001 Leuven, Belgium}
\affiliation[1]{organization={Department of Physics, 
                  University of Jyv\"askyl\"a},
                addressline={P.O. Box 35}, 
                %city={Jyv\"askyl\"a},
%               citysep={}, % Uncomment if no comma needed 
%                               between city and postcode
                postcode={FI-40014}, 
                state={Jyv\"askyl\"a},
                country={Finland}}
%\author[2]{A.~de~Roubin}[orcid=0000-0002-6817-7254]
%\author[2,3,4]{V.~A.~Sevestrean}[orcid=0009-0009-9658-2386]
\author[2,3,4]{Vasile~Alin~Sevestrean}[orcid=0009-0009-9658-2386]
\cormark[1]
\ead{sevestrean.alin@theory.nipne.ro}
%\affiliation{International Centre for Advanced Training and Research in Physics, P.O. Box MG12, 077125 Bucharest-M\u{a}gurele, Romania}%
%\affiliation{Faculty of Physics, University of Bucharest, 405 Atomiștilor, P.O. Box MG11, 077125 Bucharest-M\u{a}gurele, Romania}%
%\affiliation{“Horia Hulubei” National Institute of Physics and Nuclear Engineering, 30 Reactorului, POB MG-6, RO-077125 Bucharest-M\u{a}gurele, Romania}
%\affiliation{International Centre for Advanced Training and Research in Physics (CIFRA), P.O. Box MG12, 077125 Bucharest-M\u{a}gurele, Romania}%
\affiliation[2]{organization={International Centre for Advanced Training and Research in Physics (CIFRA)},
                addressline={POB MG-12}, 
                %city={Jyv\"askyl\"a},
%               citysep={}, % Uncomment if no comma needed 
%                               between city and postcode
                postcode={RO-077125}, 
                state={Bucharest-M\u{a}gurele},
                country={Romania}}% 
\affiliation[3]{organization={Faculty of Physics, University of Bucharest},
                addressline={405 Atomiștilor, POB MG-11}, 
                %city={Jyv\"askyl\"a},
%               citysep={}, % Uncomment if no comma needed 
%                               between city and postcode
                postcode={RO-077125}, 
                state={Bucharest-M\u{a}gurele},
                country={Romania}}% 
\affiliation[4]{organization={"Horia Hulubei" National Institute of Physics and Nuclear Engineering},
                addressline={30 Reactorului, POB MG-6}, 
                %city={Jyv\"askyl\"a},
%               citysep={}, % Uncomment if no comma needed 
%                               between city and postcode
                postcode={RO-077125}, 
                state={Bucharest-M\u{a}gurele},
                country={Romania}}
                %                                                  
%\author[3]{K.~S.~Tyrin}[orcid=0000-0003-4041-899X]      
%\affiliation{National Research Centre ``Kurchatov Institute'', Ploschad' Akademika Kurchatova 1, 123182 Moscow, Russia}%
%\affiliation{National Research Center ``Kurchatov Institute'', Pl. akademika Kurchatova 1, 123098 Moscow, Russia}%
%\affiliation{KU Leuven, Instituut voor Kern- en Stralingsfysica, B-3001 Leuven, Belgium}%
\affiliation[5]{organization={KU Leuven},
                addressline={Instituut voor Kern- en Stralingsfysica}, 
                %city={Moscow},
                postcode={B-3001}, 
                state={Leuven}, 
                country={Belgium}}
\affiliation[6]{organization={Universit\'e de Bordeaux},
                addressline={CNRS/IN2P3, UMR 5797}, 
                %city={Moscow},
                postcode={F-33170}, 
                state={Gradignan}, 
                country={France}}   
%\author[2,4]{O.~Ni\c{t}escu}[orcid=0000-0002-9598-8415]   
%\author[2]{S.~Stoica}[orcid=0000-0003-4632-7327]                
%\author[1]{M.~Ramalho}[orcid=0000-0003-3514-6678] 
%\author[1,2]{J.~Suhonen}[orcid=0000-0002-9898-660X]
%\author[2,4]{Ovidiu-Vasile~Ni\c{t}escu}[orcid=0000-0002-9598-8415]   
\author[2,4]{Ovidiu~Ni\c{t}escu}[orcid=0000-0002-9598-8415]   
\author[2]{Sabin~Stoica}[orcid=0000-0003-4632-7327]                
\author[1]{Marlom~Ramalho}[orcid=0000-0003-3514-6678] 
\author[1,2]{Jouni~Suhonen}[orcid=0000-0002-9898-660X]
\cormark[1]
\ead{jouni.t.suhonen@jyu.fi}
%\author[5,6]{A.~de Roubin}
%\author[1]{D.~A.~Nesterenko}[orcid=0000-0002-6103-2845]  
%\author[1]{A.~Kankainen}[orcid=0000-0003-1082-7602] 
%\author[6]{P.~Ascher}[orcid=0000-0002-1990-0848]
%\author[7]{S.~Ayet}[orcid=0000-0002-0053-1691]
\author[5,6]{Antoine~de Roubin}
\author[1]{Dmitrii~Nesterenko}[orcid=0000-0002-6103-2845]  
\author[1]{Anu~Kankainen}[orcid=0000-0003-1082-7602] 
\author[6]{Pauline~Ascher}[orcid=0000-0002-1990-0848]
\author[7]{Samuel~Ayet~San~Andres}[orcid=0000-0002-0053-1691]
%\author[1]{O.~Beliuskina}[orcid=0000-0003-4448-7650] 
%\author[8]{P.~Delahaye}[orcid=0000-0002-8851-7826]
\author[1]{Olga~Beliuskina}[orcid=0000-0003-4448-7650] 
\author[8]{Pierre~Delahaye}[orcid=0000-0002-8851-7826]
%\thanks{Corresponding author:  jouni.t.suhonen@jyu.fi}%
\affiliation[7]{organization={Instituto de Fisica Corpuscular},
                addressline={CSIC-UV}, 
                %city={Moscow},
                postcode={46980}, 
                state={Gradignan}, 
                country={Spain}} 
%\affiliation[8]{organization={GSI Helmholtzzentrum f\"ur Schwerionenforschung GmbH}, addressline={Planckstra\ss e 1},   postcode={64291},     state={Darmstadt},  country={Germany}} 
\affiliation[8]{organization={GANIL},
                addressline={CEA/DSM-CNRS/IN2P3, Bd Henri Becquerel}, 
                %city={Moscow},
                postcode={14000}, 
                state={Caen}, 
                country={France}}          
\affiliation[9]{organization={Universit\'e de Bordeaux, CNRS/IN2P3, LP2I Bordeaux},
                addressline={UMR 5797}, 
                %city={Moscow},
                postcode={F-33170}, 
                state={Gradignan}, 
                country={France}}            
%\author[6]{M.~Flayol}[orcid=0000-0002-1215-2269]
%\author[6]{M.~Gerbaux}
%\author[6]{S.~Gr\'evy}
\author[6]{Mathieu~Flayol}[orcid=0000-0002-1215-2269]
\author[6]{Mathias~Gerbaux}
\author[6]{St\'ephane~Gr\'evy}
%\affiliation{Universit\'e de Bordeaux, CNRS/IN2P3, LP2I Bordeaux, UMR 5797, F-33170 Gradignan, France}%
%\author[1,9]{M.~Hukkanen}[orcid=0000-0002-4317-3628]
\author[1,9]{Marjut~Hukkanen}[orcid=0000-0002-4317-3628]
%\affiliation[10]{organization={Centre d'Etudes Nucl\'eaires de Bordeaux Gradignan, UMR 5797 CNRS/IN2P3 - Universit\'e de Bordeaux},
                %addressline={19 Chemin du Solarium, CS 10120}, 
                %city={Moscow},
                %postcode={F-33170}, 
                %state={Gradignan Cedex}, 
                %country={France}}     
%\author[1]{A.~Jaries}[orcid=0000-0002-5279-0820]  
%\author[1]{A.~Jokinen}[orcid=0000-0002-0451-125X] 
\author[1]{Arthur~Jaries}[orcid=0000-0002-5279-0820]  
\author[1]{Ari~Jokinen}[orcid=0000-0002-0451-125X] 
 %\credit{Investigation, funding provider}
% \credit{Investigation, funding provider}  
%\author[1]{A.~Jokinen} 
%\author[1]{A.~Husson}[orcid=0000-0001-9798-0655]
%\author[10]{D.~Kahl}[orcid=0000-0003-3368-7307]
\author[6]{Audric~Husson}[orcid=0000-0001-9798-0655]
\author[10]{Daid~Kahl}[orcid=0000-0003-3368-7307]
\fnmark[2]
\affiliation[10]{organization={Extreme Light Infrastructure - Nuclear Physics},
                addressline={Horia Hulubei National Institute for R\&D in Physics and Nuclear Engineering (IFIN-HH)}, 
                %city={Moscow},
                postcode={077125}, 
                state={Bucharest-Magurele}, 
                country={Romania}}     
\author[11]{Joel~Kostensalo}[orcid=0000-0001-9883-1256]
%\affiliation{Department of Physics, University of Jyv\"askyl\"a, P.O. Box 35, FI-40014  Jyv\"askyl\"a, Finland}%
%  \credit{Writing manuscript, theoretical calculation, figure preparation}
\affiliation[11]{organization={Natural Resources Institute Finland},
                addressline={Yliopistokatu 6B}, 
                %city={Joensuu},
%               citysep={}, % Uncomment if no comma needed 
%                               between city and postcode
                postcode={FI-80100}, 
                state={Joensuu},
                country={Finland}}
%\author[2,12,13]{J.~Kotila}[orcid=0000-0001-9207-5824]
\author[2,12,13]{Jenni~Kotila}[orcid=0000-0001-9207-5824]
%\affiliation{Finnish Institute for Educational Research, University of Jyv\"askyl\"a, P.O. Box 35, FI-40014  Jyv\"askyl\"a, Finland}%
%\affiliation{Center for Theoretical Physics, Sloane Physics Laboratory Yale University, New Haven, Connecticut 06520-8120, USA}%
\affiliation[12]{organization={Finnish Institute for Educational Research, University of Jyv\"askyl\"a},
                addressline={P.O. Box 35}, 
                %city={Jyv\"askyl\"a},
%               citysep={}, % Uncomment if no comma needed 
%                               between city and postcode
                postcode={FI-40014}, 
                state={Jyv\"askyl\"a},
                country={Finland}}   
\affiliation[13]{organization={Center for Theoretical Physics, Sloane Physics Laboratory},addressline={Yale University}, city={New Haven}, postcode={Connecticut 06520-8120},  state={Connecticut},  country={USA}}                              
%  \credit{Writing manuscript, theoretical calculation, figure preparation}                             
%\author[3,7]{M.~I.~Krivoruchenko}[orcid=0000-0002-4450-1427]
%\cormark[1]
%National Research Center “Kurchatov Institute” - KCTEP, B. Cheremushkinskaya 25, 117218, Moscow, Russia
% \credit{Writing manuscript, theoretical calculation, figure preparation}
%\affiliation{GANIL, CEA/DSM-CNRS/IN2P3, Bd Henri Becquerel, 14000 Caen, France}%                
%\affiliation{National Research Centre ``Kurchatov Institute'', Ploschad' Akademika Kurchatova 1, 123182 Moscow, Russia}%
%\affiliation{Institute for Theoretical and Experimental Physics, NRC ``Kurchatov Institute'', B. Cheremushkinskaya 25, 117218 Moscow, Russia}
%\ead{mikhail.krivoruchenko@itep.ru}
%\author[1]{I.~D.~Moore}[orcid=0000-0003-0934-8727]   
\author[1]{Iain~Moore}[orcid=0000-0003-0934-8727]   
%\author[1]{M.~Vil\'en}[orcid=0000-0002-0375-2502]
%\affiliation[10]{organization={Experimental Physics Department, CERN},
               % addressline={CH-1211 Geneva 23}, 
               % country={Switzerland}}
\author[1]{Stylianos~Nikas}  
%\author[1]{S.~Nikas} 
%\author[1]{P.~Pirinen}  
%\author[1]{M.~Stryjczyk}[orcid=0000-0001-6515-2409]   
%\author[1]{V.~Virtanen}[orcid=0000-0003-0276-6483] 
\author[1]{Marek~Stryjczyk}[orcid=0000-0001-6515-2409]   
\author[1]{Ville~Virtanen}[orcid=0000-0003-0276-6483] 
%\fnmark[4]                
%\cormark[4]
%Present address: Experimental Physics Department, CERN, CH-1211 Geneva 23, Switzerland
%\affiliation{University of Jyv\"askyl\"a, Department of Physics, P.O. Box 35, FI-40014 University of Jyv\"askyl\"a, Finland}%
%\cormark[1]
%\ead{jouni.t.suhonen@jyu.fi}
   
%\author[1]{V.~Virtanen}
%  \credit{Writing manuscript, theoretical calculation, figure preparation}
%\cortext[cor2]{Principal corresponding author}
\cortext[cor1]{Principal corresponding authors}
%\cortext[cor2]{Corresponding author}
%\fntext[fn1]{Present address: GSI Helmholtzzentrum f\"ur Schwerionenforschung GmbH, 64291 Darmstadt, Germany}
\fntext[fn2]{Present address: Facility for Rare Isotope Beams, Michigan State University, 640 South Shaw Lane East Lansing, MI 48824, USA}
%\fntext[fn2]{Present address: University of Surrey, Department of Physics, Guildford GU2 7XH, United Kingdom}
%\fntext[fn3]{Present address: KU Leuven, Instituut voor Kern- en Stralingsfysica, B-3001 Leuven, Belgium}
%\fntext[fn4]{Present address: Experimental Physics Department, CERN, CH-1211 Geneva 23, Switzerland}
%\cortext[cor2]{Present address}
%\cortext[cor3]{Present address}
%\cortext[cor4]{Present address}
%\nonumnote{This note has no numbers. In this work we   demonstrate $a_b$ the formation Y\_1 of a new type of    polariton on the interface between a cuprous oxide slab  and a polystyrene micro-sphere placed on the slab.


\begin{abstract}%[S U M M A R Y]
A direct measurement of the ground-state-to-ground-state electron-capture  decay $Q$ value of $^{95}$Tc has been performed utilizing the double Penning trap mass spectrometer JYFLTRAP. The $Q$ value was determined to be 1695.92(13) keV by taking advantage of the high resolving power of the phase-imaging ion-cyclotron-resonance technique to resolve the low-lying isomeric state of $^{95}$Tc (excitation energy of 38.910(40) keV) from the ground state. The mass excess of  $^{95}$Tc was measured to be $-$86015.95(18) keV/c$^2$, exhibiting a precision of about 28 times higher and in agreement with the value from the newest Atomic Mass Evaluation (AME2020).
Combined with the nuclear energy-level data for the decay-daughter $^{95}$Mo, two potential ultra-low $Q$-value transitions are identified for future long-term neutrino-mass determination experiments.
The atomic self-consistent many-electron Dirac--Hartree--Fock--Slater method and the nuclear shell model have been used to predict the partial half-lives and energy-release distributions for the two transitions. The dominant correction terms  related to those processes are considered, including the exchange and overlap corrections, and the shake-up and shake-off effects. The normalized distribution of the released energy in the electron-capture decay of $^{95}$Tc to excited states of $^{95}$Mo is compared to that of $^{163}$Ho currently being used for electron-neutrino-mass determination. 
\end{abstract}
%Furthermore, the refined mass excess of $-$85871.39(25) keV/c$^2$ for $^{96}$Tc is approximately 20 times more precise and 49.4(50) keV lower than the value adopted from AME2020. This refined value agrees with the reported value from storage-ring mass spectrometry (Nuclear Physics A 756 (2005) 3–38), which provided an adjusted mass value of $-$85877(30) keV/c$^2$, corresponding to unresolved ground and isomeric states.
%\textcolor{red}
%The results show that the indium level $2p_{1/2}$ for this decay branch leads to a significant increase in the number of EC events in the energy region sensitive to the electron neutrino mass. 

%\item 
%\item 
%\item 

% Keywords

\maketitle

% Main text
%\section{}\label{}

% Use the style of numbering in square brackets.
% If nothing is used, default style will be taken.
%\begin{enumerate}[a)]
%\item 
%\item 
%\item 
%\end{enumerate}  

%\begin{itemize}
%\item 
%\item 
%\item 
%\end{itemize}  

%\begin{description}
%\item[]
%\item[] 
%\item[] 
%\end{description}  

% Figure

Neutrino oscillations in atmospheric, solar, and reactor neutrinos have confirmed that at least two neutrino mass eigenstates have non-zero rest mass. However, these oscillations cannot assess the absolute mass scale, but only the squared differences of the mass eigenstates~\cite{Fukuda1998,SNOCollaboration2002,Gerbino2018a}. Neutrinos are the second most abundant particles in the universe, and play an important role on cosmological scales~\cite{Giusarma_2023}. Accurate measurements of the total neutrino mass involve their imprint on the cosmic microwave background (CMB) as well as on structure formation in the early universe.

The most direct method to measure the absolute mass scale of antineutrinos involves studying the electron energy spectrum of $\beta^{-}$ decay. 
Though the neutrinoless double  $\beta^{-}$-decay experiments can be used to infer the effective Majorana-neutrino mass from the measured lifetime,  the exact relation depends on the mediator model and relies on the calculation of the involved transition matrix elements~\cite{Suhonen1998,Avignone2008,Ejiri2019,Agostini2023}. 
%currently
The ongoing leading experiment for the absolute neutrino mass scale determination is the Karlsruhe Tritium Neutrino (KATRIN) $\beta^-$-decay experiment which is designed to measure the electron-antineutrino mass,  $m_{{\overline\nu_e}}$, with a sensitivity of 0.2 eV/c$^{2}$ at 90\% C.L. Most recently, KATRIN has set a limit of $m_{{\overline\nu_e}}$ < 0.8 eV/c$^{2}$ (90\% C.L.)~\cite{Drexlin2013,Aker2019,Aker2022}. 
Another experiment, Project 8, takes advantage of the cyclotron radiation emission spectroscopy (CRES) technique via measurements of the tritium end-point spectrum. The new technique CRES will allow for an eventual sensitivity to m$_{{\overline\nu_e}}$  down to 0.04 eV/c$^{2}$.
The first frequency-based neutrino mass limit of electron-weighted neutrino mass < 155 eV/c$^{2}$  is extracted from the background-free measurement of the continuous tritium $\beta$ spectrum in a Bayesian (frequentist) analysis ~\cite{Ashtar23}. 
An alternative method in the ECHo~\cite{Gastaldo2014,Gastaldo2017,Velte2019,Echo2023} and HOLMES~\cite{Nucciotti2018,HOLMES23} experiments, uses electron capture (EC) on $^{163}$Ho, and has reached a current limit of 150 eV/c$^{2}$ for the electron-neutrino mass~\cite{Velte2019}. 

%Ramsey cleaning frequency scan ok-5-withErrorBar-95Nb-Tc.pdf
 %, $^{187}$Re
A $Q$ value as small as possible is desired  in these single decay experiments for electron (anti)neutrino mass determination.  The effective fraction of decays in a given energy interval $\Delta{E}$ at the endpoint area  will be larger with a lower $Q$ value~\cite{McDonald2013,Ferri2015}.
Currently, only ground-state-to-ground-state (gs-to-gs) decay cases $^{3}$H ($\beta$ decay)  and $^{163}$Ho (electron capture), are being used for direct neutrino-mass-determination experiments. Ongoing intensive searches for isotopes undergoing $\beta$/EC decays from the ground state to an excited state with a low $Q$ value are actively conducted at JYFLTRAP, LEBIT, CPT, ISOTRAP and SHIPTRAP Penning traps~\cite{Haaranen2013,Suhonen2014,Sandler2019,Karthein2019a,DeRoubin2020,ge2021,ge2021b,Ge2022a,ERONEN2022,Ge2022a,Ge2022b,Ramalho2022,Gamage22,Keblbeck2023,Ge2023}. Penning trap mass spectrometry (PTMS) is the leading technique for accurate and precise mass and $Q$ value determination, and it is hitherto the only direct method to measure the  decay  $Q$ value  to a sub-keV precision or better to verify whether a potential candidate is an ultra-low (< 1 keV) $Q$-value transition or not.

%and   $^{96}$Tc
In this article, we report on the first direct measurement of the gs-to-gs EC $Q$ value of $^{95}$Tc  with the JYFLTRAP PTMS. The precise $Q$ value obtained in this study, in conjunction with nuclear energy level data for excited states of $^{95}$Mo, is utilized to ascertain their ground-state-to-excited-state (gs-to-es) $Q$ values. 
%The possible ultra-low low $Q$-value transitions are checked with the refined experimental values.
%confirmed to be energetically negative and ruled out from the list for direct mass measurement experiment. 
%The improved value with sub-keV precision for the atomic mass of  $^{96}$Tc is important for testing the reliability of input data used in global evaluations of atomic masses, i.e., the atomic mass evaluation (AME2020)~\cite{Huang2021,Wang2021}, and improving its overall accuracy. 
%The enhanced precision, with sub-keV accuracy, in the atomic mass measurement of $^{96}$Tc holds significance for assessing the reliability of input data used in global evaluations of atomic masses, specifically in the context of the Atomic Mass Evaluation (AME2020)~\cite{Huang2021,Wang2021}. This improvement contributes to an overall enhancement in accuracy.  potential exists for use of two 
In the case of $^{95}$Tc, there are two potential low $Q$-value gs-to-es EC transitions, that could be used for neutrino-mass detection. To explore this potential, we have utilized two computational approaches, the atomic self-consistent many-electron Dirac--Hartree--Fock--Slater method and the nuclear shell model, to predict the partial half-lives and energy-release distributions for the EC-decay transitions in question.

%%%%%%%%%%%%%%%%%
\section{Experimental method}
%%%%%%%%%%%%%%%
The experiment was performed at the Ion Guide Isotope Separator On-Line facility (IGISOL) using the JYFLTRAP double Penning trap mass spectrometer~\cite{Eronen2012} at the University of Jyv\"askyl\"a, Finland~\cite{Moore2013,Kolhinen2013}. % Figure~\ref{fig:igisol} gives a schematic view of the experimental setup.

To generate $^{95}$Tc ions, a natural Mo target foil was irradiated with a few $\mu$A proton beam at 45 MeV from the K-130 cyclotron at the Accelerator Laboratory of the University of Jyv\"askyl\"a.
A helium-filled small volume gas cell was used to stop the recoils produced from the proton-induced fusion-evaporation reaction, and the ions were extracted using gas flow and guided through a sextupole ion guide~\cite{Karvonen2008} with a combination of DC and RF fields. Subsequently, the ions were accelerated with a 30 kV electric potential, followed by mass separation using a 55$^\circ$ dipole magnet with a typical mass resolving power of $M/\Delta{M}$ $\approx$ 500.
After isobaric separation for ions of $A/q = 95$, including the reaction products $^{95}$Nb$^{+}$, $^{95m}$Tc$^{+}$, $^{95}$Tc$^{+}$ and $^{95}$Mo$^{+}$, they were directed to a radiofrequency-quadrupole cooler-buncher (RFQ-CB)~\cite{Nieminen2001}, where they underwent accumulation, cooling, and bunching.

%For ions of A/q = 96, the dipole magnet selected $^{96}$Nb$^{+}$, $^{96m}$Tc$^{+}$, $^{96}$Tc$^{+}$, $^{96}$Mo$^{+}$, transporting them to the RFQ. Subsequently, the bunches from the RFQ were conveyed to the JYFLTRAP double Penning trap mass spectrometer for further purification and the ultimate mass-difference measurements.
%The samples of decay-daughter ions $^{95-96}$Mo$^{+}$ are prepared with the upstairs offline glow-discharge ion source. A 90$^\circ$ electrostatic bender is used to select the ions either from the online target station or from the offline ion source to transmit downstream. samples 
Decay-daughter ions of $^{95}$Mo$^{+}$ were prepared using the upstairs offline glow-discharge ion source. A 90$^\circ$ electrostatic bender selected ions either from the online target station or the offline ion source for downstream transmission.

\begin{figure}[!htb]
   %\flushleft
   %\includegraphics[width=250px,height=125px]{Fig2.pdf} 
   \includegraphics[width=0.99\columnwidth]{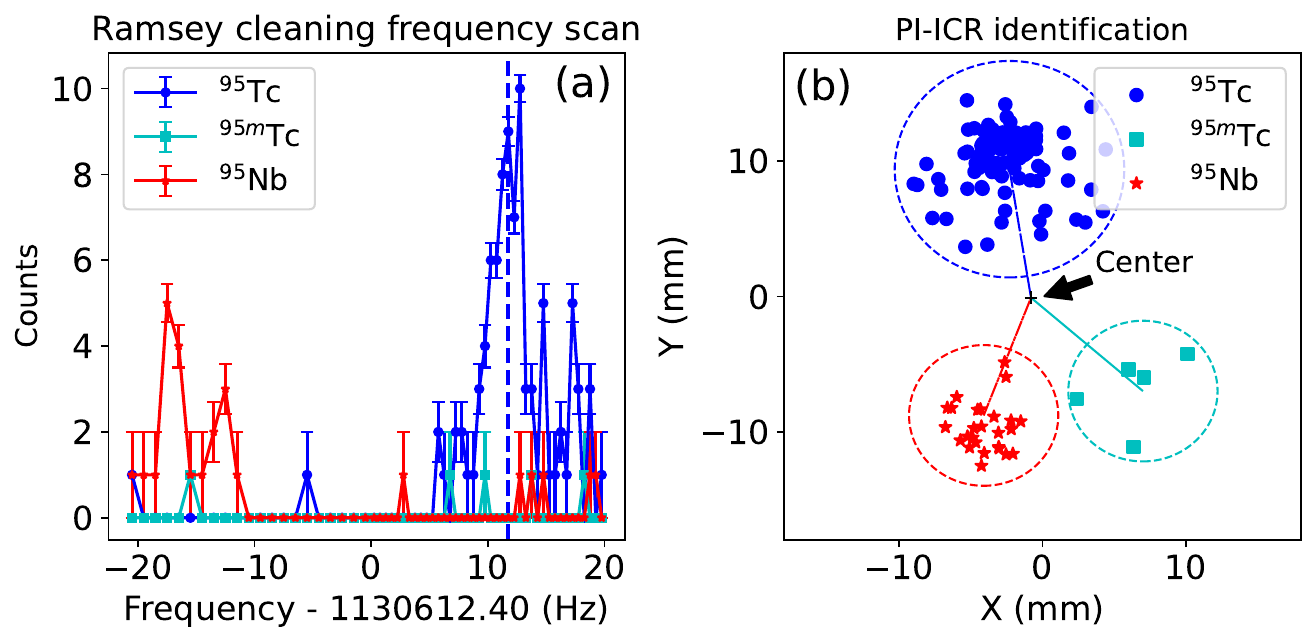}
   \caption{(Color online).  
(a) Ramsey-type dipole excitation frequency scan with a 5 ms (On) - 17 ms (Off) - 5 ms (On) excitation pattern in the second trap filtered by the positional gates shown in (b) using the PI-ICR identification (755 ms phase accumulation time) plot. The used angular gates are highlighted. The vertical dashed blue line shows the chosen optimal frequency to transmit $^{95}$Tc ions while suppressing the others.
   %{ (a) A Ramsey-type dipole excitation frequency scan  with a   5 ms (On) - 17 ms (Off) - 5 ms (On) excitation pattern in the second trap gated by  (b) the PI-ICR identification (755 ms phase accumulation time) plot. The used angular gates are highlighted. %shown on % with the measured phases (right) with the PI-ICR method
   %for selection of ions of interest. 
  %The vertical dashed lines in various colors indicate the ``survival'' excitation frequency to be applied for the selection of the corresponding ion species.  The vertical dashed line shows the optimal frequency to transmit $^{136}$Cs ions while suppressing the others.   %The dashed line indicates the ``survival''  frequency of $^{95}$Tc.}
}  
   \label{fig:Ramsey}
\end{figure}

JYFLTRAP comprises two cylindrical Penning traps in a 7-T superconducting solenoid. The first trap, functioning as a purification trap, is filled with helium buffer gas and is used for isobaric purification  through the sideband buffer gas cooling technique~\cite{Savard1991}. This method achieves purification with a mass resolving power of $\approx$ $10^{5}$.
%by mass selectively converting ion motion from magnetron to reduced cyclotron motion.
In the purification trap, all cooled and centered ions ($^{95}$Nb$^{+}$, $^{95m}$Tc$^{+}$, $^{95}$Tc$^{+}$, and $^{95}$Mo$^{+}$)
%or $^{96}$Nb, $^{96m}$Tc, $^{96}$Tc, $^{96}$Mo)
are initially excited to a large magnetron motion orbit. This is accomplished by applying a dipole excitation at the magnetron motion frequency $\nu_{-}$ for approximately 11 ms. Subsequently, a quadrupole excitation is executed for approximately 100 ms to center the ions of interest through collisions with the buffer gas. 
The buffer gas cooling technique eliminated $^{95}$Mo$^{+}$ but did not have enough mass resolving power to remove the other aforementioned ions.
%Only the centered ions of $^{95}$Tc, exhibiting a specific mass-dependent quadrupole excitation frequency, are extracted through a 1.5 mm diaphragm that separates the two traps.
%To prepare mono-isotopic samples of $^{95}$Tc$^{+}$, the procedure described in~\cite{Ge2023} was utilised. This utilized coupling of the dipolar excitation with Ramsey’s method of time-separated oscillatory fields~\cite{Eronen2008a} and the phase-imaging ion-cyclotron-resonance (PI-ICR) technique~\cite{nesterenko2021,Nesterenko2018}. procedure described in~\cite{Ge2023} was utilised
To prepare mono-isotopic samples of $^{95}$Tc$^{+}$, the coupling of the dipolar excitation with Ramsey’s method of time-separated oscillatory fields~\cite{Eronen2008a} and the phase-imaging ion-cyclotron-resonance (PI-ICR) technique~\cite{nesterenko2021,Nesterenko2018} was utilized, as described in details in~\cite{Ge2023}.
A plot of the Ramsey-type dipole excitation frequency scan with a 5 ms (On) - 17 ms (Off) - 5 ms (On) excitation pattern in the second (precision) trap, filtered by the positional gates using the PI-ICR identification with a 755 ms phase accumulation time, is shown in Fig.~\ref{fig:Ramsey}.

%$\nu_{c}=\frac{1}{2\pi}\frac{qB}{m}$
For $Q$-value measurements, the PI-ICR method is used to measure the cyclotron frequency, $\nu_{c}={qB}/({{2\pi}m})$, where $B$ is the magnetic field strength, $q$ is the charge and $m$ the mass of the stored ion. 
The PI-ICR technique~\cite{Nesterenko2018} provides around 40 times better resolving power than the conventional time-of-flight ion-cyclotron-resonance (TOF-ICR) method~\cite{Nesterenko2018,Eliseev2014,Eliseev2013}.
%, was used to determine the $\nu_{c}$. 
%The scheme of PI-ICR technique at JYFLTRAP~\cite{Nesterenko2018} relies on the direct measurements of the cyclotron motion and the cyclotron motion at the same time to project the radial ion motion onto the position-sensitive MCP detector. 
%
Two timing patterns are needed for the determination of $\nu_c$. The patterns differ only in their quadrupolar conversion pulse, separated in time by the defined phase-accumulation time, $t_{acc}$. The phase images of these two are projected onto a position-sensitive MCP detector after the trap. Additionally a center point, measured without any excitations, is needed for angle determination.
%By extracting the ions directly to project them onto the position-sensitive MCP detector after the injection into the center of the second trap storing for a few milliseconds without exciting the cyclotron motion, the center spot of the ions of interest is collected.

The angle between two phase images of the projected radial motions with respect to the center spot is denoted as $\alpha_c = \alpha_+ - \alpha_-$, where $\alpha_+$ and $\alpha_-$ represent the polar angles of the  cyclotron and magnetron motion phases. 
The cyclotron frequency $\nu_{c}$  is derived from: 
%\begin{equation}
%\label{eq:nuc2}
%$\nu_{c}=\frac{\alpha_{c}+2\pi n_{c}}{2\pi{t_{acc}}}$,
$\nu_{c}={(\alpha_{c}+2\pi n_{c})}/{2\pi{t_{acc}}}$,
% m=r(m_{ref}-m_e)+m_e,$n_{+}$ + $n_{-}$
%\end{equation}
where $n_{c}$ represents the full number of revolutions made by the measured ions during the phase accumulation time $t_{acc}$. Different accumulation times for $^{95}$Tc$^{+}$ were utilized to unambiguously assign $n_c$.
%in Eq.~\eqref{eq:nuc2}. 
An accumulation time of 574 ms was employed for the actual measurements to determine the final $\nu_{c}$ for both $^{95}$Tc$^{+}$ and $^{95}$Mo$^{+}$ ions; the choice also ensures that any leaked isobaric contaminant would not overlap with the ions of interest.
%and an accumulation time of 518 ms was utilized for $^{96}$Tc$^{+}$-$^{96}$Mo$^{+}$.
%To ensure the accuracy of the conversion frequency, more than two distinct accumulation times were applied for $^{95}$Tc$^{+}$ and $^{96}$Tc$^{+}$. 
%One fixed accumulation time 518 ms for long time of data accumulation was employed for the actual measurements to determine the final $\nu_{c}$.
%The positions of the magnetron-motion and cyclotron-motion phase spots were chosen such that the angle $\alpha_c$  did not exceed a few degrees in order to
The positions of the phase spots for magnetron and cyclotron motion were carefully selected to maintain an angle $\alpha_c$ within a few degrees. 
This choice aimed to minimize the shift in the $\nu_{c}$ ratio of the $^{95}$Tc$^{+}$-$^{95}$Mo$^{+}$ pair due to the conversion of the cyclotron motion to magnetron motion and the possible distortion of the ion-motion projection onto the detector to a level well below 10$^{-10}$~\cite{Eliseev2014}. %Figure~\ref{fig:2-phases} illustrates a representative measurement with "cyclotron" and "magnetron" phase spots collected respective to the center spot. 
The excitation of the $\nu_{+}$ delay was systematically scanned over one magnetron period, while the extraction delay varied over one cyclotron period. This accounted for any residual magnetron and cyclotron motion that might have shifted the different spots. The total data accumulation time of  interleaved measurements of  $\nu_{c}$ for $^{95}$Tc$^{+}$-$^{95}$Mo$^{+}$ ions was  $\approx$ 4.9 hours, respectively.
%The measurements were separated to 3 time slots and the total data accumulation time of  interleaved measurements of  $\nu_{c}$ of the $^{96}$Tc$^{+}$, $^{96}$Mo$^{+}$ ions were  $\approx$ 11.4 hours.

%The $Q_{EC}$ value can be given as the mass difference of the two ion species: ($^{96}$Tc$^{+}$)($^{96}$Mo$^{+}$) 
The gs-to-gs electron-capture $Q$ value, $Q_{\rm EC}$, can be derived from the mass difference of the decay pair:
\begin{equation}
\label{eq:Qec}
%M_{ioi} = R(M_{ref} - m_e)   + m_e  + (R \cdot B_{ref} - B_{ioi})/c^2,
Q_{\rm EC}=(M_p - M_d)c^2 = (R-1)(M_d - qm_e)c^2+(R \cdot B_{d} - B_{p}),
%Q_{\beta^-}=M_p - M_d = (R-1)(M_d - qm_e)+\Delta{B_{p,d}},
% m=r(m_{ref}-m_e)+m_e,\frac
\end{equation}
where $M_p$ and $M_d$ represent the masses of the parent and daughter atoms, respectively, and $R$ (=${\nu_{c,d}}/{\nu_{c,p}}$) denotes their cyclotron frequency ratio for singly charged ions ($q=1$), with $m_e$ being the mass of an electron. The electron binding energies of the parent and daughter atoms, denoted as $B_p$ and $B_d$, are neglected due to their small values (on the order of a few eVs~\cite{NIST_ASD}), and $R$ is close to 1.
%Since both the parent and daughter ions have the same $A/q$, the mass-dependent error becomes negligible compared to the statistical uncertainty achieved in the measurements. 
%Moreover, due to the fact that the mass difference between the parent and daughter is very small ($\Delta M/M < 10^{-4}$), the contribution of uncertainty to the $Q$ value from the mass uncertainty of the reference (daughter), which is 0.12 keV/c$^2$ for $^{95}$Mo~\cite{Wang2021}, can be neglected.
Since both the parent and daughter ions have the same $A/q$ and their relative mass difference $\Delta M/M < 10^{-4}$, the mass-dependent error becomes negligible compared to the statistical uncertainty achieved in the measurements. 
Also, the contribution of uncertainty to the $Q$ value from the mass uncertainty of the reference (daughter), which is 0.12 keV/c$^2$ for $^{95}$Mo~\cite{Wang2021}, can be neglected.

%Moreover, the mass difference between the parent and daughter is very small ($\Delta M/M < 10^{-4}$), rendering the contribution of uncertainty to the $Q$ value from the mass uncertainty of the reference (daughter) practically negligible. 

%-----------------------------Fig. 2-2 --------------------------------
%-----------------------------Fig. 2-2-------------------------------

%-----------------------------Table 2 --------------------------------
%%%%%%%%%%%%%%%%%%%%%%%
\begin{table*}[!htb]
%\begin{table*}[!]
%\small
 \fontsize{8}{9}\selectfont
\caption{Potential candidate transitions of initial state (ground state) of parent  nuclei $^{95}$Tc (9/2$^{+}$) to the final states (excited states) of daughter  $^{95}$Mo with ultra-low $Q$ values. The first column lists the excited final state of  $^{95}$Mo for the low $Q$ value transition. The decay type is provided in the second column. The third and fourth columns present the derived decay $Q_{\rm EC}$ values in keV, sourced from literature (Lit.)~\cite{Wang2021} and this work (T. W.), respectively. 
   The fifth column displays the experimental excitation energy $E^{*}$ with its experimental error~\cite{NNDC,Wiedeking16,Basu2010} in keV.  The sixth column shows the confidence ($\sigma$) of the $Q^*_{\rm EC}$ being positive/negative.  Columns seven to nine, denoted as $\Delta{x}$, represent the distance of $Q_{\mathrm{EC}}^{}$ values to the computed atomic relaxation energy following the electron capture $\varepsilon_x$ in the daughter atoms~\cite{Thompson2009}. 
   %The eighth to ninth  columns denoted as $\Delta_{x}$  give the distance of the $Q_{\mathrm{EC}}^{*}$  value to the binding energy $\varepsilon_x$ (from~\cite{X-Ray_Data_Booklet}) of the electrons in the daughter atoms. 
   FNU means forbidden non-unique. Spin-parity assignments and energy values enclosed in braces \{\} signify uncertain assignments or uncertainties in excitation energy, resulting in uncertainties in the decay type or decay energy. All the energies are in unit of keV.   }
%  \begin{ruledtabular}
    \begin{tabular*}{0.99\textwidth}{@{}ccccccccc@{}}
  % \begin{tabular*}{\textwidth}{lcccc}%Initial state  & 
  \toprule
 % {ccccccccc}%Initial state  & ^{i}
 Final state &Decay type & \makecell[c]{$Q^*_{\rm EC}$ \\(Lit.)} &\makecell[c]{$Q^*_{\rm EC}$ \\ (T. W.)}& E$^{*}$ & \makecell[c]{$Q/\delta Q$ \\(T. W.)} &\makecell[c]{$\Delta_{\mathrm{K}}$ \\ (T. W.)}  &\makecell[c]{$\Delta_{\mathrm{L1}}$ \\ (T. W.)}  & \makecell[c]{$\Delta_{\mathrm{L2}}$ \\ (T. W.)} \\
%&&&AME2020&AME2020&this  work&this work&&&\\
% & dQ$_{EC}$ & E$^{*}_{error}
   \midrule
%&&&AME2020&AME2020&this  work&this work&&&\\
% & d$Q_{\mathrm{EC}}$ & E^{*}_{error}
%\hline\noalign{\smallskip}
$^{95}$Mo (9/2$^{+}$) &allowed&   15.6(50)& 20.52(61) &{1675.40(60)}& 33& 0.47(61)&17.64(61)& 17.89(61)  \\
 %  $^{95}$Mo (7/2$^{+}$, 9/2$^{+}$) &allowed&   24.0(94)& 28.9(80) &\textcolor{blue}{1667.0(80) old}& 4& 8.9(80)&26.1(80)& 26.3(80)  \\
 $^{95}$Mo (\{7/2$^{+}$, 9/2$^{+}$\})&  \{allowed\}&  8.0(51)&12.9(10) &1683.0(10)& 13 & & 10.0(10)& 10.3(10) \\
   $^{95}$Mo (1/2$^{+}$)& 4th FNU &-1.0(\{51\})&3.92(\{13\})  &1692(\{\})& \{29\} &&1.04(\{13\})& 1.29(\{13\}) \\
     \bottomrule
      \end{tabular*}
   \label{table:low-Q}
%\end{ruledtabular}
%\end{tabular}
\end{table*}

%%%%%%%%%%%%%%%%%%%%%%%
%-----------------------------Table 2 --------------------------------

\section{Results and discussion}
The determination of $Q_{\rm EC}$ depends on the measured cyclotron  frequency ratio $R$ via Eq.~\ref{eq:Qec}. 
Two  data sets for $^{95}$Tc$^{+}$-$^{95}$Mo$^{+}$  
%and  three data sets for $^{96}$Tc$^{+}$-$^{96}$Mo$^{+}$  
were collected. 
%The TOF-ICR and PI-ICR data were split to 6 and 8 parts for final fitting, respectively.  with a same different accumulation time
A full scanning measurement of the magnetron phase, cyclotron phase and center spot in sequence (one cycle) was completed in less than 5 minutes for each ion species. % of the decay pair $^{95}$Tc$^{+}$-$^{95}$Mo$^{+}$. 
%or $^{96}$Tc$^{+}$-$^{96}$Mo$^{+}$.
%$^{96}$Tc$^{+}$  and $^{96}$Mo$^{+}$. 
In the analysis, the position of each spot was fit with the maximum likelihood method. A few cycles were summed to have reasonable statistics for fitting. %before determining the position of each spot.
The phase angles were calculated accordingly % based on the determined positions of the phase spots
to deduce the cyclotron frequencies of each ion species. 
The cyclotron frequency $\nu_{c}$ of the daughter  $^{95}$Mo$^{+}$ as a reference was linearly interpolated to the time of the measurement of the parent $^{95}$Tc$^{+}$ (ion of interest) to deduce the cyclotron frequency ratio $R$. Only the bunches with less than five detected ions were considered in the data analysis in order to reduce a possible cyclotron frequency shift due to ion-ion interactions~\cite{Kellerbauer2003,Roux2013}. The count-rate related frequency shifts were not observed in the analysis. 
The temporal fluctuation of the magnetic field  $\delta_B(\nu_{c})/\nu_{c}=  \Delta t \times 2.01(25)  \times  10^{-12}$/min~\cite{nesterenko2021}, where $\Delta t$ is the time interval between two consecutive reference measurements, is considered in the final results. 
Contribution of  temporal fluctuations of the magnetic field to the final frequency ratio uncertainty was less than 10$^{-10}$ since the parent-daughter measurements were interleaved with $\Delta t$ < 10 minutes. The frequency shifts in the PI-ICR measurement due to ion image distortions were well below the statistical uncertainty and thus ignored in the calculation of the final uncertainty.  Furthermore,  decay pair ions $^{95}$Tc$^{+}$-$^{95}$Mo$^{+}$, being mass doublets, cancel many of the systematic uncertainties in the cyclotron frequency ratio.

\begin{figure}[!htb]
   %\flushleft 
   \includegraphics[width=0.99\columnwidth]{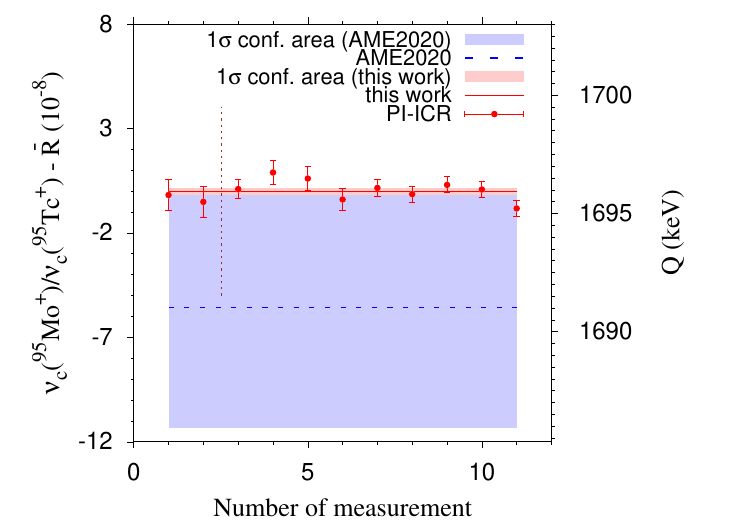}
   %{Fig3-95-95-ratio-Q-values.pdf}
   \caption{(Color online). The measured experimental results from this work compared to the literature values~\cite{Huang2021,Wang2021}. The deviations of the individually measured cyclotron frequency ratios  $R$ ($\nu_c$($^{95}$Mo$^{+}$)/$\nu_c$($^{95}$Tc$^{+}$)) from the measured value $\overline{R}$ (left axis) and $Q$ value (right axis) in this work are compared to values adopted from AME2020. The red points with uncertainties represent individual measurements using the PI-ICR method. Vertical brown dashed lines separate measurements conducted at different time slots. The weighted average value $\overline{R}$  is depicted by the solid red line, and its 1$\sigma$ uncertainty band is shaded in red. The dashed blue line illustrates the difference between our new value and the one referenced in AME2020, with its 1$\sigma$ uncertainty area shaded in blue.}
   \label{fig:ratio}
\end{figure}

The weighted mean ratio $\overline{R}$ of all single ratios was calculated along with the inner and outer errors to deduce the Birge ratio~\cite{Birge1932}.  The maximum of the inner and outer errors was taken as the weight to calculate $\overline{R}$.  
In Fig.~\ref{fig:ratio}, results of the analysis including all data with comparison to literature values are demonstrated. The final parent-to-daughter frequency ratio $\overline{R}$ with their uncertainty is determined to be  1.000 019 183 9(15). The corresponding gs-to-gs $Q$ values is 1695.92(13) keV. 
%A comparison of our results to the literature values are tabulated in  Table.~\ref{table:Q-value}. 
%The final frequency ratio $\overline{R}$ with its  uncertainty as well as the corresponding $Q$ value are $\overline{R}$ =  1.000 032 725 4(24) and Q$_{\beta^-}$ = 2923.50(21), respectively. The mass excess of 

%-----------------------------Fig. 3 --------------------------------
%(referenced in Table~\ref{table:Q-value})
%\subsection{$^{95}$Tc}
The gs-to-gs $Q_{\rm EC}$ value of 1695.92(13) keV from this work  is $\approx$ 37 times more precise than that deduced from the evaluated masses in AME2020~\cite{Huang2021,Wang2021}. The measured  $Q_{\rm EC}$ value has a deviation of 4.9(50) keV from the AME2020 value and is $\approx$1$\sigma$ larger.  The  $Q_{\rm EC}$ value in AME2020 is derived primarily from two ${\beta^+}$-decay experiments $^{95}$Tc(${\beta^+}$)$^{95}$Mo~\cite{Langer63,CRETZU65}. %and $^{95}$Ru(${\beta^+}$)$^{95}$Tc. 
Combined with the atomic mass of $^{95}$Mo (mass excess: $-$87711.87(12) keV/c$^2$) from AME2020~\cite{Wang2021,Huang2021}, we deduce the mass excess of its parent nucleus  $^{95}$Tc (9/2$^{+}$) 
%, with a half-life of 20.0(1) hours~\cite{NNDC},
to be  $-$86015.95(18) keV/c$^2$. The mass of  $^{95}$Tc in AME2020 is primarily evaluated from  ${\beta^+}$-decay experiments $^{95}$Tc(${\beta^+}$)$^{95}$Mo and $^{95}$Ru(${\beta^+}$)$^{95}$Tc with influence of 97.4\% and 2.6\%, respectively~\cite{Langer63,Anto74,Pin68}.     
% Previous studies have already demonstrated that mass values derived in indirect methods, like decay spectroscopy and nuclear reactions, may demonstrate large discrepancies with those from direct mass measurements and be inaccurate in a broad range of mass numbers~\cite{Eliseev2011a,Nesterenko2019}. \cite{Basu2010}

The high-precision electron-capture energy from this work, together with the nuclear energy level data from Ref.~\cite{NNDC,Wiedeking16,Basu2010} of the excited states of $^{95}$Mo as tabulated in Table~\ref{table:low-Q},  was used to determine the gs-to-es $Q$ value ($Q^*_{\rm EC}$) of three candidate states as shown in Table~\ref{table:low-Q}.  The newly determined $Q^*_{\rm EC}$ confirm that the decay of the ground state of $^{95}$Tc to the three excited states of interest are energetically allowed.
%as shown in Table~\ref{table:low-Q}. 
 %, see Fig~\ref {fig:level-scheme}. 
 
 In case of EC, the closer the $Q$ value of the decay to one of the ionization energies of the captured electrons, the larger the resonance enhancement of the rate near the end-point, where the effects of a non-vanishing neutrino mass are relevant. The event-rate dependence on the $Q$ value near the end-point for EC is steeper than that for $\beta^-$ decay. 
 As tabulated in Table~\ref{table:low-Q}, $\Delta_{x}$  gives the distance of the $Q_{\mathrm{EC}}^{*}$  value to the computed atomic relaxation energy $\varepsilon_x$ following
the capture of electrons in the allowed daughter atomic shells ($x=$ K, L1, L2, and other electrons from s-levels and p$_{1/2}$-levels from the third and higher shells). 
% For the total energy $Q_{\nu}$ of the emitted neutrino in EC decay, additionally the atomic binding energy {$e_x$}, where ${x}$ denotes the possible allowed atomic shell (K, L1, L2, M1, M2, N1, N2, O1, O2, P1) of the captured electron, needs to be accounted for:  $Q^{i}_{\nu,{x}} = Q^{i}_{EC} - {e_{x}}$, and this will make the EC energy even smaller, as tabulated in Table~\ref{table:low-Q}. 
For the state with the excitation energy of 1675.40(60) keV, the captures of electrons occupying the K and higher shells for the transition $^{95}$Tc (9/2$^+$) $\rightarrow$ $^{95}$Mo$^*$ are energetically allowed,
while for states with the excitation energy of 1683.0(10) keV and 1692 keV, only electrons from s-levels and p$_{1/2}$-levels from the second (L) and higher shells can possibly be captured due to angular momentum conservation and the finite overlap of their wave function with the nucleus. 
The transition  $^{95}$Tc (9/2$^-$) $\rightarrow$ $^{95}$Mo$^*$ (1692 keV), giving the values of 1.04($\{13\}$) keV and 1.29($\{13\}$) keV for the distance of $Q_{\mathrm{EC}}^{*}$ to the computed atomic relaxation energy following the electron capture $\varepsilon_{L1}$ =  2.878 keV and $\varepsilon_{L2}$ =  2.632 keV, is of the decay type of 4$^{\rm th}$ FNU (forbidden non-unique). It has a long half-life which will  result in an extremely low fraction of events landing near the endpoint. This transition is not of interest for future neutrino mass determination due to the low branching ratio. To confirm whether the emitted neutrino energy 1.04($\{13\}$) keV is ultra-low, further high-precision measurements of the excitation energy of the state are required. The parity of the 1683.0(10) keV state needs to be determined to verify the decay type of the transition to this state.
%$\approx 1 keV$ of the electrons in the daughter atoms of allowed shells (L1, L2) 
%k to be energetically forbidden.
%In the present case, captures of electrons occupying the K and L shells for the transition $^{159}$Dy(3/2$^-$) $\rightarrow$ $^{159}$Tb$^*$(5/2$^-$) are energetically forbidden. Only electrons from $s$ and $p_{1/2}$-levels from the third and higher shells (M1, M2, N1, N2, O1, O2, and P1) can possibly be captured due to angular momentum conservation and the finite overlap of their wave function with the nucleus. 
 %The refined gs-to-ge $Q_{EC}$ values were at the level of more than around 200$\sigma$ below 0 keV, thus excluding them as possible low Q-value transitions to be used for neutrino mass determination. 
% The calculated $Q_{EC}$ values of potential candidate transitions of  the  ground state of parent  nuclei $^{95}$Tc  to the excited states of daughter  $^{95}$Mo are tabulated in Table~\ref{table:low-Q}. 
The gs-to-gs  $Q$ value of $^{95}$Tc is now well refined to sub-keV uncertainty, combined with the energy level of 1675.40-keV state, an ultra-low distance (0.47 keV) of $Q_{\mathrm{EC}}^{*}$ to the computed atomic relaxation energy following the electron capture $\varepsilon_{k}$ = 20.054 keV~\cite{Thompson2009} is observed.
%, which render the 
%but the  excitation energy 1675.4(61) keV of the daughter state needs to be measured with a higher precision, along with the branching ratio of the transition. 
%and 1692 keV) are already rather well-known, due to the large uncertainty of $Q^{gs}_{EC}$ of 1.2 keV which is primarily determined from reaction data~\cite{Huang2020,NSR1968MY01,NNDC}, it is difficult to identify which transition is energetically allowed and how small the corresponding transition energy is.  

%95 Tc (AME2020) 1690.5(51) -86021.4(51)
%95 Tc (This Work) 1.000 019 183 9(15) 1695.92(13) -86015.95(18)
%96 Tc (AME2020) 2973.0(50) -85822.0(50) 
%96 Tc (This Work) 1.000 032 725 4(24) 2923.50(21) -85871.39(25)

%−86086 30
%−85877 30
%95Tc -86021 5 19:258 h 0.026 9=2+ 10 20Sz02 T 1947 b+=100
%95Tcm -85982 5

%96Tc -85822 5 4:28 d 0.07 7+ 08 1947 b+=100
%96Tcm -85788 5

\section{Theoretical predictions}
In the following we employed two calculation methods in order to predict the transition half-life and the distribution of energy released in the decay, namely atomic many-electron Dirac--Hartree--Fock--Slater (DHFS) self-consistent method and the nuclear shell model many-nucleon framework. The DHFS framework has been proven adequate for this type of calculations in our previous work \cite{SevestreanPRA2023}.

Using the DHFS method we obtained the wave-functions and the energy levels of the atomic electrons. The calculations were performed for both the initial atom and the final atom. The initial atom was in its ground state. For the final atom, we considered all possible states with the electron configuration of the initial atom having a hole in each shell from which the electron could be captured.  We assumed a central field $V(r)$ for the atomic system. The potential has three components as given in \cite{SalvatCPC2019}:

\begin{equation}
V_{\mathrm{DHFS}}(r)=V_{\mathrm{nuc}}(r)+V_{\mathrm{el}}(r)+V_{\mathrm{ex}}(r),
\end{equation}
where $V_{\mathrm{nuc}}(r)$, $V_{\mathrm{el}}(r)$, $V_{\mathrm{ex}}(r)$ are the nuclear, electronic and exchange potential. The nuclear potential takes into account a realistic Fermi charge distribution in the nucleus \cite{HahnPR1956}. The electronic potential is generated based on the charge distribution of the entire electron cloud. The procedure is iterative: after solving for the wave functions of the electrons, the charge distribution is recalculated and thus the potential as well. Then the computation is restarted until convergence. The exchange potential assures the correct asymptotic behaviour of the potential at $r\rightarrow\infty$. 
For the atomic structure calculations we made use of the RADIAL subroutine package \cite{SalvatCPC2019}, which also contains the \textsc{DHFS.F} code.

We denote the electron states as $\ket{(n, \kappa)}$, where $n$ is the principal quantum number and $\kappa$ is the relativistic quantum number. The atomic relaxation energy following the capture of an electron from the $x=(n,\kappa)$ shell is denoted as $\varepsilon_{x}$. The values for $\varepsilon_{x}$ are calculated according to the refined energy conservation in \cite{SevestreanPRA2023}:
\begin{equation}
    \varepsilon_{x}=|T_{\rm g.s.}|-|T_{x}|,
\end{equation}
where $T_{\rm g.s.}$ and $T_{x}$ are the total binding energy of the final atom in the ground state and in the excited state with a hole in the $x$ shell.

For allowed transitions the energy distribution of an EC event is calculated as a sum over all atomic shells with $\kappa=\pm1$ as
\begin{equation}
\label{eq:rho}
\rho(E)=\frac{G_{\beta}^{2}}{(2 \pi)^{2}} C \sum_{x} n_{x} \beta_{x}^{2} B_{x} S_{x} p_{\nu} E_{\nu} \frac{\Gamma_{x} /(2 \pi)}{\left(E-\varepsilon_{x}\right)^{2}+\Gamma_{x}^{2} / 4},
\end{equation}
where $B_{x}$ and $S_{x}$ are the exchange and overlap corrections, and the shake-up and shake-off corrections, respectively, as discussed in the following subsections. Here we go beyond the formalism used in \cite{ge2021b} by adding the shake-up and shake-off corrections into the energy distribution $\rho(E)$.
Here $E$ is related to the energy of the neutrino $E_{\nu}$ and the Q value as $E=Q^*_{\rm EC}-E_{\nu}$. The momentum of the neutrino is denoted as $p_{\nu}=\sqrt{E_{\nu}^2-m_{\nu}^2}$. The Coulomb amplitude is represented as $\beta_{x}$, while $n_{x}$ is the relative occupancy of the shell. The intrinsic line-widths of Breit--Wigner resonances centered at $\varepsilon_{x}$ are denoted as $\Gamma_{x}$ and are taken from \cite{CampbellADNDT2001}. The Fermi constant $G_{\rm F}$ and the Cabibbo angle $\theta_{\rm C}$ are combined in $G_\beta=G_{\rm F}\cos \theta_{\rm C}$. The nuclear structure information is contained in the shape factor $C$ in terms of the nuclear form factor ${ }^{A} F_{101}^{(0)}$ \cite{Behrens1982}:

\begin{equation}
C=\left[{ }^{A} F_{101}^{(0)}\right]^{2}=\left[-\frac{g_{\rm A}}{\sqrt{2 J_{i}+1}} M_{\mathrm{GT}}\right]^{2},
\end{equation}
where $M_{\mathrm{GT}}$ is the Gamow--Teller nuclear matrix element \cite{JSuhonen2007}. The angular momentum of the initial nucleus is denoted as $J_{i}$, while the strength of the weak axial coupling is represented as $g_{\rm A}$. For this calculation, we took values for $g_{\rm A}$ in the conservative interval 0.7 to 1 \cite{JPCSHonma2006,PRCBarea2013,Suhonen2017} and present the partial half-life corresponding to the mean decay rate in Table \ref{table:T12}. The mean rate corresponds to a $g_{\rm A}$ value equal to $0.857$. The half-life values for all $g_{\rm A}$ are between $-33.3\%$ and $+36.1\%$ of each mean value. 

We define $\lambda(E)$ as the total decay probability in the interval $(0,E)$:
\begin{equation}
\lambda(E)=\int_{0}^{E} \rho\left(E^{\prime}\right) d E^{\prime} 
\end{equation}
and the total decay constant is calculated as $\lambda \equiv \lambda\left(Q^*_{\rm EC}-m_\nu\right)$. After integration, using the narrow-width approximation, the decay rate can be written as 
\begin{equation}
\lambda =\sum_{x}\lambda_{x},
\end{equation}
with the partial decay constant defined as
\begin{equation}
\lambda_{x}=\frac{G_{\beta}^{2}}{(2 \pi)^{2}} C n_{x} \beta_{x}^{2} B_{x} S_{x} p_{\nu x}\left(Q_{\mathrm{EC}}^{*}-\varepsilon_{x}\right),
\end{equation}
where $p_{\nu x}=\sqrt{\left(Q_{\mathrm{EC}}^{*}-\varepsilon_{x}\right)^2-m_{\nu}^2}$.

While an electron is captured by the nucleus, the other electrons (the spectator electrons) can undergo some processes which affect the decay rate. In the following subsections we will present a couple of corrections related to those processes.

\begin{figure*}[!htb]
   \flushleft
   \includegraphics[width=2\columnwidth]{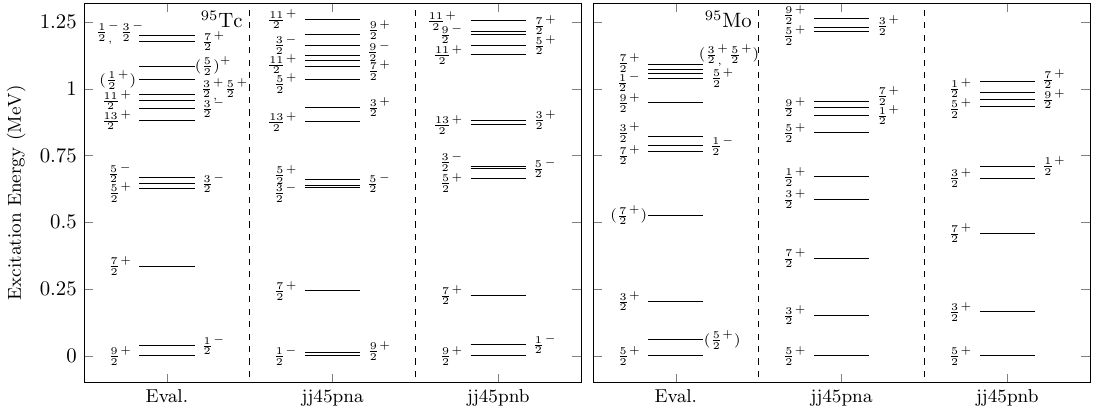}
   \caption{
Computed level schemes of $^{95}$Tc and its daughter $^{95}$Mo using the Hamiltonians \textit{jj45pna} and \textit{jj45pnb}. A comparison with the available data is performed with the parentheses denoting uncertainty in parity and spin-parity assignments. The evaluated experimental energies are gathered from \cite{Basu2010}.
   }
   \label{fig:Level-spectra}
\end{figure*}

\subsection{Exchange and overlap corrections}

In the lowest-order approximation of the EC-decay process an electron is captured from an atomic shell $x$ by the nucleus and a hole in this atomic shell is observed. However, the electrons are identical particles and one cannot distinguish between them leading to the possibility of a higher-order process where first an electron is captured from shell $y$, different from $x$, and simultaneously an electron from shell $x$ is promoted to shell $y$ and, again, the hole is observed in shell $x$.
%an electron can be capture and at the same time another electron in the place of the initially captured electron. 
In both situations the hole appears in the same atomic shell, thus making those two paths for the EC decay indistinguishable from one another. The related higher-order correction is called the exchange correction.

After the EC decay the atomic number of the nucleus decreases by one unit and a hole is left in the daughter nucleus. Those changes affect the spectator electrons and thus their initial $\ket{(m, \kappa)}$ and final $\ket{(m, \kappa)^{\prime}}$ wave functions do not perfectly overlap anymore, i.e., $\bra{(m, \kappa)^{\prime}}\ket{(m, \kappa)}\neq1$.
We use here the generalized formula from \cite{MougeotARI2018}:

\begin{equation}
	B_{n \kappa}=\left|\frac{b_{n \kappa}}{\beta_{n \kappa}}\right|^{2} ,
\end{equation}
where $\beta_{n \kappa}$ is the previously mentioned Coulomb amplitude, and
\begin{eqnarray}
	b_{n \kappa}=&&\left[\prod_{m, \mu}\bra{(m, \mu)^{\prime}}\ket{(m, \mu)}^{n_{m \mu}}\right] \bra{(n, \kappa)^{\prime}}\ket{(n, \kappa)}^{-\frac{1}{2|\kappa|}}\nonumber\\
	&&\times\left[\beta_{n \kappa}-\sum_{\substack{m \neq n}} \beta_{m \kappa} \frac{\bra{(m, \kappa)^{\prime}}\ket{(n, \kappa)}}{\bra{(m, \kappa)^{\prime}}\ket{(m, \kappa)}}\right] .
\end{eqnarray}
This higher-order correction is called the overlap correction.

\subsection{Shake-up and shake-off effects}

During the EC decay, a spectator electron can be promoted to upper unoccupied shells, and this process is called the shake-up effect. The spectator electron can also be ejected to the continuum, and this process is called the shake-off effect. We consider a maximum of one spectator electron undergoing this transition, such that in the final state we would have at most two holes in the atomic shell. We use the formalism developed in \cite{MougeotARI2018} to end up with the following expression for the $S_{x}$ factor of Eq. (\ref{eq:rho})

\begin{equation}
	S_{n \kappa}=1+\sum_{m, \mu}P_{m \mu} .
\end{equation}
Here $P_{m \mu}$ represents the probability that a spectator electron from  $\ket{(m \mu)}$ undergoes the shake-up or shake-off process. A spectator electron can remain in the initial shell, can swap shells with another spectator electron or be promoted or ejected. Thus $P_{m \mu}$ can be computed as unity minus the probability for the electron to remain in the initial shell and the probability to swap shells, as
\begin{eqnarray}
	P_{m \mu}=&&1-|\bra{(m, \mu)^{\prime}}\ket{(m, \mu)}|^{2 n_{m\mu}}\nonumber\\
	&&-\sum_{l \neq m} n_{l \mu}^{\prime} n_{m \mu}|\bra{(l, \mu)^{\prime}}\ket{(m, \mu)}|^2 ,
\end{eqnarray}
where $n_x$ represents the occupancy of the shell $x$.

We also computed the impact of the two corrections on our theoretical calculations. Both the exchange and overlap corrections and the shake-up and shake-off effects increase the total decay rate by 8.1\% and 8.6\%, respectively. The impact of exchange and overlap corrections is 7.5\% and 7.8\%. In comparison, the shake-up and shake-off effects account for 0.61\% and 0.64\%. The first value in each pair of numbers refers to the decay corresponding to $Q^*_{\rm EC}=20.52$ keV while the second one refers to $Q^*_{\rm EC}=12.9$ keV. This increase in the decay rate is the result of the fact that both corrections open new channels for the decay process.

\begin{table*}[ht]
\fontsize{8}{9}\selectfont
\caption{Comparison of the experimental excitation energies $E_{\rm exc}$ (measured in keV), electric quadrupole moments $Q$ (measured in $e$barns), and magnetic dipole moments $\mu$ (measured in nuclear magnetons, $\mu_N$), with those computed by the NSM using the \textit{jj45pna}, \textit{jj45pnb}, and \textit{glekpn} Hamiltonians. The experimental evaluated energies are sourced from \cite{Basu2010} and the nuclear moments from \cite{Hinfurtner1995,Proctor1951,Alzner1984,Stone2021}. The calculations employed effective charges of $e_{\rm eff}^p = 1.5e$ for protons and $e_{\rm eff}^n = 0.5e$ for neutrons, and the bare g-factors $g_l(p) = 1$, $g_l(n) = 0$, $g_s(p) = 5.585$, and $g_s(n) = -3.826$ were used for determining the nuclear moments.}
\label{table:EM-observables}
\begin{tabular*}{0.99\textwidth}{@{\extracolsep{\fill}}lclr|ccc|ccc|ccc}
\toprule
\multicolumn{4}{c|}{Experimental Evaluation} & \multicolumn{3}{c|}{jj45pna} & \multicolumn{3}{c|}{jj45pnb} & \multicolumn{3}{c}{glekpn} \\
\makecell[c]{$\text{Isotope}$ \\ $(J^\pi)$} & \makecell[c]{$E_{\rm exc}$ \\ (keV)} & \makecell[c]{$Q$ \\ ($e$barn)} & \makecell[c]{$\mu$ \\ ($\mu_N$)} & \makecell[c]{$E_{\rm exc}$ \\ (keV)} & \makecell[c]{$Q$ \\ ($e$barn)} & \makecell[c]{$\mu$ \\ ($\mu_N$)} & \makecell[c]{$E_{\rm exc}$ \\ (keV)} & \makecell[c]{$Q$ \\ ($e$barn)} & \makecell[c]{$\mu$ \\ ($\mu_N$)} & \makecell[c]{$E_{\rm exc}$ \\ (keV)} & \makecell[c]{$Q$ \\ ($e$barn)} & \makecell[c]{$\mu$ \\ ($\mu_N$)} \\
\midrule
$^{95}$Tc \phantom{-}$(\frac{9}{2}^{+})$ & 0 & - & 5.94(6) & 11 & -0.1392 & 6.3286 & 0 & -0.1508 & 6.3342 & 0 & 0.0784 & 6.3127 \\
$^{95}$Mo $(\frac{5}{2}^{+})$ & 0 & -0.022(1) & -0.9132(3) & 0 & -0.0379 & -0.9418 & 0 & -0.0238 & -0.9768 & 574 & -0.3359 & -0.8376 \\
$^{95}$Mo $(\frac{3}{2}^{+})$ & 204 & - & -0.404(12) & 151 & 0.0672 & -0.0734 & 165 & 0.0656 & -0.1335 & 816 & -0.0782 & 0.6095 \\
\bottomrule
\end{tabular*}
\end{table*}

\subsection{Nuclear Matrix Elements}

For the nuclear-structure calculation we employed the \textit{NuShellX@MSU} \cite{Brown2014} code using the shell-model formalism. We used interactions \textit{jj45pna} \cite{Machleidt2001}, a two-nucleon potential with a perturbative G-matrix approach with the single particle energies adjusted in the Coulomb part to reproduce the recent results in \cite{Vaquero2020} and \textit{jj45pnb} \cite{Lisetskiy2004}, both sharing the same jj45pn model space consisting of the proton $0f_{5/2}$-$1p_{3/2}$-$1p_{1/2}$-$1g_{9/2}$ orbitals and the neutron $0g_{7/2}$-$1d_{5/2}$-$1d_{3/2}$-$2s_{1/2}$-$0h_{11/2}$ orbitals with no truncations. Additionally, we employed the \textit{glekpn} \cite{Mach1990} interaction with the model space consisting of $0f_{7/2}$-$0f_{5/2}$-$1p_{3/2}$-$1p_{1/2}$-$1g_{9/2}$ proton orbitals and $0g_{9/2}$-$0g_{7/2}$-$1d_{5/2}$-$1d_{3/2}$-$2s_{1/2}$ neutron orbitals. For this model space, we applied truncations by fixing the occupation of the $0f_{7/2}$ proton orbital to 8 protons and the $0g_{9/2}$ neutron orbital to 10 neutrons, resulting in a shell-model closed core up to the magic numbers 28 for protons and 50 for neutrons. Lastly, for this interaction three sets of single-particle energies are available, we utilized the one suitable for the mass region $A=94-98$. To assess the validity of the applied interactions, we have computed the energy-level schemes for both jj45pn model-space interactions as can be seen in Figure \ref{fig:Level-spectra}. The \textit{glekpn} interaction did not reproduce the level schemes correctly so that the related energy spectrum is not shown here.
The results indicate that the \textit{jj45pnb} interaction provides better agreement with the evaluated data. 

Further, we have compared the nuclear moments obtained using the three interactions with available evaluated experimental data, as presented in Table \ref{table:EM-observables}. Both the \textit{jj45pna} and \textit{jj45pnb} interaction showed a higher level of agreement with the observed values, showcasing their applicability and efficiency for the current study, whereas the Hamiltonian \textit{glekpn} performs quite poorly. 
%didn not reproduce the level schemes correctly, as seen even in the few states in Table \ref{table:EM-observables}. Frequently, with energy differences of more than 400 keV between the computed and experimental energies. 
Nevertheless, engaging the \textit{glekpn} interaction demonstrates the importance of exploring different interactions to fully understand their performance and limitations in this nuclear mass region.

%We computed energy levels of the daughter nucleus in order to identify the theoretical states corresponding to the experimental states of interest ($1675.4$ keV, $1683$ keV). We compared the theoretical energies with the experimental ones. We concluded that the theoretical states closest to the experimental ones are the following for the \textit{jj45pnb} interaction: a $J_{f}=9/2$ state with the energy $E^{*}_{\rm th}=1897$ keV corresponds closest to the experimental state $E^{*}=1675.4$ keV, and the $J_{f}=7/2$ state at $E^{*}_{\rm th}=1642$ keV matches best the $E^{*}=1683$ keV experimental state, while for the \textit{glekpn} interaction: a state with the energy $E^{*}_{\rm th}=1703$ keV and $J_{f}=9/2$ corresponds best with the experimental state $E^{*}=1675.4$ keV, and the $J_{f}=7/2$ state at $E^{*}_{\rm th}=1707$ keV matches the closest to the $E^{*}=1683$ keV experimental state.

\subsection{Results of the calculations}

We computed the energy levels of the daughter nucleus to identify the theoretical states corresponding to the experimental states of interest ($1675.4$ keV, $1683$ keV). We compared the theoretical energies with the experimental ones and concluded that the following are the theoretical states closest to the experimental ones: for the experimental state with $J_{f}=9/2$ and $E^{*}=1675.4$ keV the best matches were $E^{*}_{\rm th}=1748$ keV for \textit{jj45pna}, $E^{*}_{\rm th}=1897$ keV for \textit{jj45pnb}, and $E^{*}_{\rm th}=1703$ keV for \textit{glekpn}. All the mentioned theoretical states have $J_{f}=9/2$. For the experimental state having the energy $E^{*}=1683$ keV and with the angular momentum and parity uncertain \{7/2$^{+}$, 9/2$^{+}$\}, the closest correspondence is for \textit{jj45pna} the energy $E^{*}_{\rm th}=1584$ keV, for \textit{jj45pnb} the energy $E^{*}_{\rm th}=1642$ keV, and for \textit{glekpn} the energy $E^{*}_{\rm th}=1707$ keV. The mentioned three theoretical states have the angular momentum and parity $7/2^{+}$.

%An important nucleus in the search of neutrino mass through EC capture is $^{163}$Ho \cite{Gastaldo2017}. The $Q^*_{EC}$ of $^{163}$Ho is 2.831 keV \cite{Wang2021} and thus the normalized distributions of released energy in the EC decay of $^{163}$Ho is 4 orders of magnitudes bigger than the spectra of $^{95}$Tc close to $E-Q^{*}_{EC}=0$

%-----------------------------Table 3--------------------------------
%%%%%%%%%%%%%%%%%%%%%%%
\begin{table*}[!htb]
%\begin{table*}[!]
%\small
 \fontsize{8}{9}\selectfont
\caption{Computed mean half-lives using $g_{\rm A}=0.857$ (see the main text) for the EC decay of $^{95}$Tc to the two excited states in $^{95}$Mo (with experimental energies $E^{*}=1675.4$ keV and $1683$ keV), using three shell-model interactions for the Gamow-Teller matrix element, with their experimental $Q$ values shown in column 1. The second column indicates the used interactions and the third column the Gamow--Teller nuclear matrix element \cite{JSuhonen2007}. The computed total half-life and partial half-lives are demonstrated in columns 4-12. The atomic subshells are denoted using the X-ray notation.
   }
    \begin{tabular*}{\textwidth}{@{\extracolsep{\fill}} lccccccccccc}
  \toprule
 $Q^*_{\rm EC}$ & interaction & $M_{\rm GT}$
 & Total half-life &$\mathrm{~K} $ & $\mathrm{~L} 1$ & $\mathrm{~L}2$ & $\mathrm{M}1$ & $\mathrm{M} 2$ & $\mathrm{N} 1$ & $\mathrm{N} 2$ & $\mathrm{O} 1$\\
 (keV)& & & ($10^{3}$yr) & ($10^{7}$yr) & ($10^{4}$yr) & ($10^{5}$yr) & ($10^{4}$yr) & ($10^{6}$yr) & ($10^{5}$yr) & ($10^{7}$yr) & ($10^{6}$yr)\\
   \midrule
%&&&AME2020&AME2020&this  work&this work&&&\\
% & d$Q_{\mathrm{EC}}$ & E^{*}_{error}
%\hline\noalign{\smallskip}
 $20.52$ & jj45pna & $-0.00696$ & $877$ & $20.0$ & $117$ & $708$ & $640$ & $267$ & $223$ & $146$ & $491$\\
& jj45pnb & -0.016533 & $155$ & $3.55$ & $20.8$ & $125$ & $81.6$ & $47.4$ & $39.4$ & $25.8$ & $87.1$ \\
& glekpn & 0.0070667 & $850$ & $19.4$& $114$ & $587$ & $446$ & $259$ & $216$ & $141$ & $477$\\
\hline
12.9 & jj45pna & $-0.02412$ & $173$ & - & $24.1$ & $143$ & $79.7$ & $46.0$ & $37.5$ & $24.5$ & $82.6$\\
&  jj45pnb & -0.0198667 & $255$ & - & $35.6$ & $210$ & $117$ & $67.9$ & $55.3$ & $36.1$ & $122$ \\
 & glekpn & 0.2296 & $1.9$ & - & $0.266$ & $1.57$ & $0.880$ & $0.508$ & $0.414$& $0.271$ & $0.911$ \\
     \bottomrule
      \end{tabular*}
   \label{table:T12}
%\end{ruledtabular}
%\end{tabular}
\end{table*}

%%%%%%%%%%%%%%%%%%%%%%%
%-----------------------------Table 3 --------------------------------

\begin{figure}[!htb]
   %\flushleft
   \includegraphics[width=0.99\columnwidth]{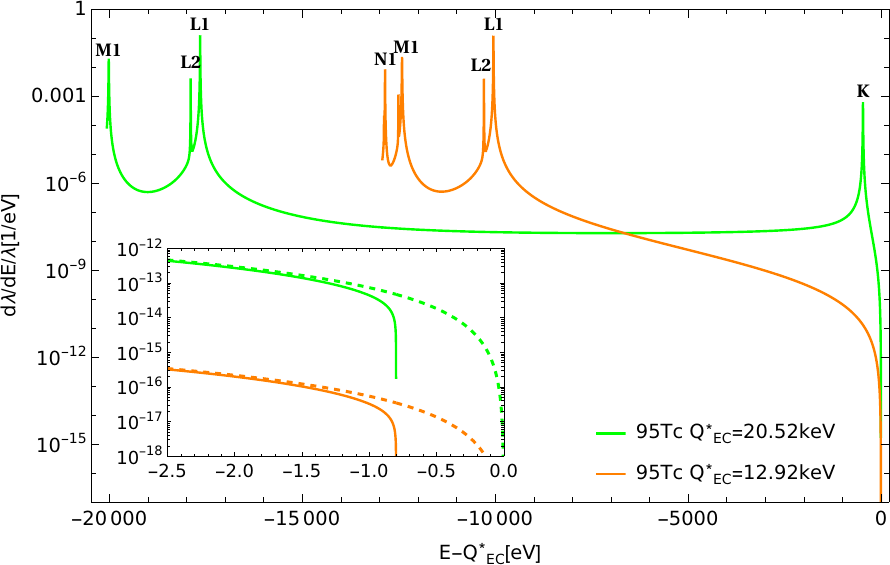}
   %{Fig4.png}95Tc_Q_pm_dQ.png  is the one standard deviation from this work
   \caption{
   Normalized distributions of released energy in the EC decay of $^{95}$Tc in the transitions to the excited states of $^{95}$Mo, as functions of $E-Q^{*}_{\rm EC}$. The experimental excitation energies are $E^{*}=1675.40$ keV and $E^{*}=1683.0$ keV, while the corresponding $Q$ values are $Q^{*}_{\rm EC}=20.52$ keV (green) and $Q^{*}_{\rm EC}=12.9$ keV (orange).
   %The black line depicts the transition with $Q^{*}_{EC}=12.9$ keV. The magenta line corresponds to the decay using the experimental $Q^{*}_{EC}$ of 20.52 keV, while the green and the orange line are associated to the $Q^{*}_{EC}$ of 19.91 keV and 21.13 keV ($Q^{*}_{EC} \pm \sigma (Q^{*}_{EC}$); $\sigma (Q^{*}_{EC}$) = 0.61 keV). 
   The K, L1, L2, M1 and N1 notations indicate sub-shells from which the electron was captured. The M2, N1 N2 and O1 sub-shells are harder to distinguish and are not labeled. The inset indicates an enlarged endpoint region showing the effect of neutrino masses of 0.8 eV and 0 eV. The dotted lines depict the spectra for a massless neutrino, while the solid lines correspond to a neutrino mass of 0.8 eV. 
  % \textcolor{red}{Vasile: could you add the K, L1, L2, M1, and N1 labels in the figures? If the color of yellow would be changed to black, it will be nice.}
   %Normalized distributions of released energy in the EC decay of $^{95}$Tc in the transitions to the two excited states of $^{95}$Mo, of interest here, as functions of $E-Q^{*}_{EC}$. The blue line corresponds to the decay to the excited final state of energy $E^{*}=1675.4$ keV and with  $Q^{*}_{EC}=28.9$ keV, while the orange line is associated to the $E^{*}=1683$ keV energy level with $Q^{*}_{EC}=12.9$ keV. K, L1, L2, M1, and N1 indicate subshells from which the electron was captured. The M2, N2 and O1 subshells are harder to distinguish and are not labeled.
   }
   \label{fig:4-spectra}
\end{figure}

\begin{figure}[!htb]
   %\flushleft
\includegraphics[width=0.99\columnwidth]{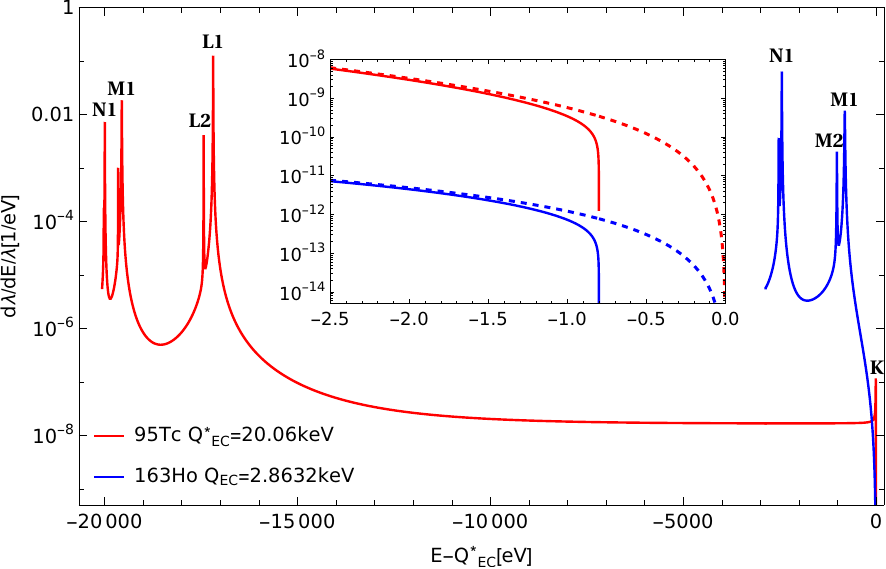}
   %{Fig5.png}95Tc_163Ho.png
\caption{
Normalized distribution of released energy in the EC decay of $^{95}$Tc in the transition to the excited state $E^{*}=1675.40$ keV of $^{95}$Mo, with $Q^{*}_{\rm EC}=20.06$ keV, as function of $E-Q^{*}_{\rm EC}$ in comparison to that of $^{163}$Ho with gs-to-gs $Q^{}_{\rm EC}=2.8632$ keV \cite{Schweiger2024}.
The red line corresponds to the on-resonance EC decay using the experimental $Q^{*}_{EC}$ of 20.06 keV within 1$\sigma$ range of the central value 20.52 keV, while the blue line corresponds to the $Q^{*}_{\rm EC}$ of 2.8632 keV. 
%The K, L1, L2, M1, and N1 notation indicate subshells from which the electron was captured. The M2, N2 and O1 subshells are harder to distinguish and are not labeled. The inset indicates the enlarged detail showing the effect of neutrino masses of 0.8 eV (full lines) and 0 eV (dotted lines).
For the sub-shell notation and the inset the reader is referred to the caption of Fig. \ref{fig:4-spectra}
%\textcolor{red}{ A recent nature paper give the  $^{163}$Ho Q value of 2863.2±0.6 eV. See: https://www.nature.com/articles/s41567-024-02461-9. Likely we need to change 2.555 keV to 2.863 keV }
%Enlarged detail of Fig. \ref{fig:4-spectra} showing the effect of neutrino masses of 1 and 0 eV. The dotted lines depict the spectra for a massless neutrino, while the solid lines correspond to a neutrino mass of 1 eV. The colors have the same interpretation as in Fig.\ref{fig:4-spectra}.
}
   \label{fig:resonance-vs-163Ho}
\end{figure}

In Table \ref{table:T12}, we present the predicted half-lives for the decay of $^{95}$Tc to the two excited states of $^{95}$Mo with the experimental energies $E^{*}=1675.4$ keV and $E^{*}=1683$ keV for all relevant atomic shells and in Fig. \ref{fig:4-spectra}, the normalized distribution of the released energy in the EC decay of $^{95}$Tc to excited states of $^{95}$Mo 
%as functions of $E-Q^{*}_{EC}$,  
is demonstrated. 
The transition spectrum of $^{95}$Tc (9/2$^+$) $\rightarrow$ $^{95}$Mo$^*$ (1675.4 keV) is indicated in green, with $Q_{\mathrm{EC}}^{*}$ 
%associated to the $Q^{*}_{EC}$ of 19.91 keV and 21.13 keV ($Q^{*}_{EC} \pm \sigma (Q^{*}_{EC}$), $\sigma (Q^{*}_{EC}$) = 0.61 keV).
of 20.52 keV, situated 0.47 keV relative to the computed atomic relaxation energy following the electron capture in the allowed K shell. 
%, is compared to 
%those of $Q^{*}_{EC} \pm 0.61$ keV for which . 
%(21.13 keV and 19.91 keV) 
In contrast, the transition  $^{95}$Tc (9/2$^+$) $\rightarrow$ $^{95}$Mo$^*$ (1683.0 keV), shown in orange, with a $Q_{\mathrm{EC}}^{*}$  value of 12.9 keV, is relatively farther from the computed atomic relaxation energy following the electron
capture of the allowed L1 shell (giving a value of 10 keV for the distance).
As illustrated in the inset of Fig. \ref{fig:4-spectra}, a more pronounced resonance enhancement in the last 0.8 eV region near the endpoint for the former transition is observed, suggesting a preference for choosing this transition as a candidate for determining a non-vanishing neutrino mass. 
%\ref{fig:5-zoom-spectra}
%\textcolor{red}{ORIGINAL: This phenomenon guides us to search for as low a distance of $Q_{\mathrm{EC}}^{*}$ to the highest ionization energy of the captured electron of all allowed shells for candidate low $Q$-value EC transitions using for neutrino mass determination experiments. [I don't understand this sentence. Do you try to say that we want to find a decay branch that has such a Q-value that is very close to one of the capture lines? ] }
%\textcolor{red} this is a better description and clear^^
{This phenomenon guides us to search for cases that have the smallest distance of $Q_{\mathrm{EC}}^{*}$ to the highest ionization energy of the captured electron of all allowed shells for neutrino-mass determination experiments.}

The decay rate close to the endpoint is highly sensitive to small variations of the $Q$ value as demonstrated by the $1s$ level, K, manifesting as a resonance itself, and thus can radically increase the number of recorded events near the endpoint. The  current accuracy of measurement of the $Q_{\mathrm{EC}}^{*}$ value does not allow to make an unambiguous conclusion about the position of the $1s$ level, relative to the endpoint. 
Assuming $Q_{\mathrm{EC}^{*}} = 20.06$ keV, which is consistent with the 1 $\sigma$ range of experimental error and is shifted by -0.46 keV relative to the central value, the resonance  at the endpoint provides the highest EC event rate in the neutrino-mass sensitive region.
% significantly increases the EC counting 
Figure \ref{fig:resonance-vs-163Ho} shows the normalized EC energy spectrum as a function of the energy, $E-Q_{\mathrm{EC}}^{*}$, deposited in a calorimeter through the de-excitation of atomic shells for $Q_{\mathrm{EC}}^{*} = 20.06$ keV, in comparison to the EC energy spectra in  $^{163}$Ho atom. The $s1$ level, K, has a significant EC counting-rate enhancement for the transition $^{95}$Tc (9/2$^{+}$) $\rightarrow$ $^{95}$Mo$^{*}$ (1675.40 keV) in the neutrino-mass sensitive region, as shown on an enlarged scale in the inset of Fig. \ref{fig:resonance-vs-163Ho}. Assuming a $Q$ value of $Q_{\mathrm{EC}}^{*} = 20.06$ keV, technetium is about three orders of magnitude more effective than holmium. Based on these findings, it can be conjectured that we have found a potentially strong transition for direct electron-neutrino mass determination. However, the short half-life of about 1 day for the $^{95}$Tc can prove to be a challenge experimentally.
%However, $^{95}$Tc is radioactive with a half-life of about 1 day, rendering a usage of this atom for measurement of the electron neutrino mass a challenge.
%with $Q_{\mathrm{EC}^{}} = 2.8632$ eV.
%from Refs. ~\cite{ge2021b,Gastaldo2017}
 %can enter the physical region
% the accuracy of measurement of the $Q$ value does not allow to make an unambiguous conclusion about the position of the $2p_{1/2}$ level relative to the endpoint.
%with a QEC value of  keV, on the other hand, has the neutrino energy Eνx= keV for the electron M1-capture and a promisingly short decay half-life in order to become a potential candidate for future neutrino-mass measurements.
%low emitted neutrino energy is ultra-low

\section{Conclusion and outlook}

A direct high-precision gs-to-gs  EC-decay $Q$-value measurement of $^{95}$Tc (9/2$^{+}$) $\rightarrow$ $^{95}$Mo (5/2$^{+}$)  %and $^{96}$Tc (7$^{+}$)$\rightarrow $$^{96}$Mo (0$^{+}$) 
was performed using the PI-ICR technique at the JYFLTRAP Penning trap mass spectrometer.  A $Q$ value of 1695.92(13)~keV  was obtained and the  precision  was improved by a factor of around 37 compared to literature. The measurement also improved the mass excess of $^{95}$Tc by a factor of 28 compared to previous experiments.
%A discrepancy of around 10 standard deviations was found compared to the adopted value in the AME2020. the smallest  distance 
Three candidate transitions of $^{95}$Tc (9/2$^{+}$) $\rightarrow$ $^{95}$Mo$^{*}$ were validated to be energetically allowed. 
%The gs-to-gs  $Q$ value of $^{95}$Tc is now well refined to sub-keV uncertainty, combined with the energy level of 1675.40-keV state, an ultra-low distance (0.52 keV) of $Q_{\mathrm{EC}}^{*}$ to the allowed binding energy $\varepsilon_{k}$ = 20.000	keV \cite{X-Ray_Data_Booklet} is discovered.
The refined sub-keV precision of the gs-to-gs  $Q$ value of $^{95}$Tc allows us to find an ultra-low energy difference (0.47 keV) between $Q_{\mathrm{EC}}^{*}$ and the atomic relaxation energy  $\varepsilon_{k}$ = 20.054 keV in the K capture for the allowed gs-to-es transition to the 1675.40-keV state.  
%The allowed transition to the 1675.4-keV state is still of high interest, but the excitation energy is to be measured with a higher precision. 
%The transition to the 1692-keV state with an 4$^{\rm th}$ FNU decay type, has a small 
%distance of $\approx$  1 keV for the gs-to-es $Q_{\mathrm{EC}}^{*}$  to the computed atomic relaxation energy following the electron capture in the allowed daughter atomic shell L1. 
%For further confirmation of whether this low emitted neutrino energy is ultra-low, the excitation energy needs to be determined accurately.  
The spin of the 1683.0-keV state needs to be determined along with its energy with higher precision in order to see if the related transition is allowed and of low $Q$ value.
%Furthermore, We confirmed that both of the two potential ultra-low $Q$-value $\beta^-$-decay transitions, $^{96}$Tc (7$^{+}$)$\rightarrow $$^{96}$Mo$^{*}$ (5$^{+}$, 2975.280(70) keV; 8$^{+}$, 2978.370(80) keV),  are energetically forbidden at the $\approx$200$\sigma$ level. 
%This findings underline the need to measure the $Q$-values to high precision before attempts to detect such possible low $Q$-value decay branches is made, with the goal to realize these decays for neutrino mass determination.  %Moreover, with the high resolving power of PI-ICR technique to have identify the isomeric state of  $^{96}$Tc, we verify that the measurement from the storage ring mass spectrometry by additional adjustment of mixture of two states, is correct and in good agreement with his work thus raise doubts on the reliability of input data of $^{96}$Tc used in global evaluations of atomic masses AME2020. 
%\section{Acknowledgements}

The atomic self-consistent many-electron Dirac--Hartree--Fock--Slater method and three nuclear shell-model interactions were utilized to predict the partial decay half-lives and energy distributions of gs-to-es EC transitions in $^{95}$Tc with low $Q$ values. We computed the energy levels of the parent and daughter nuclei and a few electromagnetic moments to assess the validity of the three shell-model interactions (\textit{jj45pna}, \textit{jj45pnb}, \textit{glekpn}). Multiple corrections, such as exchange, overlap, shake-up, and shake-off effects, were accounted for in these predictions. 
%The energy-release distribution during the EC decay of $^{95}$Tc to excited states of $^{95}$Mo was normalized and compared to that of $^{163}$Ho.
%The normalized energy release distributions in the EC decay of $^{95}$Tc of interest are found to be four orders of magnitude smaller compared to the spectra of $^{163}$Ho near $E-Q^{}_{EC}=0$, resulting this transition less competitive.
%The event ratio at the last 1eV of the normalized energy release distributions in the the allowed gs-es EC transition of $^{95}$Tc to the 1675.40-keV state, are found to be about 2 orders of magnitude larger compared to the spectra of $^{163}$Ho near $E-Q^{}_{EC}=0$, resulting this transition very competitive. 1p1/2 Proximity of QEC and atomic lines K
From the calculations, the ultra-low distance to the atomic line K, level 1s, for $^{95}$Tc (9/2$^{+}$) $\rightarrow$ $^{95}$Mo$^{*}$ (1675.40 keV) results in a significant increase in the number of EC events in the energy region sensitive to the electron neutrino mass. These findings confirm a potentially powerful transition for direct electron-neutrino mass determination.

%\acknowledgments 

\bibliography{my-final-bib-from-jabref}

\begin{thebibliography}{79}
\providecommand{\natexlab}[1]{#1}
\providecommand{\url}[1]{\texttt{#1}}
\providecommand{\href}[2]{#2}
\providecommand{\path}[1]{#1}
\providecommand{\DOIprefix}{doi:}
\providecommand{\ArXivprefix}{arXiv:}
\providecommand{\URLprefix}{URL: }
\providecommand{\Pubmedprefix}{pmid:}
\providecommand{\doi}[1]{\href{http://dx.doi.org/#1}{\path{#1}}}
\providecommand{\Pubmed}[1]{\href{pmid:#1}{\path{#1}}}
\providecommand{\BIBand}{and}
\providecommand{\bibinfo}[2]{#2}
\ifx\xfnm\undefined \def\xfnm[#1]{\unskip,\space#1}\fi
\makeatletter\def\@biblabel#1{#1.}\makeatother
%Type = Article
\bibitem[{Fukuda et~al.(1998)Fukuda, Hayakawa, Ichihara, Inoue, Ishihara,
  Ishino, Itow, Kajita, Kameda, Kasuga, Kobayashi, Kobayashi, Koshio, Miura,
  Nakahata, Nakayama, Okada, Okumura, Sakurai, Shiozawa, Suzuki, Takeuchi,
  Totsuka, Yamada, Earl, Habig, Kearns, Messier, Scholberg, Stone, Sulak,
  Walter, Goldhaber, Barszczxak, Casper, Gajewski, Halverson, Hsu, Kropp,
  Price, Reines, Smy, Sobel, Vagins, Ganezer, Keig, Ellsworth, Tasaka,
  Flanagan, Kibayashi, Learned, Matsuno, Stenger, Takemori, Ishii, Kanzaki,
  Kobayashi, Mine, Nakamura, Nishikawa, Oyama, Sakai, Sakuda, Sasaki, Echigo,
  Kohama, Suzuki, Haines, Blaufuss, Kim, Sanford, Svoboda, Chen, Conner,
  Goodman, Sullivan, Hill, Jung, Martens, Mauger, {Mc Grew}, Sharkey, Viren,
  Yanagisawa, Doki, Miyano, Okazawa, Saji, Takahata, Nagashima, Takita,
  Yamaguchi, Yoshida, Kim, Etoh, Fujita, Hasegawa, Hasegawa, Hatakeyama,
  Iwamoto, Koga, Maruyama, Ogawa, Shirai, Suzuki, Tsushima, Koshiba, Nemoto,
  Nishijima, Futagami, Hayato, Kanaya, Kaneyuki, Watanabe, Kielczewska, Doyle,
  George, Stachyra, Wai, Wilkes and Young}]{Fukuda1998}
\bibinfo{author}{Fukuda\xfnm[ Y.]}, \bibinfo{author}{Hayakawa\xfnm[ T.]},
  \bibinfo{author}{Ichihara\xfnm[ E.]}, \bibinfo{author}{Inoue\xfnm[ K.]},
  \bibinfo{author}{Ishihara\xfnm[ K.]}, \bibinfo{author}{Ishino\xfnm[ H.]},
  \bibinfo{author}{Itow\xfnm[ Y.]}, \bibinfo{author}{Kajita\xfnm[ T.]},
  \bibinfo{author}{Kameda\xfnm[ J.]}, \bibinfo{author}{Kasuga\xfnm[ S.]},
  \bibinfo{author}{Kobayashi\xfnm[ K.]}, \bibinfo{author}{Kobayashi\xfnm[ Y.]},
  \bibinfo{author}{Koshio\xfnm[ Y.]}, \bibinfo{author}{Miura\xfnm[ M.]},
  \bibinfo{author}{Nakahata\xfnm[ M.]}, \bibinfo{author}{Nakayama\xfnm[ S.]},
  \bibinfo{author}{Okada\xfnm[ A.]}, \bibinfo{author}{Okumura\xfnm[ K.]},
  \bibinfo{author}{Sakurai\xfnm[ N.]}, \bibinfo{author}{Shiozawa\xfnm[ M.]},
  \bibinfo{author}{Suzuki\xfnm[ Y.]}, \bibinfo{author}{Takeuchi\xfnm[ Y.]},
  \bibinfo{author}{Totsuka\xfnm[ Y.]}, \bibinfo{author}{Yamada\xfnm[ S.]},
  \bibinfo{author}{Earl\xfnm[ M.]}, \bibinfo{author}{Habig\xfnm[ A.]},
  \bibinfo{author}{Kearns\xfnm[ E.]}, \bibinfo{author}{Messier\xfnm[ M.D.]},
  \bibinfo{author}{Scholberg\xfnm[ K.]}, \bibinfo{author}{Stone\xfnm[ J.L.]},
  \bibinfo{author}{Sulak\xfnm[ L.R.]}, \bibinfo{author}{Walter\xfnm[ C.W.]},
  \bibinfo{author}{Goldhaber\xfnm[ M.]}, \bibinfo{author}{Barszczxak\xfnm[
  T.]}, \bibinfo{author}{Casper\xfnm[ D.]}, \bibinfo{author}{Gajewski\xfnm[
  W.]}, \bibinfo{author}{Halverson\xfnm[ P.G.]}, \bibinfo{author}{Hsu\xfnm[
  J.]}, \bibinfo{author}{Kropp\xfnm[ W.R.]}, \bibinfo{author}{Price\xfnm[
  L.R.]}, \bibinfo{author}{Reines\xfnm[ F.]}, \bibinfo{author}{Smy\xfnm[ M.]},
  \bibinfo{author}{Sobel\xfnm[ H.W.]}, \bibinfo{author}{Vagins\xfnm[ M.R.]},
  \bibinfo{author}{Ganezer\xfnm[ K.S.]}, \bibinfo{author}{Keig\xfnm[ W.E.]},
  \bibinfo{author}{Ellsworth\xfnm[ R.W.]}, \bibinfo{author}{Tasaka\xfnm[ S.]},
  \bibinfo{author}{Flanagan\xfnm[ J.W.]}, \bibinfo{author}{Kibayashi\xfnm[
  A.]}, \bibinfo{author}{Learned\xfnm[ J.G.]}, \bibinfo{author}{Matsuno\xfnm[
  S.]}, \bibinfo{author}{Stenger\xfnm[ V.J.]}, \bibinfo{author}{Takemori\xfnm[
  D.]}, \bibinfo{author}{Ishii\xfnm[ T.]}, \bibinfo{author}{Kanzaki\xfnm[ J.]},
  \bibinfo{author}{Kobayashi\xfnm[ T.]}, \bibinfo{author}{Mine\xfnm[ S.]},
  \bibinfo{author}{Nakamura\xfnm[ K.]}, \bibinfo{author}{Nishikawa\xfnm[ K.]},
  \bibinfo{author}{Oyama\xfnm[ Y.]}, \bibinfo{author}{Sakai\xfnm[ A.]},
  \bibinfo{author}{Sakuda\xfnm[ M.]}, \bibinfo{author}{Sasaki\xfnm[ O.]},
  \bibinfo{author}{Echigo\xfnm[ S.]}, \bibinfo{author}{Kohama\xfnm[ M.]},
  \bibinfo{author}{Suzuki\xfnm[ A.T.]}, \bibinfo{author}{Haines\xfnm[ T.J.]},
  \bibinfo{author}{Blaufuss\xfnm[ E.]}, \bibinfo{author}{Kim\xfnm[ B.K.]},
  \bibinfo{author}{Sanford\xfnm[ R.]}, \bibinfo{author}{Svoboda\xfnm[ R.]},
  \bibinfo{author}{Chen\xfnm[ M.L.]}, \bibinfo{author}{Conner\xfnm[ Z.]},
  \bibinfo{author}{Goodman\xfnm[ J.A.]}, \bibinfo{author}{Sullivan\xfnm[
  G.W.]}, \bibinfo{author}{Hill\xfnm[ J.]}, \bibinfo{author}{Jung\xfnm[ C.K.]},
  \bibinfo{author}{Martens\xfnm[ K.]}, \bibinfo{author}{Mauger\xfnm[ C.]},
  \bibinfo{author}{{Mc Grew}\xfnm[ C.]}, \bibinfo{author}{Sharkey\xfnm[ E.]},
  \bibinfo{author}{Viren\xfnm[ B.]}, \bibinfo{author}{Yanagisawa\xfnm[ C.]},
  \bibinfo{author}{Doki\xfnm[ W.]}, \bibinfo{author}{Miyano\xfnm[ K.]},
  \bibinfo{author}{Okazawa\xfnm[ H.]}, \bibinfo{author}{Saji\xfnm[ C.]},
  \bibinfo{author}{Takahata\xfnm[ M.]}, \bibinfo{author}{Nagashima\xfnm[ Y.]},
  \bibinfo{author}{Takita\xfnm[ M.]}, \bibinfo{author}{Yamaguchi\xfnm[ T.]},
  \bibinfo{author}{Yoshida\xfnm[ M.]}, \bibinfo{author}{Kim\xfnm[ S.B.]},
  \bibinfo{author}{Etoh\xfnm[ M.]}, \bibinfo{author}{Fujita\xfnm[ K.]},
  \bibinfo{author}{Hasegawa\xfnm[ A.]}, \bibinfo{author}{Hasegawa\xfnm[ T.]},
  \bibinfo{author}{Hatakeyama\xfnm[ S.]}, \bibinfo{author}{Iwamoto\xfnm[ T.]},
  \bibinfo{author}{Koga\xfnm[ M.]}, \bibinfo{author}{Maruyama\xfnm[ T.]},
  \bibinfo{author}{Ogawa\xfnm[ H.]}, \bibinfo{author}{Shirai\xfnm[ J.]},
  \bibinfo{author}{Suzuki\xfnm[ A.]}, \bibinfo{author}{Tsushima\xfnm[ F.]},
  \bibinfo{author}{Koshiba\xfnm[ M.]}, \bibinfo{author}{Nemoto\xfnm[ M.]},
  \bibinfo{author}{Nishijima\xfnm[ K.]}, \bibinfo{author}{Futagami\xfnm[ T.]},
  \bibinfo{author}{Hayato\xfnm[ Y.]}, \bibinfo{author}{Kanaya\xfnm[ Y.]},
  \bibinfo{author}{Kaneyuki\xfnm[ K.]}, \bibinfo{author}{Watanabe\xfnm[ Y.]},
  \bibinfo{author}{Kielczewska\xfnm[ D.]}, \bibinfo{author}{Doyle\xfnm[ R.A.]},
  \bibinfo{author}{George\xfnm[ J.S.]}, \bibinfo{author}{Stachyra\xfnm[ A.L.]},
  \bibinfo{author}{Wai\xfnm[ L.L.]}, \bibinfo{author}{Wilkes\xfnm[ R.J.]},
  \bibinfo{author}{Young\xfnm[ K.K.]}.
\newblock \bibinfo{title}{{Evidence for oscillation of atmospheric neutrinos}}.
\newblock \emph{\bibinfo{journal}{Physical Review Letters}}
  \bibinfo{year}{1998};\bibinfo{volume}{81}(\bibinfo{number}{8}):\bibinfo{pages}{1562--1567}.
\newblock \URLprefix \url{http://link.aps.org/doi/10.1103/PhysRevLett.81.1562}.
  \DOIprefix\doi{10.1103/PhysRevLett.81.1562}.
  \href{http://arxiv.org/abs/9807003}{\tt arXiv:9807003}.
%Type = Article
\bibitem[{{SNO Collaboration}(2002)}]{SNOCollaboration2002}
\bibinfo{author}{{SNO Collaboration}\xfnm[]}.
\newblock \bibinfo{title}{{Direct Evidence for Neutrino Flavor Transformation
  from Neutral-Current Interactions in the Sudbury Neutrino Observatory}}.
\newblock \emph{\bibinfo{journal}{Physical Review Letters}}
  \bibinfo{year}{2002};\bibinfo{volume}{89}(\bibinfo{number}{1}):\bibinfo{pages}{1--6}.
\newblock \URLprefix \url{http://dx.doi.org/10.1103/PhysRevLett.89.011301}.
  \DOIprefix\doi{10.1103/PhysRevLett.89.011301}.
  \href{http://arxiv.org/abs/0204008}{\tt arXiv:0204008}.
%Type = Article
\bibitem[{Gerbino and Lattanzi(2018)}]{Gerbino2018a}
\bibinfo{author}{Gerbino\xfnm[ M.]}, \bibinfo{author}{Lattanzi\xfnm[ M.]}.
\newblock \bibinfo{title}{Status of neutrino properties and future
  prospects{\textemdash}cosmological and astrophysical constraints}.
\newblock \emph{\bibinfo{journal}{Frontiers in Physics}}
  \bibinfo{year}{2018};\bibinfo{volume}{5}.
\newblock \DOIprefix\doi{10.3389/fphy.2017.00070}.
%Type = Article
\bibitem[{Giusarma et~al.(2023)Giusarma, Reyes, Villaescusa-Navarro, He, Ho and
  Hahn}]{Giusarma_2023}
\bibinfo{author}{Giusarma\xfnm[ E.]}, \bibinfo{author}{Reyes\xfnm[ M.]},
  \bibinfo{author}{Villaescusa-Navarro\xfnm[ F.]}, \bibinfo{author}{He\xfnm[
  S.]}, \bibinfo{author}{Ho\xfnm[ S.]}, \bibinfo{author}{Hahn\xfnm[ C.]}.
\newblock \bibinfo{title}{Learning neutrino effects in cosmology with
  convolutional neural network}.
\newblock \emph{\bibinfo{journal}{The Astrophysical Journal}}
  \bibinfo{year}{2023};\bibinfo{volume}{950}(\bibinfo{number}{1}):\bibinfo{pages}{70}.
\newblock \URLprefix \url{https://dx.doi.org/10.3847/1538-4357/accd61}.
  \DOIprefix\doi{10.3847/1538-4357/accd61}.
%Type = Article
\bibitem[{Suhonen and Civitarese(1998)}]{Suhonen1998}
\bibinfo{author}{Suhonen\xfnm[ J.]}, \bibinfo{author}{Civitarese\xfnm[ O.]}.
\newblock \bibinfo{title}{{Weak-interaction and nuclear-structure aspects of
  nuclear double beta decay}}.
\newblock \emph{\bibinfo{journal}{Physics Reports}}
  \bibinfo{year}{1998};\bibinfo{volume}{300}(\bibinfo{number}{3-4}):\bibinfo{pages}{123--214}.
\newblock \URLprefix \url{http://dx.doi.org/10.1016/S0370-1573(97)00087-2}.
  \DOIprefix\doi{10.1016/S0370-1573(97)00087-2}.
%Type = Article
\bibitem[{Avignone et~al.(2008)Avignone, Elliott and Engel}]{Avignone2008}
\bibinfo{author}{Avignone\xfnm[ F.T.]}, \bibinfo{author}{Elliott\xfnm[ S.R.]},
  \bibinfo{author}{Engel\xfnm[ J.]}.
\newblock \bibinfo{title}{{Double beta decay, Majorana neutrinos, and neutrino
  mass}}.
\newblock \emph{\bibinfo{journal}{Reviews of Modern Physics}}
  \bibinfo{year}{2008};\bibinfo{volume}{80}(\bibinfo{number}{2}):\bibinfo{pages}{481--516}.
\newblock \URLprefix \url{http://link.aps.org/doi/10.1103/RevModPhys.80.481}.
  \DOIprefix\doi{10.1103/RevModPhys.80.481}.
  \href{http://arxiv.org/abs/0708.1033}{\tt arXiv:0708.1033}.
%Type = Article
\bibitem[{Ejiri et~al.(2019)Ejiri, Suhonen and Zuber}]{Ejiri2019}
\bibinfo{author}{Ejiri\xfnm[ H.]}, \bibinfo{author}{Suhonen\xfnm[ J.]},
  \bibinfo{author}{Zuber\xfnm[ K.]}.
\newblock \bibinfo{title}{{Neutrino–nuclear responses for astro-neutrinos,
  single beta decays and double beta decays}}.
\newblock \emph{\bibinfo{journal}{Physics Reports}}
  \bibinfo{year}{2019};\bibinfo{volume}{797}:\bibinfo{pages}{1--102}.
\newblock \URLprefix \url{https://doi.org/10.1016/j.physrep.2018.12.001
  https://dx.doi.org/10.1016/j.physrep.2018.12.001}.
  \DOIprefix\doi{10.1016/j.physrep.2018.12.001}.
%Type = Article
\bibitem[{Agostini et~al.(2023)Agostini, Benato, Detwiler, Men\'endez and
  Vissani}]{Agostini2023}
\bibinfo{author}{Agostini\xfnm[ M.]}, \bibinfo{author}{Benato\xfnm[ G.]},
  \bibinfo{author}{Detwiler\xfnm[ J.A.]}, \bibinfo{author}{Men\'endez\xfnm[
  J.]}, \bibinfo{author}{Vissani\xfnm[ F.]}.
\newblock \bibinfo{title}{Toward the discovery of matter creation with
  neutrinoless $\ensuremath{\beta}\ensuremath{\beta}$ decay}.
\newblock \emph{\bibinfo{journal}{Rev Mod Phys}}
  \bibinfo{year}{2023};\bibinfo{volume}{95}:\bibinfo{pages}{025002}.
\newblock \URLprefix
  \url{https://link.aps.org/doi/10.1103/RevModPhys.95.025002}.
  \DOIprefix\doi{10.1103/RevModPhys.95.025002}.
%Type = Article
\bibitem[{Drexlin et~al.(2013)Drexlin, Hannen, Mertens and
  Weinheimer}]{Drexlin2013}
\bibinfo{author}{Drexlin\xfnm[ G.]}, \bibinfo{author}{Hannen\xfnm[ V.]},
  \bibinfo{author}{Mertens\xfnm[ S.]}, \bibinfo{author}{Weinheimer\xfnm[ C.]}.
\newblock \bibinfo{title}{{Current direct neutrino mass experiments}}.
\newblock \emph{\bibinfo{journal}{Advances in High Energy Physics}}
  \bibinfo{year}{2013};\bibinfo{volume}{2013}(\bibinfo{number}{i}):\bibinfo{pages}{1--39}.
\newblock \URLprefix \url{http://www.hindawi.com/journals/ahep/2013/293986/}.
  \DOIprefix\doi{10.1155/2013/293986}.
  \href{http://arxiv.org/abs/1307.0101}{\tt arXiv:1307.0101}.
%Type = Article
\bibitem[{Aker et~al.(2019)Aker, Altenm{\"{u}}ller, Arenz, Babutzka, Barrett,
  Bauer, Beck, Beglarian, Behrens, Bergmann, Besserer, Blaum, Block, Bobien,
  Bokeloh, Bonn, Bornschein, Bornschein, Bouquet, Brunst, Caldwell, {La
  Cascio}, Chilingaryan, Choi, Corona, Debowski, Deffert, Descher, Doe,
  Dragoun, Drexlin, Dunmore, Dyba, Edzards, Eisenbl{\"{a}}tter, Eitel,
  Ellinger, Engel, Enomoto, Erhard, Eversheim, Fedkevych, Felden, Fischer,
  Flatt, Formaggio, Fr{\"{a}}nkle, Franklin, Frankrone, Friedel, Fuchs, Fulst,
  Furse, Gauda, Gemmeke, Gil, Gl{\"{u}}ck, G{\"{o}}rhardt, Groh, Grohmann,
  Gr{\"{o}}ssle, Gumbsheimer, {Ha Minh}, Hackenjos, Hannen, Harms, Hartmann,
  Hau{\ss}mann, Heizmann, Helbing, Hickford, Hilk, Hillen, Hillesheimer, Hinz,
  H{\"{o}}hn, Holzapfel, Holzmann, Houdy, Howe, Huber, James, Jansen, Kaboth,
  Karl, Kazachenko, Kellerer, Kernert, Kippenbrock, Kleesiek, Klein,
  K{\"{o}}hler, K{\"{o}}llenberger, Kopmann, Korzeczek, Kosmider,
  Koval{\'{i}}k, Krasch, Kraus, Krause, Kuckert, Kuffner, Kunka, Lasserre, Le,
  Lebeda, Leber, Lehnert, Letnev, Leven, Lichter, Lobashev, Lokhov,
  MacHatschek, Malcherek, M{\"{u}}ller, Mark, Marsteller, Martin, Melzer,
  Menshikov, Mertens, Minter, Mirz, Monreal, {Morales Guzm{\'{a}}n},
  M{\"{u}}ller, Naumann, Ndeke, Neumann, Niemes, Noe, Oblath, Ortjohann,
  Osipowicz, Ostrick, Otten, Parno, Phillips, Plischke, Pollithy, Poon,
  Pouryamout, Prall, Priester, R{\"{o}}llig, R{\"{o}}ttele, Ranitzsch, Rest,
  Rinderspacher, Robertson, Rodenbeck, Rohr, Roll, Rupp, Ry{\v{s}}av{\'{y}},
  Sack, Saenz, Sch{\"{a}}fer, Schimpf, Schl{\"{o}}sser, Schl{\"{o}}sser,
  Schl{\"{u}}ter, Sch{\"{o}}n, Sch{\"{o}}nung, Schrank, Schulz, Schwarz,
  Seitz-Moskaliuk, Seller, Sibille, Siegmann, Skasyrskaya, Slez{\'{a}}k,
  {\v{S}}palek, Spanier, Steidl, Steinbrink, Sturm, Suesser, Sun,
  Tcherniakhovski, Telle, Th{\"{u}}mmler, Thorne, Titov, Tkachev, Trost, Urban,
  V{\'{e}}nos, Valerius, Vandevender, Vianden, {Vizcaya Hern{\'{a}}ndez}, Wall,
  W{\"{u}}stling, Weber, Weinheimer, Weiss, Welte, Wendel, Wierman, Wilkerson,
  Wolf, Xu, Yen, Zacher, Zadorozhny, Zbořil and Zeller}]{Aker2019}
\bibinfo{author}{Aker\xfnm[ M.]}, \bibinfo{author}{Altenm{\"{u}}ller\xfnm[
  K.]}, \bibinfo{author}{Arenz\xfnm[ M.]}, \bibinfo{author}{Babutzka\xfnm[
  M.]}, \bibinfo{author}{Barrett\xfnm[ J.]}, \bibinfo{author}{Bauer\xfnm[ S.]},
  \bibinfo{author}{Beck\xfnm[ M.]}, \bibinfo{author}{Beglarian\xfnm[ A.]},
  \bibinfo{author}{Behrens\xfnm[ J.]}, \bibinfo{author}{Bergmann\xfnm[ T.]},
  \bibinfo{author}{Besserer\xfnm[ U.]}, \bibinfo{author}{Blaum\xfnm[ K.]},
  \bibinfo{author}{Block\xfnm[ F.]}, \bibinfo{author}{Bobien\xfnm[ S.]},
  \bibinfo{author}{Bokeloh\xfnm[ K.]}, \bibinfo{author}{Bonn\xfnm[ J.]},
  \bibinfo{author}{Bornschein\xfnm[ B.]}, \bibinfo{author}{Bornschein\xfnm[
  L.]}, \bibinfo{author}{Bouquet\xfnm[ H.]}, \bibinfo{author}{Brunst\xfnm[
  T.]}, \bibinfo{author}{Caldwell\xfnm[ T.S.]}, \bibinfo{author}{{La
  Cascio}\xfnm[ L.]}, \bibinfo{author}{Chilingaryan\xfnm[ S.]},
  \bibinfo{author}{Choi\xfnm[ W.]}, \bibinfo{author}{Corona\xfnm[ T.J.]},
  \bibinfo{author}{Debowski\xfnm[ K.]}, \bibinfo{author}{Deffert\xfnm[ M.]},
  \bibinfo{author}{Descher\xfnm[ M.]}, \bibinfo{author}{Doe\xfnm[ P.J.]},
  \bibinfo{author}{Dragoun\xfnm[ O.]}, \bibinfo{author}{Drexlin\xfnm[ G.]},
  \bibinfo{author}{Dunmore\xfnm[ J.A.]}, \bibinfo{author}{Dyba\xfnm[ S.]},
  \bibinfo{author}{Edzards\xfnm[ F.]},
  \bibinfo{author}{Eisenbl{\"{a}}tter\xfnm[ L.]}, \bibinfo{author}{Eitel\xfnm[
  K.]}, \bibinfo{author}{Ellinger\xfnm[ E.]}, \bibinfo{author}{Engel\xfnm[
  R.]}, \bibinfo{author}{Enomoto\xfnm[ S.]}, \bibinfo{author}{Erhard\xfnm[
  M.]}, \bibinfo{author}{Eversheim\xfnm[ D.]}, \bibinfo{author}{Fedkevych\xfnm[
  M.]}, \bibinfo{author}{Felden\xfnm[ A.]}, \bibinfo{author}{Fischer\xfnm[
  S.]}, \bibinfo{author}{Flatt\xfnm[ B.]}, \bibinfo{author}{Formaggio\xfnm[
  J.A.]}, \bibinfo{author}{Fr{\"{a}}nkle\xfnm[ F.M.]},
  \bibinfo{author}{Franklin\xfnm[ G.B.]}, \bibinfo{author}{Frankrone\xfnm[
  H.]}, \bibinfo{author}{Friedel\xfnm[ F.]}, \bibinfo{author}{Fuchs\xfnm[ D.]},
  \bibinfo{author}{Fulst\xfnm[ A.]}, \bibinfo{author}{Furse\xfnm[ D.]},
  \bibinfo{author}{Gauda\xfnm[ K.]}, \bibinfo{author}{Gemmeke\xfnm[ H.]},
  \bibinfo{author}{Gil\xfnm[ W.]}, \bibinfo{author}{Gl{\"{u}}ck\xfnm[ F.]},
  \bibinfo{author}{G{\"{o}}rhardt\xfnm[ S.]}, \bibinfo{author}{Groh\xfnm[ S.]},
  \bibinfo{author}{Grohmann\xfnm[ S.]}, \bibinfo{author}{Gr{\"{o}}ssle\xfnm[
  R.]}, \bibinfo{author}{Gumbsheimer\xfnm[ R.]}, \bibinfo{author}{{Ha
  Minh}\xfnm[ M.]}, \bibinfo{author}{Hackenjos\xfnm[ M.]},
  \bibinfo{author}{Hannen\xfnm[ V.]}, \bibinfo{author}{Harms\xfnm[ F.]},
  \bibinfo{author}{Hartmann\xfnm[ J.]}, \bibinfo{author}{Hau{\ss}mann\xfnm[
  N.]}, \bibinfo{author}{Heizmann\xfnm[ F.]}, \bibinfo{author}{Helbing\xfnm[
  K.]}, \bibinfo{author}{Hickford\xfnm[ S.]}, \bibinfo{author}{Hilk\xfnm[ D.]},
  \bibinfo{author}{Hillen\xfnm[ B.]}, \bibinfo{author}{Hillesheimer\xfnm[ D.]},
  \bibinfo{author}{Hinz\xfnm[ D.]}, \bibinfo{author}{H{\"{o}}hn\xfnm[ T.]},
  \bibinfo{author}{Holzapfel\xfnm[ B.]}, \bibinfo{author}{Holzmann\xfnm[ S.]},
  \bibinfo{author}{Houdy\xfnm[ T.]}, \bibinfo{author}{Howe\xfnm[ M.A.]},
  \bibinfo{author}{Huber\xfnm[ A.]}, \bibinfo{author}{James\xfnm[ T.M.]},
  \bibinfo{author}{Jansen\xfnm[ A.]}, \bibinfo{author}{Kaboth\xfnm[ A.]},
  \bibinfo{author}{Karl\xfnm[ C.]}, \bibinfo{author}{Kazachenko\xfnm[ O.]},
  \bibinfo{author}{Kellerer\xfnm[ J.]}, \bibinfo{author}{Kernert\xfnm[ N.]},
  \bibinfo{author}{Kippenbrock\xfnm[ L.]}, \bibinfo{author}{Kleesiek\xfnm[
  M.]}, \bibinfo{author}{Klein\xfnm[ M.]}, \bibinfo{author}{K{\"{o}}hler\xfnm[
  C.]}, \bibinfo{author}{K{\"{o}}llenberger\xfnm[ L.]},
  \bibinfo{author}{Kopmann\xfnm[ A.]}, \bibinfo{author}{Korzeczek\xfnm[ M.]},
  \bibinfo{author}{Kosmider\xfnm[ A.]}, \bibinfo{author}{Koval{\'{i}}k\xfnm[
  A.]}, \bibinfo{author}{Krasch\xfnm[ B.]}, \bibinfo{author}{Kraus\xfnm[ M.]},
  \bibinfo{author}{Krause\xfnm[ H.]}, \bibinfo{author}{Kuckert\xfnm[ L.]},
  \bibinfo{author}{Kuffner\xfnm[ B.]}, \bibinfo{author}{Kunka\xfnm[ N.]},
  \bibinfo{author}{Lasserre\xfnm[ T.]}, \bibinfo{author}{Le\xfnm[ T.L.]},
  \bibinfo{author}{Lebeda\xfnm[ O.]}, \bibinfo{author}{Leber\xfnm[ M.]},
  \bibinfo{author}{Lehnert\xfnm[ B.]}, \bibinfo{author}{Letnev\xfnm[ J.]},
  \bibinfo{author}{Leven\xfnm[ F.]}, \bibinfo{author}{Lichter\xfnm[ S.]},
  \bibinfo{author}{Lobashev\xfnm[ V.M.]}, \bibinfo{author}{Lokhov\xfnm[ A.]},
  \bibinfo{author}{MacHatschek\xfnm[ M.]}, \bibinfo{author}{Malcherek\xfnm[
  E.]}, \bibinfo{author}{M{\"{u}}ller\xfnm[ K.]}, \bibinfo{author}{Mark\xfnm[
  M.]}, \bibinfo{author}{Marsteller\xfnm[ A.]}, \bibinfo{author}{Martin\xfnm[
  E.L.]}, \bibinfo{author}{Melzer\xfnm[ C.]}, \bibinfo{author}{Menshikov\xfnm[
  A.]}, \bibinfo{author}{Mertens\xfnm[ S.]}, \bibinfo{author}{Minter\xfnm[
  L.I.]}, \bibinfo{author}{Mirz\xfnm[ S.]}, \bibinfo{author}{Monreal\xfnm[
  B.]}, \bibinfo{author}{{Morales Guzm{\'{a}}n}\xfnm[ P.I.]},
  \bibinfo{author}{M{\"{u}}ller\xfnm[ K.]}, \bibinfo{author}{Naumann\xfnm[
  U.]}, \bibinfo{author}{Ndeke\xfnm[ W.]}, \bibinfo{author}{Neumann\xfnm[ H.]},
  \bibinfo{author}{Niemes\xfnm[ S.]}, \bibinfo{author}{Noe\xfnm[ M.]},
  \bibinfo{author}{Oblath\xfnm[ N.S.]}, \bibinfo{author}{Ortjohann\xfnm[
  H.W.]}, \bibinfo{author}{Osipowicz\xfnm[ A.]}, \bibinfo{author}{Ostrick\xfnm[
  B.]}, \bibinfo{author}{Otten\xfnm[ E.]}, \bibinfo{author}{Parno\xfnm[ D.S.]},
  \bibinfo{author}{Phillips\xfnm[ D.G.]}, \bibinfo{author}{Plischke\xfnm[ P.]},
  \bibinfo{author}{Pollithy\xfnm[ A.]}, \bibinfo{author}{Poon\xfnm[ A.W.]},
  \bibinfo{author}{Pouryamout\xfnm[ J.]}, \bibinfo{author}{Prall\xfnm[ M.]},
  \bibinfo{author}{Priester\xfnm[ F.]}, \bibinfo{author}{R{\"{o}}llig\xfnm[
  M.]}, \bibinfo{author}{R{\"{o}}ttele\xfnm[ C.]},
  \bibinfo{author}{Ranitzsch\xfnm[ P.C.]}, \bibinfo{author}{Rest\xfnm[ O.]},
  \bibinfo{author}{Rinderspacher\xfnm[ R.]}, \bibinfo{author}{Robertson\xfnm[
  R.G.]}, \bibinfo{author}{Rodenbeck\xfnm[ C.]}, \bibinfo{author}{Rohr\xfnm[
  P.]}, \bibinfo{author}{Roll\xfnm[ C.]}, \bibinfo{author}{Rupp\xfnm[ S.]},
  \bibinfo{author}{Ry{\v{s}}av{\'{y}}\xfnm[ M.]}, \bibinfo{author}{Sack\xfnm[
  R.]}, \bibinfo{author}{Saenz\xfnm[ A.]}, \bibinfo{author}{Sch{\"{a}}fer\xfnm[
  P.]}, \bibinfo{author}{Schimpf\xfnm[ L.]},
  \bibinfo{author}{Schl{\"{o}}sser\xfnm[ K.]},
  \bibinfo{author}{Schl{\"{o}}sser\xfnm[ M.]},
  \bibinfo{author}{Schl{\"{u}}ter\xfnm[ L.]},
  \bibinfo{author}{Sch{\"{o}}n\xfnm[ H.]},
  \bibinfo{author}{Sch{\"{o}}nung\xfnm[ K.]}, \bibinfo{author}{Schrank\xfnm[
  M.]}, \bibinfo{author}{Schulz\xfnm[ B.]}, \bibinfo{author}{Schwarz\xfnm[
  J.]}, \bibinfo{author}{Seitz-Moskaliuk\xfnm[ H.]},
  \bibinfo{author}{Seller\xfnm[ W.]}, \bibinfo{author}{Sibille\xfnm[ V.]},
  \bibinfo{author}{Siegmann\xfnm[ D.]}, \bibinfo{author}{Skasyrskaya\xfnm[
  A.]}, \bibinfo{author}{Slez{\'{a}}k\xfnm[ M.]},
  \bibinfo{author}{{\v{S}}palek\xfnm[ A.]}, \bibinfo{author}{Spanier\xfnm[
  F.]}, \bibinfo{author}{Steidl\xfnm[ M.]}, \bibinfo{author}{Steinbrink\xfnm[
  N.]}, \bibinfo{author}{Sturm\xfnm[ M.]}, \bibinfo{author}{Suesser\xfnm[ M.]},
  \bibinfo{author}{Sun\xfnm[ M.]}, \bibinfo{author}{Tcherniakhovski\xfnm[ D.]},
  \bibinfo{author}{Telle\xfnm[ H.H.]}, \bibinfo{author}{Th{\"{u}}mmler\xfnm[
  T.]}, \bibinfo{author}{Thorne\xfnm[ L.A.]}, \bibinfo{author}{Titov\xfnm[
  N.]}, \bibinfo{author}{Tkachev\xfnm[ I.]}, \bibinfo{author}{Trost\xfnm[ N.]},
  \bibinfo{author}{Urban\xfnm[ K.]}, \bibinfo{author}{V{\'{e}}nos\xfnm[ D.]},
  \bibinfo{author}{Valerius\xfnm[ K.]}, \bibinfo{author}{Vandevender\xfnm[
  B.A.]}, \bibinfo{author}{Vianden\xfnm[ R.]}, \bibinfo{author}{{Vizcaya
  Hern{\'{a}}ndez}\xfnm[ A.P.]}, \bibinfo{author}{Wall\xfnm[ B.L.]},
  \bibinfo{author}{W{\"{u}}stling\xfnm[ S.]}, \bibinfo{author}{Weber\xfnm[
  M.]}, \bibinfo{author}{Weinheimer\xfnm[ C.]}, \bibinfo{author}{Weiss\xfnm[
  C.]}, \bibinfo{author}{Welte\xfnm[ S.]}, \bibinfo{author}{Wendel\xfnm[ J.]},
  \bibinfo{author}{Wierman\xfnm[ K.J.]}, \bibinfo{author}{Wilkerson\xfnm[
  J.F.]}, \bibinfo{author}{Wolf\xfnm[ J.]}, \bibinfo{author}{Xu\xfnm[ W.]},
  \bibinfo{author}{Yen\xfnm[ Y.R.]}, \bibinfo{author}{Zacher\xfnm[ M.]},
  \bibinfo{author}{Zadorozhny\xfnm[ S.]}, \bibinfo{author}{Zbořil\xfnm[ M.]},
  \bibinfo{author}{Zeller\xfnm[ G.]}.
\newblock \bibinfo{title}{{Improved Upper Limit on the Neutrino Mass from a
  Direct Kinematic Method by KATRIN}}.
\newblock \emph{\bibinfo{journal}{Physical Review Letters}}
  \bibinfo{year}{2019};\bibinfo{volume}{123}(\bibinfo{number}{22}):\bibinfo{pages}{1--11}.
\newblock \URLprefix \url{https://doi.org/10.1103/PhysRevLett.123.221802}.
  \DOIprefix\doi{10.1103/PhysRevLett.123.221802}.
  \href{http://arxiv.org/abs/1909.06048}{\tt arXiv:1909.06048}.
%Type = Article
\bibitem[{Aker et~al.(2022)Aker, Beglarian, Behrens, Berlev, Besserer,
  Bieringer, Block, Bornschein, Bornschein, Böttcher, Brunst, Caldwell,
  Carney, Cascio, Chilingaryan, Choi, Debowski, Deffert, Descher, Barrero, Doe,
  Dragoun, Drexlin, Eitel, Ellinger, Engel, Enomoto, Felden, Formaggio,
  Fränkle, Franklin, Friedel, Fulst, Gauda, Gil, Glück, Grössle,
  Gumbsheimer, Gupta, Höhn, Hannen, Haußmann, Helbing, Hickford, Hiller,
  Hillesheimer, Hinz, Houdy, Huber, Jansen, Karl, Kellerer, Kellerer, Klein,
  Köhler, Köllenberger, Kopmann, Korzeczek, Koval\'ik, Krasch, Krause, Kunka,
  Lasserre, Le, Lebeda, Lehnert, Lokhov, Machatschek, Malcherek, Mark,
  Marsteller, Martin, Melzer, Menshikov, Mertens, Mostafa, Müller, Niemes,
  Oelpmann, Parno, Poon, Poyato, Priester, Röllig, Röttele, Robertson,
  Rodejohann, Rodenbeck, Ryšavý, Sack, Saenz, Schäfer, Schaller, Schimpf,
  Schlösser, Schlösser, Schlüter, Schneidewind, Schrank, Schulz, Schwemmer,
  Šef\^c\'ik, Sibille, Siegmann, Slezák, Steidl, Sturm, Sun, Tcherniakhovski,
  Telle, Thorne, Thümmler, Titov, Tkachev, Urban, Valerius, Vénos,
  Hernández, Weinheimer, Welte, Wendel, Wilkerson, Wolf, Wüstling, Xu, Yen,
  Zadoroghny and Zeller}]{Aker2022}
\bibinfo{author}{Aker\xfnm[ M.]}, \bibinfo{author}{Beglarian\xfnm[ A.]},
  \bibinfo{author}{Behrens\xfnm[ J.]}, \bibinfo{author}{Berlev\xfnm[ A.]},
  \bibinfo{author}{Besserer\xfnm[ U.]}, \bibinfo{author}{Bieringer\xfnm[ B.]},
  \bibinfo{author}{Block\xfnm[ F.]}, \bibinfo{author}{Bornschein\xfnm[ B.]},
  \bibinfo{author}{Bornschein\xfnm[ L.]}, \bibinfo{author}{Böttcher\xfnm[
  M.]}, \bibinfo{author}{Brunst\xfnm[ T.]}, \bibinfo{author}{Caldwell\xfnm[
  T.S.]}, \bibinfo{author}{Carney\xfnm[ R.M.D.]}, \bibinfo{author}{Cascio\xfnm[
  L.L.]}, \bibinfo{author}{Chilingaryan\xfnm[ S.]}, \bibinfo{author}{Choi\xfnm[
  W.]}, \bibinfo{author}{Debowski\xfnm[ K.]}, \bibinfo{author}{Deffert\xfnm[
  M.]}, \bibinfo{author}{Descher\xfnm[ M.]}, \bibinfo{author}{Barrero\xfnm[
  D.D.]}, \bibinfo{author}{Doe\xfnm[ P.J.]}, \bibinfo{author}{Dragoun\xfnm[
  O.]}, \bibinfo{author}{Drexlin\xfnm[ G.]}, \bibinfo{author}{Eitel\xfnm[ K.]},
  \bibinfo{author}{Ellinger\xfnm[ E.]}, \bibinfo{author}{Engel\xfnm[ R.]},
  \bibinfo{author}{Enomoto\xfnm[ S.]}, \bibinfo{author}{Felden\xfnm[ A.]},
  \bibinfo{author}{Formaggio\xfnm[ J.A.]}, \bibinfo{author}{Fränkle\xfnm[
  F.M.]}, \bibinfo{author}{Franklin\xfnm[ G.B.]},
  \bibinfo{author}{Friedel\xfnm[ F.]}, \bibinfo{author}{Fulst\xfnm[ A.]},
  \bibinfo{author}{Gauda\xfnm[ K.]}, \bibinfo{author}{Gil\xfnm[ W.]},
  \bibinfo{author}{Glück\xfnm[ F.]}, \bibinfo{author}{Grössle\xfnm[ R.]},
  \bibinfo{author}{Gumbsheimer\xfnm[ R.]}, \bibinfo{author}{Gupta\xfnm[ V.]},
  \bibinfo{author}{Höhn\xfnm[ T.]}, \bibinfo{author}{Hannen\xfnm[ V.]},
  \bibinfo{author}{Haußmann\xfnm[ N.]}, \bibinfo{author}{Helbing\xfnm[ K.]},
  \bibinfo{author}{Hickford\xfnm[ S.]}, \bibinfo{author}{Hiller\xfnm[ R.]},
  \bibinfo{author}{Hillesheimer\xfnm[ D.]}, \bibinfo{author}{Hinz\xfnm[ D.]},
  \bibinfo{author}{Houdy\xfnm[ T.]}, \bibinfo{author}{Huber\xfnm[ A.]},
  \bibinfo{author}{Jansen\xfnm[ A.]}, \bibinfo{author}{Karl\xfnm[ C.]},
  \bibinfo{author}{Kellerer\xfnm[ F.]}, \bibinfo{author}{Kellerer\xfnm[ J.]},
  \bibinfo{author}{Klein\xfnm[ M.]}, \bibinfo{author}{Köhler\xfnm[ C.]},
  \bibinfo{author}{Köllenberger\xfnm[ L.]}, \bibinfo{author}{Kopmann\xfnm[
  A.]}, \bibinfo{author}{Korzeczek\xfnm[ M.]}, \bibinfo{author}{Koval\'ik\xfnm[
  A.]}, \bibinfo{author}{Krasch\xfnm[ B.]}, \bibinfo{author}{Krause\xfnm[ H.]},
  \bibinfo{author}{Kunka\xfnm[ N.]}, \bibinfo{author}{Lasserre\xfnm[ T.]},
  \bibinfo{author}{Le\xfnm[ T.L.]}, \bibinfo{author}{Lebeda\xfnm[ O.]},
  \bibinfo{author}{Lehnert\xfnm[ B.]}, \bibinfo{author}{Lokhov\xfnm[ A.]},
  \bibinfo{author}{Machatschek\xfnm[ M.]}, \bibinfo{author}{Malcherek\xfnm[
  E.]}, \bibinfo{author}{Mark\xfnm[ M.]}, \bibinfo{author}{Marsteller\xfnm[
  A.]}, \bibinfo{author}{Martin\xfnm[ E.L.]}, \bibinfo{author}{Melzer\xfnm[
  C.]}, \bibinfo{author}{Menshikov\xfnm[ A.]}, \bibinfo{author}{Mertens\xfnm[
  S.]}, \bibinfo{author}{Mostafa\xfnm[ J.]}, \bibinfo{author}{Müller\xfnm[
  K.]}, \bibinfo{author}{Niemes\xfnm[ S.]}, \bibinfo{author}{Oelpmann\xfnm[
  P.]}, \bibinfo{author}{Parno\xfnm[ D.S.]}, \bibinfo{author}{Poon\xfnm[
  A.W.P.]}, \bibinfo{author}{Poyato\xfnm[ J.M.L.]},
  \bibinfo{author}{Priester\xfnm[ F.]}, \bibinfo{author}{Röllig\xfnm[ M.]},
  \bibinfo{author}{Röttele\xfnm[ C.]}, \bibinfo{author}{Robertson\xfnm[
  R.G.H.]}, \bibinfo{author}{Rodejohann\xfnm[ W.]},
  \bibinfo{author}{Rodenbeck\xfnm[ C.]}, \bibinfo{author}{Ryšavý\xfnm[ M.]},
  \bibinfo{author}{Sack\xfnm[ R.]}, \bibinfo{author}{Saenz\xfnm[ A.]},
  \bibinfo{author}{Schäfer\xfnm[ P.]}, \bibinfo{author}{Schaller\xfnm[ A.]},
  \bibinfo{author}{Schimpf\xfnm[ L.]}, \bibinfo{author}{Schlösser\xfnm[ K.]},
  \bibinfo{author}{Schlösser\xfnm[ M.]}, \bibinfo{author}{Schlüter\xfnm[
  L.]}, \bibinfo{author}{Schneidewind\xfnm[ S.]},
  \bibinfo{author}{Schrank\xfnm[ M.]}, \bibinfo{author}{Schulz\xfnm[ B.]},
  \bibinfo{author}{Schwemmer\xfnm[ A.]}, \bibinfo{author}{Šef\^c\'ik\xfnm[
  M.]}, \bibinfo{author}{Sibille\xfnm[ V.]}, \bibinfo{author}{Siegmann\xfnm[
  D.]}, \bibinfo{author}{Slezák\xfnm[ M.]}, \bibinfo{author}{Steidl\xfnm[
  M.]}, \bibinfo{author}{Sturm\xfnm[ M.]}, \bibinfo{author}{Sun\xfnm[ M.]},
  \bibinfo{author}{Tcherniakhovski\xfnm[ D.]}, \bibinfo{author}{Telle\xfnm[
  H.H.]}, \bibinfo{author}{Thorne\xfnm[ L.A.]},
  \bibinfo{author}{Thümmler\xfnm[ T.]}, \bibinfo{author}{Titov\xfnm[ N.]},
  \bibinfo{author}{Tkachev\xfnm[ I.]}, \bibinfo{author}{Urban\xfnm[ K.]},
  \bibinfo{author}{Valerius\xfnm[ K.]}, \bibinfo{author}{Vénos\xfnm[ D.]},
  \bibinfo{author}{Hernández\xfnm[ A.P.V.]}, \bibinfo{author}{Weinheimer\xfnm[
  C.]}, \bibinfo{author}{Welte\xfnm[ S.]}, \bibinfo{author}{Wendel\xfnm[ J.]},
  \bibinfo{author}{Wilkerson\xfnm[ J.F.]}, \bibinfo{author}{Wolf\xfnm[ J.]},
  \bibinfo{author}{Wüstling\xfnm[ S.]}, \bibinfo{author}{Xu\xfnm[ W.]},
  \bibinfo{author}{Yen\xfnm[ Y.R.]}, \bibinfo{author}{Zadoroghny\xfnm[ S.]},
  \bibinfo{author}{Zeller\xfnm[ G.]}.
\newblock \bibinfo{title}{Direct neutrino-mass measurement with
  sub-electronvolt sensitivity}.
\newblock \emph{\bibinfo{journal}{Nature Physics}}
  \bibinfo{year}{2022};\bibinfo{volume}{18}:\bibinfo{pages}{160}.
\newblock \URLprefix \url{https://doi.org/10.1038/s41567-021-01463-1}.
  \DOIprefix\doi{https://10.1038/s41567-021-01463-1}.
%Type = Article
\bibitem[{Ashtari~Esfahani et~al.(2023)Ashtari~Esfahani, B\"oser, Buzinsky,
  Carmona-Benitez, Claessens, de~Viveiros, Doe, Fertl, Formaggio, Gaison,
  Gladstone, Grando, Guigue, Hartse, Heeger, Huyan, Johnston, Jones, Kazkaz,
  LaRoque, Li, Lindman, Machado, Marsteller, Matth\'e, Mohiuddin, Monreal,
  Mueller, Nikkel, Novitski, Oblath, Pe\~na, Pettus, Reimann, Robertson, Rosa
  De~Jes\'us, Rybka, Salda\~na, Schram, Slocum, Stachurska, Sun, Surukuchi,
  Tedeschi, Telles, Thomas, Thomas, Thorne, Th\"ummler, Tvrznikova, Van
  De~Pontseele, VanDevender, Weintroub, Weiss, Wendler, Young, Zayas and
  Ziegler}]{Ashtar23}
\bibinfo{author}{Ashtari~Esfahani\xfnm[ A.]}, \bibinfo{author}{B\"oser\xfnm[
  S.]}, \bibinfo{author}{Buzinsky\xfnm[ N.]},
  \bibinfo{author}{Carmona-Benitez\xfnm[ M.C.]},
  \bibinfo{author}{Claessens\xfnm[ C.]}, \bibinfo{author}{de~Viveiros\xfnm[
  L.]}, \bibinfo{author}{Doe\xfnm[ P.J.]}, \bibinfo{author}{Fertl\xfnm[ M.]},
  \bibinfo{author}{Formaggio\xfnm[ J.A.]}, \bibinfo{author}{Gaison\xfnm[
  J.K.]}, \bibinfo{author}{Gladstone\xfnm[ L.]}, \bibinfo{author}{Grando\xfnm[
  M.]}, \bibinfo{author}{Guigue\xfnm[ M.]}, \bibinfo{author}{Hartse\xfnm[ J.]},
  \bibinfo{author}{Heeger\xfnm[ K.M.]}, \bibinfo{author}{Huyan\xfnm[ X.]},
  \bibinfo{author}{Johnston\xfnm[ J.]}, \bibinfo{author}{Jones\xfnm[ A.M.]},
  \bibinfo{author}{Kazkaz\xfnm[ K.]}, \bibinfo{author}{LaRoque\xfnm[ B.H.]},
  \bibinfo{author}{Li\xfnm[ M.]}, \bibinfo{author}{Lindman\xfnm[ A.]},
  \bibinfo{author}{Machado\xfnm[ E.]}, \bibinfo{author}{Marsteller\xfnm[ A.]},
  \bibinfo{author}{Matth\'e\xfnm[ C.]}, \bibinfo{author}{Mohiuddin\xfnm[ R.]},
  \bibinfo{author}{Monreal\xfnm[ B.]}, \bibinfo{author}{Mueller\xfnm[ R.]},
  \bibinfo{author}{Nikkel\xfnm[ J.A.]}, \bibinfo{author}{Novitski\xfnm[ E.]},
  \bibinfo{author}{Oblath\xfnm[ N.S.]}, \bibinfo{author}{Pe\~na\xfnm[ J.I.]},
  \bibinfo{author}{Pettus\xfnm[ W.]}, \bibinfo{author}{Reimann\xfnm[ R.]},
  \bibinfo{author}{Robertson\xfnm[ R.G.H.]}, \bibinfo{author}{Rosa
  De~Jes\'us\xfnm[ D.]}, \bibinfo{author}{Rybka\xfnm[ G.]},
  \bibinfo{author}{Salda\~na\xfnm[ L.]}, \bibinfo{author}{Schram\xfnm[ M.]},
  \bibinfo{author}{Slocum\xfnm[ P.L.]}, \bibinfo{author}{Stachurska\xfnm[ J.]},
  \bibinfo{author}{Sun\xfnm[ Y.H.]}, \bibinfo{author}{Surukuchi\xfnm[ P.T.]},
  \bibinfo{author}{Tedeschi\xfnm[ J.R.]}, \bibinfo{author}{Telles\xfnm[ A.B.]},
  \bibinfo{author}{Thomas\xfnm[ F.]}, \bibinfo{author}{Thomas\xfnm[ M.]},
  \bibinfo{author}{Thorne\xfnm[ L.A.]}, \bibinfo{author}{Th\"ummler\xfnm[ T.]},
  \bibinfo{author}{Tvrznikova\xfnm[ L.]}, \bibinfo{author}{Van
  De~Pontseele\xfnm[ W.]}, \bibinfo{author}{VanDevender\xfnm[ B.A.]},
  \bibinfo{author}{Weintroub\xfnm[ J.]}, \bibinfo{author}{Weiss\xfnm[ T.E.]},
  \bibinfo{author}{Wendler\xfnm[ T.]}, \bibinfo{author}{Young\xfnm[ A.]},
  \bibinfo{author}{Zayas\xfnm[ E.]}, \bibinfo{author}{Ziegler\xfnm[ A.]}
  (\bibinfo{collaboration}{Project 8 Collaboration}).
\newblock \bibinfo{title}{Tritium beta spectrum measurement and neutrino mass
  limit from cyclotron radiation emission spectroscopy}.
\newblock \emph{\bibinfo{journal}{Phys Rev Lett}}
  \bibinfo{year}{2023};\bibinfo{volume}{131}:\bibinfo{pages}{102502}.
\newblock \URLprefix
  \url{https://link.aps.org/doi/10.1103/PhysRevLett.131.102502}.
  \DOIprefix\doi{10.1103/PhysRevLett.131.102502}.
%Type = Article
\bibitem[{Gastaldo et~al.(2014)Gastaldo, Blaum, Doerr, D{\"{u}}llmann,
  Eberhardt, Eliseev, Enss, Faessler, Fleischmann, Kempf, Krivoruchenko,
  Lahiri, Maiti, Novikov, Ranitzsch, Simkovic, Szusc and Wegner}]{Gastaldo2014}
\bibinfo{author}{Gastaldo\xfnm[ L.]}, \bibinfo{author}{Blaum\xfnm[ K.]},
  \bibinfo{author}{Doerr\xfnm[ A.]}, \bibinfo{author}{D{\"{u}}llmann\xfnm[
  C.E.]}, \bibinfo{author}{Eberhardt\xfnm[ K.]}, \bibinfo{author}{Eliseev\xfnm[
  S.]}, \bibinfo{author}{Enss\xfnm[ C.]}, \bibinfo{author}{Faessler\xfnm[ A.]},
  \bibinfo{author}{Fleischmann\xfnm[ A.]}, \bibinfo{author}{Kempf\xfnm[ S.]},
  \bibinfo{author}{Krivoruchenko\xfnm[ M.]}, \bibinfo{author}{Lahiri\xfnm[
  S.]}, \bibinfo{author}{Maiti\xfnm[ M.]}, \bibinfo{author}{Novikov\xfnm[
  Y.N.]}, \bibinfo{author}{Ranitzsch\xfnm[ P.C.]},
  \bibinfo{author}{Simkovic\xfnm[ F.]}, \bibinfo{author}{Szusc\xfnm[ Z.]},
  \bibinfo{author}{Wegner\xfnm[ M.]}.
\newblock \bibinfo{title}{{The electron capture 163Ho experiment ECHo}}.
\newblock \emph{\bibinfo{journal}{Journal of Low Temperature Physics}}
  \bibinfo{year}{2014};\bibinfo{volume}{176}(\bibinfo{number}{5-6}):\bibinfo{pages}{876--884}.
\newblock \URLprefix \url{http://dx.doi.org/10.1007/s10909-014-1187-4}.
  \DOIprefix\doi{10.1007/s10909-014-1187-4}.
  \href{http://arxiv.org/abs/1306.2655}{\tt arXiv:1306.2655}.
%Type = Article
\bibitem[{Gastaldo et~al.(2017)Gastaldo, Blaum, Chrysalidis, {Day Goodacre},
  Domula, Door, Dorrer, D{\"{u}}llmann, Eberhardt, Eliseev, Enss, Faessler,
  Filianin, Fleischmann, Fonnesu, Gamer, Haas, Hassel, Hengstler, Jochum,
  Johnston, Kebschull, Kempf, Kieck, K{\"{o}}ster, Lahiri, Maiti, Mantegazzini,
  Marsh, Neroutsos, Novikov, Ranitzsch, Rothe, Rischka, Saenz, Sander,
  Schneider, Scholl, Sch{\"{u}}ssler, Schweiger, Simkovic, Stora, Sz{\"{u}}cs,
  T{\"{u}}rler, Veinhard, Weber, Wegner, Wendt and Zuber}]{Gastaldo2017}
\bibinfo{author}{Gastaldo\xfnm[ L.]}, \bibinfo{author}{Blaum\xfnm[ K.]},
  \bibinfo{author}{Chrysalidis\xfnm[ K.]}, \bibinfo{author}{{Day
  Goodacre}\xfnm[ T.]}, \bibinfo{author}{Domula\xfnm[ A.]},
  \bibinfo{author}{Door\xfnm[ M.]}, \bibinfo{author}{Dorrer\xfnm[ H.]},
  \bibinfo{author}{D{\"{u}}llmann\xfnm[ C.E.]},
  \bibinfo{author}{Eberhardt\xfnm[ K.]}, \bibinfo{author}{Eliseev\xfnm[ S.]},
  \bibinfo{author}{Enss\xfnm[ C.]}, \bibinfo{author}{Faessler\xfnm[ A.]},
  \bibinfo{author}{Filianin\xfnm[ P.]}, \bibinfo{author}{Fleischmann\xfnm[
  A.]}, \bibinfo{author}{Fonnesu\xfnm[ D.]}, \bibinfo{author}{Gamer\xfnm[ L.]},
  \bibinfo{author}{Haas\xfnm[ R.]}, \bibinfo{author}{Hassel\xfnm[ C.]},
  \bibinfo{author}{Hengstler\xfnm[ D.]}, \bibinfo{author}{Jochum\xfnm[ J.]},
  \bibinfo{author}{Johnston\xfnm[ K.]}, \bibinfo{author}{Kebschull\xfnm[ U.]},
  \bibinfo{author}{Kempf\xfnm[ S.]}, \bibinfo{author}{Kieck\xfnm[ T.]},
  \bibinfo{author}{K{\"{o}}ster\xfnm[ U.]}, \bibinfo{author}{Lahiri\xfnm[ S.]},
  \bibinfo{author}{Maiti\xfnm[ M.]}, \bibinfo{author}{Mantegazzini\xfnm[ F.]},
  \bibinfo{author}{Marsh\xfnm[ B.]}, \bibinfo{author}{Neroutsos\xfnm[ P.]},
  \bibinfo{author}{Novikov\xfnm[ Y.N.]}, \bibinfo{author}{Ranitzsch\xfnm[
  P.C.]}, \bibinfo{author}{Rothe\xfnm[ S.]}, \bibinfo{author}{Rischka\xfnm[
  A.]}, \bibinfo{author}{Saenz\xfnm[ A.]}, \bibinfo{author}{Sander\xfnm[ O.]},
  \bibinfo{author}{Schneider\xfnm[ F.]}, \bibinfo{author}{Scholl\xfnm[ S.]},
  \bibinfo{author}{Sch{\"{u}}ssler\xfnm[ R.X.]},
  \bibinfo{author}{Schweiger\xfnm[ C.]}, \bibinfo{author}{Simkovic\xfnm[ F.]},
  \bibinfo{author}{Stora\xfnm[ T.]}, \bibinfo{author}{Sz{\"{u}}cs\xfnm[ Z.]},
  \bibinfo{author}{T{\"{u}}rler\xfnm[ A.]}, \bibinfo{author}{Veinhard\xfnm[
  M.]}, \bibinfo{author}{Weber\xfnm[ M.]}, \bibinfo{author}{Wegner\xfnm[ M.]},
  \bibinfo{author}{Wendt\xfnm[ K.]}, \bibinfo{author}{Zuber\xfnm[ K.]}.
\newblock \bibinfo{title}{{The electron capture in 163Ho experiment – ECHo}}.
\newblock \emph{\bibinfo{journal}{European Physical Journal: Special Topics}}
  \bibinfo{year}{2017};\bibinfo{volume}{226}(\bibinfo{number}{8}):\bibinfo{pages}{1623--1694}.
\newblock \URLprefix
  \url{http://link.springer.com/10.1140/epjst/e2017-70071-y}.
  \DOIprefix\doi{10.1140/epjst/e2017-70071-y}.
%Type = Article
\bibitem[{Velte et~al.(2019)Velte, Ahrens, Barth, Blaum, Bra{\ss}, Door,
  Dorrer, Düllmann, Eliseev, Enss, Filianin, Fleischmann, Gastaldo,
  Goeggelmann, Goodacre, Haverkort, Hengstler, Jochum, Johnston, Keller, Kempf,
  Kieck, König, Köster, Kromer, Mantegazzini, Marsh, Novikov, Piquemal,
  Riccio, Richter, Rischka, Rothe, Schüssler, Schweiger, Stora, Wegner, Wendt,
  Zampaolo and Zuber}]{Velte2019}
\bibinfo{author}{Velte\xfnm[ C.]}, \bibinfo{author}{Ahrens\xfnm[ F.]},
  \bibinfo{author}{Barth\xfnm[ A.]}, \bibinfo{author}{Blaum\xfnm[ K.]},
  \bibinfo{author}{Bra{\ss}\xfnm[ M.]}, \bibinfo{author}{Door\xfnm[ M.]},
  \bibinfo{author}{Dorrer\xfnm[ H.]}, \bibinfo{author}{Düllmann\xfnm[ C.E.]},
  \bibinfo{author}{Eliseev\xfnm[ S.]}, \bibinfo{author}{Enss\xfnm[ C.]},
  \bibinfo{author}{Filianin\xfnm[ P.]}, \bibinfo{author}{Fleischmann\xfnm[
  A.]}, \bibinfo{author}{Gastaldo\xfnm[ L.]},
  \bibinfo{author}{Goeggelmann\xfnm[ A.]}, \bibinfo{author}{Goodacre\xfnm[
  T.D.]}, \bibinfo{author}{Haverkort\xfnm[ M.W.]},
  \bibinfo{author}{Hengstler\xfnm[ D.]}, \bibinfo{author}{Jochum\xfnm[ J.]},
  \bibinfo{author}{Johnston\xfnm[ K.]}, \bibinfo{author}{Keller\xfnm[ M.]},
  \bibinfo{author}{Kempf\xfnm[ S.]}, \bibinfo{author}{Kieck\xfnm[ T.]},
  \bibinfo{author}{König\xfnm[ C.M.]}, \bibinfo{author}{Köster\xfnm[ U.]},
  \bibinfo{author}{Kromer\xfnm[ K.]}, \bibinfo{author}{Mantegazzini\xfnm[ F.]},
  \bibinfo{author}{Marsh\xfnm[ B.]}, \bibinfo{author}{Novikov\xfnm[ Y.N.]},
  \bibinfo{author}{Piquemal\xfnm[ F.]}, \bibinfo{author}{Riccio\xfnm[ C.]},
  \bibinfo{author}{Richter\xfnm[ D.]}, \bibinfo{author}{Rischka\xfnm[ A.]},
  \bibinfo{author}{Rothe\xfnm[ S.]}, \bibinfo{author}{Schüssler\xfnm[ R.X.]},
  \bibinfo{author}{Schweiger\xfnm[ C.]}, \bibinfo{author}{Stora\xfnm[ T.]},
  \bibinfo{author}{Wegner\xfnm[ M.]}, \bibinfo{author}{Wendt\xfnm[ K.]},
  \bibinfo{author}{Zampaolo\xfnm[ M.]}, \bibinfo{author}{Zuber\xfnm[ K.]}.
\newblock \bibinfo{title}{{High-resolution and low-background $^{163}$Ho
  spectrum: interpretation of the resonance tails}}.
\newblock \emph{\bibinfo{journal}{The European Physical Journal C}}
  \bibinfo{year}{2019};\bibinfo{volume}{79}(\bibinfo{number}{12}).
\newblock \URLprefix
  \url{http://link.springer.com/10.1140/epjc/s10052-019-7513-x}.
  \DOIprefix\doi{10.1140/epjc/s10052-019-7513-x}.
%Type = Article
\bibitem[{Mantegazzini et~al.(2023)Mantegazzini, Kovac, Enss, Fleischmann,
  Griedel and Gastaldo}]{Echo2023}
\bibinfo{author}{Mantegazzini\xfnm[ F.]}, \bibinfo{author}{Kovac\xfnm[ N.]},
  \bibinfo{author}{Enss\xfnm[ C.]}, \bibinfo{author}{Fleischmann\xfnm[ A.]},
  \bibinfo{author}{Griedel\xfnm[ M.]}, \bibinfo{author}{Gastaldo\xfnm[ L.]}.
\newblock \bibinfo{title}{Development and characterisation of high-resolution
  microcalorimeter detectors for the echo-100k experiment}.
\newblock \emph{\bibinfo{journal}{Nuclear Instruments and Methods in Physics
  Research Section A: Accelerators, Spectrometers, Detectors and Associated
  Equipment}}
  \bibinfo{year}{2023};\bibinfo{volume}{1055}:\bibinfo{pages}{168564}.
\newblock \URLprefix
  \url{https://www.sciencedirect.com/science/article/pii/S0168900223005545}.
  \DOIprefix\doi{https://doi.org/10.1016/j.nima.2023.168564}.
%Type = Article
\bibitem[{Nucciotti et~al.(2018)Nucciotti, Alpert, Balata, Becker, Bennett,
  Bevilacqua, Biasotti, Ceriale, Ceruti, Corsini, {De Gerone}, Dressler,
  Faverzani, Ferri, Fowler, Gallucci, Gard, Gatti, Giachero, Hays-Wehle,
  Heinitz, Hilton, K{\"{o}}ster, Lusignoli, Mates, Nisi, Orlando, Parodi,
  Pessina, Puiu, Ragazzi, Reintsema, Ribeiro-Gomez, Schmidt, Schuman, Siccardi,
  Swetz, Ullom and Vale}]{Nucciotti2018}
\bibinfo{author}{Nucciotti\xfnm[ A.]}, \bibinfo{author}{Alpert\xfnm[ B.]},
  \bibinfo{author}{Balata\xfnm[ M.]}, \bibinfo{author}{Becker\xfnm[ D.]},
  \bibinfo{author}{Bennett\xfnm[ D.]}, \bibinfo{author}{Bevilacqua\xfnm[ A.]},
  \bibinfo{author}{Biasotti\xfnm[ M.]}, \bibinfo{author}{Ceriale\xfnm[ V.]},
  \bibinfo{author}{Ceruti\xfnm[ G.]}, \bibinfo{author}{Corsini\xfnm[ D.]},
  \bibinfo{author}{{De Gerone}\xfnm[ M.]}, \bibinfo{author}{Dressler\xfnm[
  R.]}, \bibinfo{author}{Faverzani\xfnm[ M.]}, \bibinfo{author}{Ferri\xfnm[
  E.]}, \bibinfo{author}{Fowler\xfnm[ J.]}, \bibinfo{author}{Gallucci\xfnm[
  G.]}, \bibinfo{author}{Gard\xfnm[ J.]}, \bibinfo{author}{Gatti\xfnm[ F.]},
  \bibinfo{author}{Giachero\xfnm[ A.]}, \bibinfo{author}{Hays-Wehle\xfnm[ J.]},
  \bibinfo{author}{Heinitz\xfnm[ S.]}, \bibinfo{author}{Hilton\xfnm[ G.]},
  \bibinfo{author}{K{\"{o}}ster\xfnm[ U.]}, \bibinfo{author}{Lusignoli\xfnm[
  M.]}, \bibinfo{author}{Mates\xfnm[ J.]}, \bibinfo{author}{Nisi\xfnm[ S.]},
  \bibinfo{author}{Orlando\xfnm[ A.]}, \bibinfo{author}{Parodi\xfnm[ L.]},
  \bibinfo{author}{Pessina\xfnm[ G.]}, \bibinfo{author}{Puiu\xfnm[ A.]},
  \bibinfo{author}{Ragazzi\xfnm[ S.]}, \bibinfo{author}{Reintsema\xfnm[ C.]},
  \bibinfo{author}{Ribeiro-Gomez\xfnm[ M.]}, \bibinfo{author}{Schmidt\xfnm[
  D.]}, \bibinfo{author}{Schuman\xfnm[ D.]}, \bibinfo{author}{Siccardi\xfnm[
  F.]}, \bibinfo{author}{Swetz\xfnm[ D.]}, \bibinfo{author}{Ullom\xfnm[ J.]},
  \bibinfo{author}{Vale\xfnm[ L.]}.
\newblock \bibinfo{title}{{Status of the HOLMES Experiment to Directly Measure
  the Neutrino Mass}}.
\newblock \emph{\bibinfo{journal}{Journal of Low Temperature Physics}}
  \bibinfo{year}{2018};\bibinfo{volume}{193}(\bibinfo{number}{5-6}):\bibinfo{pages}{1137--1145}.
\newblock \DOIprefix\doi{10.1007/s10909-018-2025-x}.
  \href{http://arxiv.org/abs/1807.09269}{\tt arXiv:1807.09269}.
%Type = Article
\bibitem[{Borghesi et~al.(2023)Borghesi, Alpert, Balata, Becker, Bennet,
  Celasco, Cerboni, {De Gerone}, Dressler, Faverzani, Fedkevych, Ferri, Fowler,
  Gallucci, Gard, Gatti, Giachero, Hilton, Koster, Labranca, Lusignoli, Mates,
  Maugeri, Nisi, Nucciotti, Origo, Pessina, Ragazzi, Reintsema, Schmidt,
  Schumann, Swetz, Ullom and Vale}]{HOLMES23}
\bibinfo{author}{Borghesi\xfnm[ M.]}, \bibinfo{author}{Alpert\xfnm[ B.]},
  \bibinfo{author}{Balata\xfnm[ M.]}, \bibinfo{author}{Becker\xfnm[ D.]},
  \bibinfo{author}{Bennet\xfnm[ D.]}, \bibinfo{author}{Celasco\xfnm[ E.]},
  \bibinfo{author}{Cerboni\xfnm[ N.]}, \bibinfo{author}{{De Gerone}\xfnm[ M.]},
  \bibinfo{author}{Dressler\xfnm[ R.]}, \bibinfo{author}{Faverzani\xfnm[ M.]},
  \bibinfo{author}{Fedkevych\xfnm[ M.]}, \bibinfo{author}{Ferri\xfnm[ E.]},
  \bibinfo{author}{Fowler\xfnm[ J.]}, \bibinfo{author}{Gallucci\xfnm[ G.]},
  \bibinfo{author}{Gard\xfnm[ J.]}, \bibinfo{author}{Gatti\xfnm[ F.]},
  \bibinfo{author}{Giachero\xfnm[ A.]}, \bibinfo{author}{Hilton\xfnm[ G.]},
  \bibinfo{author}{Koster\xfnm[ U.]}, \bibinfo{author}{Labranca\xfnm[ D.]},
  \bibinfo{author}{Lusignoli\xfnm[ M.]}, \bibinfo{author}{Mates\xfnm[ J.]},
  \bibinfo{author}{Maugeri\xfnm[ E.]}, \bibinfo{author}{Nisi\xfnm[ S.]},
  \bibinfo{author}{Nucciotti\xfnm[ A.]}, \bibinfo{author}{Origo\xfnm[ L.]},
  \bibinfo{author}{Pessina\xfnm[ G.]}, \bibinfo{author}{Ragazzi\xfnm[ S.]},
  \bibinfo{author}{Reintsema\xfnm[ C.]}, \bibinfo{author}{Schmidt\xfnm[ D.]},
  \bibinfo{author}{Schumann\xfnm[ D.]}, \bibinfo{author}{Swetz\xfnm[ D.]},
  \bibinfo{author}{Ullom\xfnm[ J.]}, \bibinfo{author}{Vale\xfnm[ L.]}.
\newblock \bibinfo{title}{An updated overview of the holmes status}.
\newblock \emph{\bibinfo{journal}{Nuclear Instruments and Methods in Physics
  Research Section A: Accelerators, Spectrometers, Detectors and Associated
  Equipment}}
  \bibinfo{year}{2023};\bibinfo{volume}{1051}:\bibinfo{pages}{168205}.
\newblock \URLprefix
  \url{https://www.sciencedirect.com/science/article/pii/S016890022300195X}.
  \DOIprefix\doi{https://doi.org/10.1016/j.nima.2023.168205}.
%Type = Article
\bibitem[{McDonald et~al.(2013)McDonald, Drexlin, Hannen, Mertens and
  Weinheimer}]{McDonald2013}
\bibinfo{author}{McDonald\xfnm[ A.B.]}, \bibinfo{author}{Drexlin\xfnm[ G.]},
  \bibinfo{author}{Hannen\xfnm[ V.]}, \bibinfo{author}{Mertens\xfnm[ S.]},
  \bibinfo{author}{Weinheimer\xfnm[ C.]}.
\newblock \bibinfo{title}{Current direct neutrino mass experiments}.
\newblock \emph{\bibinfo{journal}{Advances in High Energy Physics}}
  \bibinfo{year}{2013};\bibinfo{volume}{2013}:\bibinfo{pages}{293986}.
\newblock \URLprefix \url{https://doi.org/10.1155/2013/293986}.
  \DOIprefix\doi{10.1155/2013/293986}.
%Type = Article
\bibitem[{Ferri et~al.(2015)Ferri, Bagliani, Biasotti, Ceruti, Corsini,
  Faverzani, Gatti, Giachero, Gotti, Kilbourne, Kling, Maino, Manfrinetti,
  Nucciotti, Pessina, Pizzigoni, {Ribeiro Gomes} and Sisti}]{Ferri2015}
\bibinfo{author}{Ferri\xfnm[ E.]}, \bibinfo{author}{Bagliani\xfnm[ D.]},
  \bibinfo{author}{Biasotti\xfnm[ M.]}, \bibinfo{author}{Ceruti\xfnm[ G.]},
  \bibinfo{author}{Corsini\xfnm[ D.]}, \bibinfo{author}{Faverzani\xfnm[ M.]},
  \bibinfo{author}{Gatti\xfnm[ F.]}, \bibinfo{author}{Giachero\xfnm[ A.]},
  \bibinfo{author}{Gotti\xfnm[ C.]}, \bibinfo{author}{Kilbourne\xfnm[ C.]},
  \bibinfo{author}{Kling\xfnm[ A.]}, \bibinfo{author}{Maino\xfnm[ M.]},
  \bibinfo{author}{Manfrinetti\xfnm[ P.]}, \bibinfo{author}{Nucciotti\xfnm[
  A.]}, \bibinfo{author}{Pessina\xfnm[ G.]}, \bibinfo{author}{Pizzigoni\xfnm[
  G.]}, \bibinfo{author}{{Ribeiro Gomes}\xfnm[ M.]},
  \bibinfo{author}{Sisti\xfnm[ M.]}.
\newblock \bibinfo{title}{{The Status of the MARE Experiment with 187Re and
  163Ho Isotopes}}.
\newblock \emph{\bibinfo{journal}{Physics Procedia}}
  \bibinfo{year}{2015};\bibinfo{volume}{61}(\bibinfo{number}{August}):\bibinfo{pages}{227--231}.
\newblock \URLprefix \url{http://dx.doi.org/10.1016/j.phpro.2014.12.037}.
  \DOIprefix\doi{10.1016/j.phpro.2014.12.037}.
%Type = Article
\bibitem[{Haaranen and Suhonen(2013)}]{Haaranen2013}
\bibinfo{author}{Haaranen\xfnm[ M.]}, \bibinfo{author}{Suhonen\xfnm[ J.]}.
\newblock \bibinfo{title}{{Beta decay of 115Cd and its possible ultra-low
  Q-value branch}}.
\newblock \emph{\bibinfo{journal}{The European Physical Journal A}}
  \bibinfo{year}{2013};\bibinfo{volume}{49}(\bibinfo{number}{7}):\bibinfo{pages}{1--9}.
\newblock \URLprefix \url{http://dx.doi.org/10.1140/epja/i2013-13093-8}.
  \DOIprefix\doi{10.1140/epja/i2013-13093-8}.
%Type = Article
\bibitem[{Suhonen(2014)}]{Suhonen2014}
\bibinfo{author}{Suhonen\xfnm[ J.]}.
\newblock \bibinfo{title}{{Theoretical studies of rare weak processes in
  nuclei}}.
\newblock \emph{\bibinfo{journal}{Physica Scripta}}
  \bibinfo{year}{2014};\bibinfo{volume}{89}(\bibinfo{number}{5}):\bibinfo{pages}{54032}.
\newblock \URLprefix \url{http://stacks.iop.org/1402-4896/89/i=5/a=054032}.
  \DOIprefix\doi{10.1088/0031-8949/89/5/054032}.
%Type = Article
\bibitem[{Sandler et~al.(2019)Sandler, Bollen, Gamage, Hamaker, Izzo, Puentes,
  Redshaw, Ringle and Yandow}]{Sandler2019}
\bibinfo{author}{Sandler\xfnm[ R.]}, \bibinfo{author}{Bollen\xfnm[ G.]},
  \bibinfo{author}{Gamage\xfnm[ N.D.]}, \bibinfo{author}{Hamaker\xfnm[ A.]},
  \bibinfo{author}{Izzo\xfnm[ C.]}, \bibinfo{author}{Puentes\xfnm[ D.]},
  \bibinfo{author}{Redshaw\xfnm[ M.]}, \bibinfo{author}{Ringle\xfnm[ R.]},
  \bibinfo{author}{Yandow\xfnm[ I.]}.
\newblock \bibinfo{title}{{Investigation of the potential ultralow Q-value
  $\beta$-decay candidates Sr 89 and Ba 139 using Penning trap mass
  spectrometry}}.
\newblock \emph{\bibinfo{journal}{Physical Review C}}
  \bibinfo{year}{2019};\bibinfo{volume}{100}(\bibinfo{number}{2}):\bibinfo{pages}{1--5}.
\newblock \URLprefix
  \url{https://link.aps.org/doi/10.1103/PhysRevC.100.024309}.
  \DOIprefix\doi{10.1103/PhysRevC.100.024309}.
  \href{http://arxiv.org/abs/1906.03335}{\tt arXiv:1906.03335}.
%Type = Article
\bibitem[{Karthein et~al.(2019)Karthein, Atanasov, Blaum, Eliseev, Filianin,
  Lunney, Manea, Mougeot, Neidherr, Novikov, Schweikhard, Welker, Wienholtz and
  Zuber}]{Karthein2019a}
\bibinfo{author}{Karthein\xfnm[ J.]}, \bibinfo{author}{Atanasov\xfnm[ D.]},
  \bibinfo{author}{Blaum\xfnm[ K.]}, \bibinfo{author}{Eliseev\xfnm[ S.]},
  \bibinfo{author}{Filianin\xfnm[ P.]}, \bibinfo{author}{Lunney\xfnm[ D.]},
  \bibinfo{author}{Manea\xfnm[ V.]}, \bibinfo{author}{Mougeot\xfnm[ M.]},
  \bibinfo{author}{Neidherr\xfnm[ D.]}, \bibinfo{author}{Novikov\xfnm[ Y.]},
  \bibinfo{author}{Schweikhard\xfnm[ L.]}, \bibinfo{author}{Welker\xfnm[ A.]},
  \bibinfo{author}{Wienholtz\xfnm[ F.]}, \bibinfo{author}{Zuber\xfnm[ K.]}.
\newblock \bibinfo{title}{{Direct decay-energy measurement as a route to the
  neutrino mass}}.
\newblock \emph{\bibinfo{journal}{Hyperfine Interactions}}
  \bibinfo{year}{2019};\bibinfo{volume}{240}(\bibinfo{number}{1}):\bibinfo{pages}{1--9}.
\newblock \URLprefix \url{http://link.springer.com/10.1007/s10751-019-1601-z}.
  \DOIprefix\doi{10.1007/s10751-019-1601-z}.
  \href{http://arxiv.org/abs/1905.05510}{\tt arXiv:1905.05510}.
%Type = Article
\bibitem[{{De Roubin} et~al.(2020){De Roubin}, Kostensalo, Eronen, Canete, {De
  Groote}, Jokinen, Kankainen, Nesterenko, Moore, Rinta-Antila, Suhonen and
  Vil{\'{e}}n}]{DeRoubin2020}
\bibinfo{author}{{De Roubin}\xfnm[ A.]}, \bibinfo{author}{Kostensalo\xfnm[
  J.]}, \bibinfo{author}{Eronen\xfnm[ T.]}, \bibinfo{author}{Canete\xfnm[ L.]},
  \bibinfo{author}{{De Groote}\xfnm[ R.P.]}, \bibinfo{author}{Jokinen\xfnm[
  A.]}, \bibinfo{author}{Kankainen\xfnm[ A.]},
  \bibinfo{author}{Nesterenko\xfnm[ D.A.]}, \bibinfo{author}{Moore\xfnm[
  I.D.]}, \bibinfo{author}{Rinta-Antila\xfnm[ S.]},
  \bibinfo{author}{Suhonen\xfnm[ J.]}, \bibinfo{author}{Vil{\'{e}}n\xfnm[ M.]}.
\newblock \bibinfo{title}{{High-Precision Q -Value Measurement Confirms the
  Potential of Cs 135 for Absolute Antineutrino Mass Scale Determination}}.
\newblock \emph{\bibinfo{journal}{Physical Review Letters}}
  \bibinfo{year}{2020};\bibinfo{volume}{124}(\bibinfo{number}{22}):\bibinfo{pages}{1--5}.
\newblock \URLprefix \url{https://doi.org/10.1103/PhysRevLett.124.222503}.
  \DOIprefix\doi{10.1103/PhysRevLett.124.222503}.
  \href{http://arxiv.org/abs/2002.08282}{\tt arXiv:2002.08282}.
%Type = Article
\bibitem[{Ge et~al.(2021{\natexlab{a}})Ge, Eronen, de~Roubin, Nesterenko,
  Hukkanen, Beliuskina, de~Groote, Geldhof, Gins, Kankainen, Koszor\'us,
  Kotila, Kostensalo, Moore, Raggio, Rinta-Antila, Suhonen, Virtanen, Weaver,
  Zadvornaya and Jokinen}]{ge2021}
\bibinfo{author}{Ge\xfnm[ Z.]}, \bibinfo{author}{Eronen\xfnm[ T.]},
  \bibinfo{author}{de~Roubin\xfnm[ A.]}, \bibinfo{author}{Nesterenko\xfnm[
  D.A.]}, \bibinfo{author}{Hukkanen\xfnm[ M.]},
  \bibinfo{author}{Beliuskina\xfnm[ O.]}, \bibinfo{author}{de~Groote\xfnm[
  R.]}, \bibinfo{author}{Geldhof\xfnm[ S.]}, \bibinfo{author}{Gins\xfnm[ W.]},
  \bibinfo{author}{Kankainen\xfnm[ A.]}, \bibinfo{author}{Koszor\'us\xfnm[
  A.]}, \bibinfo{author}{Kotila\xfnm[ J.]}, \bibinfo{author}{Kostensalo\xfnm[
  J.]}, \bibinfo{author}{Moore\xfnm[ I.D.]}, \bibinfo{author}{Raggio\xfnm[
  A.]}, \bibinfo{author}{Rinta-Antila\xfnm[ S.]},
  \bibinfo{author}{Suhonen\xfnm[ J.]}, \bibinfo{author}{Virtanen\xfnm[ V.]},
  \bibinfo{author}{Weaver\xfnm[ A.P.]}, \bibinfo{author}{Zadvornaya\xfnm[ A.]},
  \bibinfo{author}{Jokinen\xfnm[ A.]}.
\newblock \bibinfo{title}{Direct measurement of the mass difference of
  $^{72}\mathrm{As}\text{\ensuremath{-}}^{72}\mathrm{Ge}$ rules out
  $^{72}\mathrm{As}$ as a promising $\ensuremath{\beta}$-decay candidate to
  determine the neutrino mass}.
\newblock \emph{\bibinfo{journal}{Physical Review C}}
  \bibinfo{year}{2021}{\natexlab{a}};\bibinfo{volume}{103}:\bibinfo{pages}{065502}.
\newblock \URLprefix
  \url{https://link.aps.org/doi/10.1103/PhysRevC.103.065502}.
  \DOIprefix\doi{10.1103/PhysRevC.103.065502}.
%Type = Article
\bibitem[{Ge et~al.(2021{\natexlab{b}})Ge, Eronen, Tyrin, Kotila, Kostensalo,
  Nesterenko, Beliuskina, de~Groote, de~Roubin, Geldhof, Gins, Hukkanen,
  Jokinen, Kankainen, Koszor\'us, Krivoruchenko, Kujanp\"a\"a, Moore, Raggio,
  Rinta-Antila, Suhonen, Virtanen, Weaver and Zadvornaya}]{ge2021b}
\bibinfo{author}{Ge\xfnm[ Z.]}, \bibinfo{author}{Eronen\xfnm[ T.]},
  \bibinfo{author}{Tyrin\xfnm[ K.S.]}, \bibinfo{author}{Kotila\xfnm[ J.]},
  \bibinfo{author}{Kostensalo\xfnm[ J.]}, \bibinfo{author}{Nesterenko\xfnm[
  D.A.]}, \bibinfo{author}{Beliuskina\xfnm[ O.]},
  \bibinfo{author}{de~Groote\xfnm[ R.]}, \bibinfo{author}{de~Roubin\xfnm[ A.]},
  \bibinfo{author}{Geldhof\xfnm[ S.]}, \bibinfo{author}{Gins\xfnm[ W.]},
  \bibinfo{author}{Hukkanen\xfnm[ M.]}, \bibinfo{author}{Jokinen\xfnm[ A.]},
  \bibinfo{author}{Kankainen\xfnm[ A.]}, \bibinfo{author}{Koszor\'us\xfnm[
  A.]}, \bibinfo{author}{Krivoruchenko\xfnm[ M.I.]},
  \bibinfo{author}{Kujanp\"a\"a\xfnm[ S.]}, \bibinfo{author}{Moore\xfnm[
  I.D.]}, \bibinfo{author}{Raggio\xfnm[ A.]},
  \bibinfo{author}{Rinta-Antila\xfnm[ S.]}, \bibinfo{author}{Suhonen\xfnm[
  J.]}, \bibinfo{author}{Virtanen\xfnm[ V.]}, \bibinfo{author}{Weaver\xfnm[
  A.P.]}, \bibinfo{author}{Zadvornaya\xfnm[ A.]}.
\newblock \bibinfo{title}{$^{159}\mathrm{Dy}$ electron-capture: A new candidate
  for neutrino mass determination}.
\newblock \emph{\bibinfo{journal}{Phys Rev Lett}}
  \bibinfo{year}{2021}{\natexlab{b}};\bibinfo{volume}{127}:\bibinfo{pages}{272301}.
\newblock \URLprefix
  \url{https://link.aps.org/doi/10.1103/PhysRevLett.127.272301}.
  \DOIprefix\doi{10.1103/PhysRevLett.127.272301}.
%Type = Article
\bibitem[{Ge et~al.(2022{\natexlab{a}})Ge, Eronen, {de Roubin}, Tyrin, Canete,
  Geldhof, Jokinen, Kankainen, Kostensalo, Kotila, Krivoruchenko, Moore,
  Nesterenko, Suhonen and Vilén}]{Ge2022a}
\bibinfo{author}{Ge\xfnm[ Z.]}, \bibinfo{author}{Eronen\xfnm[ T.]},
  \bibinfo{author}{{de Roubin}\xfnm[ A.]}, \bibinfo{author}{Tyrin\xfnm[ K.]},
  \bibinfo{author}{Canete\xfnm[ L.]}, \bibinfo{author}{Geldhof\xfnm[ S.]},
  \bibinfo{author}{Jokinen\xfnm[ A.]}, \bibinfo{author}{Kankainen\xfnm[ A.]},
  \bibinfo{author}{Kostensalo\xfnm[ J.]}, \bibinfo{author}{Kotila\xfnm[ J.]},
  \bibinfo{author}{Krivoruchenko\xfnm[ M.]}, \bibinfo{author}{Moore\xfnm[ I.]},
  \bibinfo{author}{Nesterenko\xfnm[ D.]}, \bibinfo{author}{Suhonen\xfnm[ J.]},
  \bibinfo{author}{Vilén\xfnm[ M.]}.
\newblock \bibinfo{title}{High-precision electron-capture q value measurement
  of 111in for electron-neutrino mass determination}.
\newblock \emph{\bibinfo{journal}{Physics Letters B}}
  \bibinfo{year}{2022}{\natexlab{a}};\bibinfo{volume}{832}:\bibinfo{pages}{137226}.
\newblock \URLprefix
  \url{https://www.sciencedirect.com/science/article/pii/S0370269322003604}.
  \DOIprefix\doi{https://doi.org/10.1016/j.physletb.2022.137226}.
%Type = Article
\bibitem[{Eronen et~al.(2022)Eronen, Ge, {de Roubin}, Ramalho, Kostensalo,
  Kotila, Beliushkina, Delafosse, Geldhof, Gins, Hukkanen, Jokinen, Kankainen,
  Moore, Nesterenko, Stryjczyk and Suhonen}]{ERONEN2022}
\bibinfo{author}{Eronen\xfnm[ T.]}, \bibinfo{author}{Ge\xfnm[ Z.]},
  \bibinfo{author}{{de Roubin}\xfnm[ A.]}, \bibinfo{author}{Ramalho\xfnm[ M.]},
  \bibinfo{author}{Kostensalo\xfnm[ J.]}, \bibinfo{author}{Kotila\xfnm[ J.]},
  \bibinfo{author}{Beliushkina\xfnm[ O.]}, \bibinfo{author}{Delafosse\xfnm[
  C.]}, \bibinfo{author}{Geldhof\xfnm[ S.]}, \bibinfo{author}{Gins\xfnm[ W.]},
  \bibinfo{author}{Hukkanen\xfnm[ M.]}, \bibinfo{author}{Jokinen\xfnm[ A.]},
  \bibinfo{author}{Kankainen\xfnm[ A.]}, \bibinfo{author}{Moore\xfnm[ I.]},
  \bibinfo{author}{Nesterenko\xfnm[ D.]}, \bibinfo{author}{Stryjczyk\xfnm[
  M.]}, \bibinfo{author}{Suhonen\xfnm[ J.]}.
\newblock \bibinfo{title}{High-precision measurement of a low q value for
  allowed beta-decay of 131i related to neutrino mass determination}.
\newblock \emph{\bibinfo{journal}{Physics Letters B}}
  \bibinfo{year}{2022};\bibinfo{volume}{830}:\bibinfo{pages}{137135}.
\newblock \URLprefix
  \url{https://www.sciencedirect.com/science/article/pii/S0370269322002696}.
  \DOIprefix\doi{https://doi.org/10.1016/j.physletb.2022.137135}.
%Type = Article
\bibitem[{Ge et~al.(2022{\natexlab{b}})Ge, Eronen, de~Roubin, Kostensalo,
  Suhonen, Nesterenko, Beliuskina, de~Groote, Delafosse, Geldhof, Gins,
  Hukkanen, Jokinen, Kankainen, Kotila, Koszor\'us, Moore, Raggio,
  Rinta-Antila, Virtanen, Weaver and Zadvornaya}]{Ge2022b}
\bibinfo{author}{Ge\xfnm[ Z.]}, \bibinfo{author}{Eronen\xfnm[ T.]},
  \bibinfo{author}{de~Roubin\xfnm[ A.]}, \bibinfo{author}{Kostensalo\xfnm[
  J.]}, \bibinfo{author}{Suhonen\xfnm[ J.]}, \bibinfo{author}{Nesterenko\xfnm[
  D.A.]}, \bibinfo{author}{Beliuskina\xfnm[ O.]},
  \bibinfo{author}{de~Groote\xfnm[ R.]}, \bibinfo{author}{Delafosse\xfnm[ C.]},
  \bibinfo{author}{Geldhof\xfnm[ S.]}, \bibinfo{author}{Gins\xfnm[ W.]},
  \bibinfo{author}{Hukkanen\xfnm[ M.]}, \bibinfo{author}{Jokinen\xfnm[ A.]},
  \bibinfo{author}{Kankainen\xfnm[ A.]}, \bibinfo{author}{Kotila\xfnm[ J.]},
  \bibinfo{author}{Koszor\'us\xfnm[ A.]}, \bibinfo{author}{Moore\xfnm[ I.D.]},
  \bibinfo{author}{Raggio\xfnm[ A.]}, \bibinfo{author}{Rinta-Antila\xfnm[ S.]},
  \bibinfo{author}{Virtanen\xfnm[ V.]}, \bibinfo{author}{Weaver\xfnm[ A.P.]},
  \bibinfo{author}{Zadvornaya\xfnm[ A.]}.
\newblock \bibinfo{title}{Direct determination of the atomic mass difference of
  the pairs $^{76}\mathrm{As}\text{\ensuremath{-}}^{76}\mathrm{Se}$ and
  $^{155}\mathrm{Tb}\text{\ensuremath{-}}^{155}\mathrm{Gd}$ rules out
  $^{76}\mathrm{As}$ and $^{155}\mathrm{Tb}$ as possible candidates for
  electron (anti)neutrino mass measurements}.
\newblock \emph{\bibinfo{journal}{Phys Rev C}}
  \bibinfo{year}{2022}{\natexlab{b}};\bibinfo{volume}{106}:\bibinfo{pages}{015502}.
\newblock \URLprefix
  \url{https://link.aps.org/doi/10.1103/PhysRevC.106.015502}.
  \DOIprefix\doi{10.1103/PhysRevC.106.015502}.
%Type = Article
\bibitem[{Ramalho et~al.(2022)Ramalho, Ge, Eronen, Nesterenko, Jaatinen,
  Jokinen, Kankainen, Kostensalo, Kotila, Krivoruchenko, Suhonen, Tyrin and
  Virtanen}]{Ramalho2022}
\bibinfo{author}{Ramalho\xfnm[ M.]}, \bibinfo{author}{Ge\xfnm[ Z.]},
  \bibinfo{author}{Eronen\xfnm[ T.]}, \bibinfo{author}{Nesterenko\xfnm[ D.A.]},
  \bibinfo{author}{Jaatinen\xfnm[ J.]}, \bibinfo{author}{Jokinen\xfnm[ A.]},
  \bibinfo{author}{Kankainen\xfnm[ A.]}, \bibinfo{author}{Kostensalo\xfnm[
  J.]}, \bibinfo{author}{Kotila\xfnm[ J.]},
  \bibinfo{author}{Krivoruchenko\xfnm[ M.I.]}, \bibinfo{author}{Suhonen\xfnm[
  J.]}, \bibinfo{author}{Tyrin\xfnm[ K.S.]}, \bibinfo{author}{Virtanen\xfnm[
  V.]}.
\newblock \bibinfo{title}{Observation of an ultralow-$q$-value electron-capture
  channel decaying to $^{75}\mathrm{As}$ via a high-precision mass
  measurement}.
\newblock \emph{\bibinfo{journal}{Phys Rev C}}
  \bibinfo{year}{2022};\bibinfo{volume}{106}:\bibinfo{pages}{015501}.
\newblock \URLprefix
  \url{https://link.aps.org/doi/10.1103/PhysRevC.106.015501}.
  \DOIprefix\doi{10.1103/PhysRevC.106.015501}.
%Type = Article
\bibitem[{Gamage et~al.(2022)Gamage, Sandler, Buchinger, Clark, Ray, Orford,
  Porter, Redshaw, Savard, Sharma and Valverde}]{Gamage22}
\bibinfo{author}{Gamage\xfnm[ N.D.]}, \bibinfo{author}{Sandler\xfnm[ R.]},
  \bibinfo{author}{Buchinger\xfnm[ F.]}, \bibinfo{author}{Clark\xfnm[ J.A.]},
  \bibinfo{author}{Ray\xfnm[ D.]}, \bibinfo{author}{Orford\xfnm[ R.]},
  \bibinfo{author}{Porter\xfnm[ W.S.]}, \bibinfo{author}{Redshaw\xfnm[ M.]},
  \bibinfo{author}{Savard\xfnm[ G.]}, \bibinfo{author}{Sharma\xfnm[ K.S.]},
  \bibinfo{author}{Valverde\xfnm[ A.A.]}.
\newblock \bibinfo{title}{Precise $q$-value measurements of
  $^{112,113}\mathrm{Ag}$ and $^{115}\mathrm{Cd}$ with the canadian penning
  trap for evaluation of potential ultralow $q$-value $\ensuremath{\beta}$
  decays}.
\newblock \emph{\bibinfo{journal}{Phys Rev C}}
  \bibinfo{year}{2022};\bibinfo{volume}{106}:\bibinfo{pages}{045503}.
\newblock \URLprefix
  \url{https://link.aps.org/doi/10.1103/PhysRevC.106.045503}.
  \DOIprefix\doi{10.1103/PhysRevC.106.045503}.
%Type = Article
\bibitem[{Keblbeck et~al.(2023)Keblbeck, Bhandari, Gamage, Horana~Gamage,
  Leach, Mougeot and Redshaw}]{Keblbeck2023}
\bibinfo{author}{Keblbeck\xfnm[ D.K.]}, \bibinfo{author}{Bhandari\xfnm[ R.]},
  \bibinfo{author}{Gamage\xfnm[ N.D.]}, \bibinfo{author}{Horana~Gamage\xfnm[
  M.]}, \bibinfo{author}{Leach\xfnm[ K.G.]}, \bibinfo{author}{Mougeot\xfnm[
  X.]}, \bibinfo{author}{Redshaw\xfnm[ M.]}.
\newblock \bibinfo{title}{Updated evaluation of potential ultralow $q$-value
  $\ensuremath{\beta}$-decay candidates}.
\newblock \emph{\bibinfo{journal}{Phys Rev C}}
  \bibinfo{year}{2023};\bibinfo{volume}{107}:\bibinfo{pages}{015504}.
\newblock \URLprefix
  \url{https://link.aps.org/doi/10.1103/PhysRevC.107.015504}.
  \DOIprefix\doi{10.1103/PhysRevC.107.015504}.
%Type = Article
\bibitem[{Ge et~al.(2023)Ge, Eronen, de~Roubin, Ramalho, Kostensalo, Kotila,
  Suhonen, Nesterenko, Kankainen, Ascher, Beliuskina, Flayol, Gerbaux, Gr\'evy,
  Hukkanen, Husson, Jaries, Jokinen, Moore, Pirinen, Romero, Stryjczyk,
  Virtanen and Zadvornaya}]{Ge2023}
\bibinfo{author}{Ge\xfnm[ Z.]}, \bibinfo{author}{Eronen\xfnm[ T.]},
  \bibinfo{author}{de~Roubin\xfnm[ A.]}, \bibinfo{author}{Ramalho\xfnm[ M.]},
  \bibinfo{author}{Kostensalo\xfnm[ J.]}, \bibinfo{author}{Kotila\xfnm[ J.]},
  \bibinfo{author}{Suhonen\xfnm[ J.]}, \bibinfo{author}{Nesterenko\xfnm[
  D.A.]}, \bibinfo{author}{Kankainen\xfnm[ A.]}, \bibinfo{author}{Ascher\xfnm[
  P.]}, \bibinfo{author}{Beliuskina\xfnm[ O.]}, \bibinfo{author}{Flayol\xfnm[
  M.]}, \bibinfo{author}{Gerbaux\xfnm[ M.]}, \bibinfo{author}{Gr\'evy\xfnm[
  S.]}, \bibinfo{author}{Hukkanen\xfnm[ M.]}, \bibinfo{author}{Husson\xfnm[
  A.]}, \bibinfo{author}{Jaries\xfnm[ A.]}, \bibinfo{author}{Jokinen\xfnm[
  A.]}, \bibinfo{author}{Moore\xfnm[ I.D.]}, \bibinfo{author}{Pirinen\xfnm[
  P.]}, \bibinfo{author}{Romero\xfnm[ J.]}, \bibinfo{author}{Stryjczyk\xfnm[
  M.]}, \bibinfo{author}{Virtanen\xfnm[ V.]}, \bibinfo{author}{Zadvornaya\xfnm[
  A.]}.
\newblock \bibinfo{title}{${\ensuremath{\beta}}^{\ensuremath{-}}$ decay
  $q$-value measurement of $^{136}\mathrm{Cs}$ and its implications for
  neutrino studies}.
\newblock \emph{\bibinfo{journal}{Phys Rev C}}
  \bibinfo{year}{2023};\bibinfo{volume}{108}:\bibinfo{pages}{045502}.
\newblock \URLprefix
  \url{https://link.aps.org/doi/10.1103/PhysRevC.108.045502}.
  \DOIprefix\doi{10.1103/PhysRevC.108.045502}.
%Type = Article
\bibitem[{Eronen and Hardy(2012)}]{Eronen2012}
\bibinfo{author}{Eronen\xfnm[ T.]}, \bibinfo{author}{Hardy\xfnm[ J.C.]}.
\newblock \bibinfo{title}{{High-precision $Q_{EC}$-value measurements for
  superallowed decays}}.
\newblock \emph{\bibinfo{journal}{European Physical Journal A}}
  \bibinfo{year}{2012};\bibinfo{volume}{48}(\bibinfo{number}{4}):\bibinfo{pages}{1--8}.
\newblock \URLprefix \url{http://dx.doi.org/10.1140/epja/i2012-12048-y}.
  \DOIprefix\doi{10.1140/epja/i2012-12048-y}.
%Type = Article
\bibitem[{Moore et~al.(2013)Moore, Eronen, Gorelov, Hakala, Jokinen, Kankainen,
  Kolhinen, Koponen, Penttil{\"{a}}, Pohjalainen, Reponen, Rissanen,
  Saastamoinen, Rinta-Antila, Sonnenschein and {\"{A}}yst{\"{o}}}]{Moore2013}
\bibinfo{author}{Moore\xfnm[ I.D.]}, \bibinfo{author}{Eronen\xfnm[ T.]},
  \bibinfo{author}{Gorelov\xfnm[ D.]}, \bibinfo{author}{Hakala\xfnm[ J.]},
  \bibinfo{author}{Jokinen\xfnm[ A.]}, \bibinfo{author}{Kankainen\xfnm[ A.]},
  \bibinfo{author}{Kolhinen\xfnm[ V.S.]}, \bibinfo{author}{Koponen\xfnm[ J.]},
  \bibinfo{author}{Penttil{\"{a}}\xfnm[ H.]},
  \bibinfo{author}{Pohjalainen\xfnm[ I.]}, \bibinfo{author}{Reponen\xfnm[ M.]},
  \bibinfo{author}{Rissanen\xfnm[ J.]}, \bibinfo{author}{Saastamoinen\xfnm[
  A.]}, \bibinfo{author}{Rinta-Antila\xfnm[ S.]},
  \bibinfo{author}{Sonnenschein\xfnm[ V.]},
  \bibinfo{author}{{\"{A}}yst{\"{o}}\xfnm[ J.]}.
\newblock \bibinfo{title}{{Towards commissioning the new IGISOL-4 facility}}.
\newblock \emph{\bibinfo{journal}{Nuclear Instruments and Methods in Physics
  Research, Section B: Beam Interactions with Materials and Atoms}}
  \bibinfo{year}{2013};\bibinfo{volume}{317}(\bibinfo{number}{PART
  B}):\bibinfo{pages}{208--213}.
\newblock \URLprefix
  \url{http://www.sciencedirect.com/science/article/pii/S0168583X13007143
  http://dx.doi.org/10.1016/j.nimb.2013.06.036}.
  \DOIprefix\doi{10.1016/j.nimb.2013.06.036}.
%Type = Article
\bibitem[{Kolhinen et~al.(2013)Kolhinen, Eronen, Gorelov, Hakala, Jokinen,
  Jokiranta, Kankainen, Koikkalainen, Koponen, Kulmala, Lantz, Mattera, Moore,
  Penttil{\"{a}}, Pikkarainen, Pohjlainen, Reponen, Rinta-Antila, Rissanen,
  {Rodr{\'{i}}guez Triguero}, Rytk{\"{o}}nen, Saastamoinen, Solders,
  Sonnenschein and {\"{A}}yst{\"{o}}}]{Kolhinen2013}
\bibinfo{author}{Kolhinen\xfnm[ V.S.]}, \bibinfo{author}{Eronen\xfnm[ T.]},
  \bibinfo{author}{Gorelov\xfnm[ D.]}, \bibinfo{author}{Hakala\xfnm[ J.]},
  \bibinfo{author}{Jokinen\xfnm[ A.]}, \bibinfo{author}{Jokiranta\xfnm[ K.]},
  \bibinfo{author}{Kankainen\xfnm[ A.]}, \bibinfo{author}{Koikkalainen\xfnm[
  M.]}, \bibinfo{author}{Koponen\xfnm[ J.]}, \bibinfo{author}{Kulmala\xfnm[
  H.]}, \bibinfo{author}{Lantz\xfnm[ M.]}, \bibinfo{author}{Mattera\xfnm[ A.]},
  \bibinfo{author}{Moore\xfnm[ I.D.]}, \bibinfo{author}{Penttil{\"{a}}\xfnm[
  H.]}, \bibinfo{author}{Pikkarainen\xfnm[ T.]},
  \bibinfo{author}{Pohjlainen\xfnm[ I.]}, \bibinfo{author}{Reponen\xfnm[ M.]},
  \bibinfo{author}{Rinta-Antila\xfnm[ S.]}, \bibinfo{author}{Rissanen\xfnm[
  J.]}, \bibinfo{author}{{Rodr{\'{i}}guez Triguero}\xfnm[ C.]},
  \bibinfo{author}{Rytk{\"{o}}nen\xfnm[ K.]},
  \bibinfo{author}{Saastamoinen\xfnm[ A.]}, \bibinfo{author}{Solders\xfnm[
  A.]}, \bibinfo{author}{Sonnenschein\xfnm[ V.]},
  \bibinfo{author}{{\"{A}}yst{\"{o}}\xfnm[ J.]}.
\newblock \bibinfo{title}{{Recommissioning of JYFLTRAP at the new IGISOL-4
  facility}}.
\newblock \emph{\bibinfo{journal}{Nuclear Instruments and Methods in Physics
  Research, Section B: Beam Interactions with Materials and Atoms}}
  \bibinfo{year}{2013};\bibinfo{volume}{317}(\bibinfo{number}{PART
  B}):\bibinfo{pages}{506--509}.
\newblock \URLprefix
  \url{http://www.sciencedirect.com/science/article/pii/S0168583X13008641}.
  \DOIprefix\doi{10.1016/j.nimb.2013.07.050}.
%Type = Article
\bibitem[{Karvonen et~al.(2008)Karvonen, Moore, Sonoda, Kessler,
  Penttil{\"{a}}, Per{\"{a}}j{\"{a}}rvi, Ronkanen and
  {\"{A}}yst{\"{o}}}]{Karvonen2008}
\bibinfo{author}{Karvonen\xfnm[ P.]}, \bibinfo{author}{Moore\xfnm[ I.D.]},
  \bibinfo{author}{Sonoda\xfnm[ T.]}, \bibinfo{author}{Kessler\xfnm[ T.]},
  \bibinfo{author}{Penttil{\"{a}}\xfnm[ H.]},
  \bibinfo{author}{Per{\"{a}}j{\"{a}}rvi\xfnm[ K.]},
  \bibinfo{author}{Ronkanen\xfnm[ P.]},
  \bibinfo{author}{{\"{A}}yst{\"{o}}\xfnm[ J.]}.
\newblock \bibinfo{title}{{A sextupole ion beam guide to improve the efficiency
  and beam quality at IGISOL}}.
\newblock \emph{\bibinfo{journal}{Nuclear Instruments and Methods in Physics
  Research, Section B: Beam Interactions with Materials and Atoms}}
  \bibinfo{year}{2008};\bibinfo{volume}{266}(\bibinfo{number}{21}):\bibinfo{pages}{4794--4807}.
\newblock \URLprefix
  \url{http://www.sciencedirect.com/science/article/B6TJN-4T2S8KR-1/2/1d7624cd369335096dcb1fd81a410fea}.
  \DOIprefix\doi{10.1016/j.nimb.2008.07.022}.
%Type = Article
\bibitem[{Nieminen et~al.(2001)Nieminen, Huikari, Jokinen, {\"{A}}yst{\"{o}},
  Campbell and Cochrane}]{Nieminen2001}
\bibinfo{author}{Nieminen\xfnm[ A.]}, \bibinfo{author}{Huikari\xfnm[ J.]},
  \bibinfo{author}{Jokinen\xfnm[ A.]}, \bibinfo{author}{{\"{A}}yst{\"{o}}\xfnm[
  J.]}, \bibinfo{author}{Campbell\xfnm[ P.]}, \bibinfo{author}{Cochrane\xfnm[
  E.C.]}.
\newblock \bibinfo{title}{{Beam cooler for low-energy radioactive ions}}.
\newblock \emph{\bibinfo{journal}{Nuclear Instruments and Methods in Physics
  Research, Section A: Accelerators, Spectrometers, Detectors and Associated
  Equipment}}
  \bibinfo{year}{2001};\bibinfo{volume}{469}(\bibinfo{number}{2}):\bibinfo{pages}{244--253}.
\newblock \URLprefix
  \url{http://www.sciencedirect.com/science/article/B6TJM-43PGJKX-C/1/93d5587efba5cfe8571b63228952dab8}.
  \DOIprefix\doi{10.1016/S0168-9002(00)00750-6}.
%Type = Article
\bibitem[{Savard et~al.(1991)Savard, Becker, Bollen, Kluge, Moore, Otto,
  Schweikhard, Stolzenberg and Wiess}]{Savard1991}
\bibinfo{author}{Savard\xfnm[ G.]}, \bibinfo{author}{Becker\xfnm[ S.]},
  \bibinfo{author}{Bollen\xfnm[ G.]}, \bibinfo{author}{Kluge\xfnm[ H.J.]},
  \bibinfo{author}{Moore\xfnm[ R.B.]}, \bibinfo{author}{Otto\xfnm[ T.]},
  \bibinfo{author}{Schweikhard\xfnm[ L.]}, \bibinfo{author}{Stolzenberg\xfnm[
  H.]}, \bibinfo{author}{Wiess\xfnm[ U.]}.
\newblock \bibinfo{title}{{A new cooling technique for heavy ions in a Penning
  trap}}.
\newblock \emph{\bibinfo{journal}{Physics Letters A}}
  \bibinfo{year}{1991};\bibinfo{volume}{158}(\bibinfo{number}{5}):\bibinfo{pages}{247--252}.
\newblock \URLprefix \url{https://doi.org/10.1016/0375-9601(91)91008-2}.
  \DOIprefix\doi{10.1016/0375-9601(91)91008-2}.
%Type = Article
\bibitem[{Eronen et~al.(2008)Eronen, Elomaa, Hager, Hakala, Jokinen, Kankainen,
  Rahaman, Rissanen, Weber and {\"{A}}yst{\"{o}}}]{Eronen2008a}
\bibinfo{author}{Eronen\xfnm[ T.]}, \bibinfo{author}{Elomaa\xfnm[ V.V.]},
  \bibinfo{author}{Hager\xfnm[ U.]}, \bibinfo{author}{Hakala\xfnm[ J.]},
  \bibinfo{author}{Jokinen\xfnm[ A.]}, \bibinfo{author}{Kankainen\xfnm[ A.]},
  \bibinfo{author}{Rahaman\xfnm[ S.]}, \bibinfo{author}{Rissanen\xfnm[ J.]},
  \bibinfo{author}{Weber\xfnm[ C.]}, \bibinfo{author}{{\"{A}}yst{\"{o}}\xfnm[
  J.]}.
\newblock \bibinfo{title}{{JYFLTRAP: Mass spectrometry and isomerically clean
  beams}}.
\newblock \emph{\bibinfo{journal}{Acta Physica Polonica B}}
  \bibinfo{year}{2008};\bibinfo{volume}{39}(\bibinfo{number}{2}):\bibinfo{pages}{445--455}.
\newblock \URLprefix \url{https://www.actaphys.uj.edu.pl/R/39/2/445/pdf}.
%Type = Article
\bibitem[{Nesterenko et~al.(2021)Nesterenko, Eronen, Ge, Kankainen and
  Vilen}]{nesterenko2021}
\bibinfo{author}{Nesterenko\xfnm[ D.A.]}, \bibinfo{author}{Eronen\xfnm[ T.]},
  \bibinfo{author}{Ge\xfnm[ Z.]}, \bibinfo{author}{Kankainen\xfnm[ A.]},
  \bibinfo{author}{Vilen\xfnm[ M.]}.
\newblock \bibinfo{title}{Study of radial motion phase advance during motion
  excitations in a penning trap and accuracy of jyfltrap mass spectrometer}.
\newblock \emph{\bibinfo{journal}{Eur Phys J A}}
  \bibinfo{year}{2021};\bibinfo{volume}{57}:\bibinfo{pages}{302}.
\newblock \URLprefix \url{https://doi.org/10.1140/epja/s10050-021-00608-3}.
%Type = Article
\bibitem[{Nesterenko et~al.(2018)Nesterenko, Eronen, Kankainen, Canete,
  Jokinen, Moore, Penttil{\"{a}}, Rinta-Antila, de~Roubin and
  Vilen}]{Nesterenko2018}
\bibinfo{author}{Nesterenko\xfnm[ D.A.]}, \bibinfo{author}{Eronen\xfnm[ T.]},
  \bibinfo{author}{Kankainen\xfnm[ A.]}, \bibinfo{author}{Canete\xfnm[ L.]},
  \bibinfo{author}{Jokinen\xfnm[ A.]}, \bibinfo{author}{Moore\xfnm[ I.D.]},
  \bibinfo{author}{Penttil{\"{a}}\xfnm[ H.]},
  \bibinfo{author}{Rinta-Antila\xfnm[ S.]}, \bibinfo{author}{de~Roubin\xfnm[
  A.]}, \bibinfo{author}{Vilen\xfnm[ M.]}.
\newblock \bibinfo{title}{{Phase-Imaging Ion-Cyclotron-Resonance technique at
  the JYFLTRAP double Penning trap mass spectrometer}}.
\newblock \emph{\bibinfo{journal}{European Physical Journal A}}
  \bibinfo{year}{2018};\bibinfo{volume}{54}(\bibinfo{number}{9}):\bibinfo{pages}{0--13}.
\newblock \URLprefix \url{https://dx.doi.org/10.1140/epja/i2018-12589-y}.
  \DOIprefix\doi{10.1140/epja/i2018-12589-y}.
%Type = Article
\bibitem[{Eliseev et~al.(2014)Eliseev, Blaum, Block, D{\"{o}}rr, Droese,
  Eronen, Goncharov, H{\"{o}}cker, Ketter, Ramirez, Nesterenko, Novikov and
  Schweikhard}]{Eliseev2014}
\bibinfo{author}{Eliseev\xfnm[ S.]}, \bibinfo{author}{Blaum\xfnm[ K.]},
  \bibinfo{author}{Block\xfnm[ M.]}, \bibinfo{author}{D{\"{o}}rr\xfnm[ A.]},
  \bibinfo{author}{Droese\xfnm[ C.]}, \bibinfo{author}{Eronen\xfnm[ T.]},
  \bibinfo{author}{Goncharov\xfnm[ M.]}, \bibinfo{author}{H{\"{o}}cker\xfnm[
  M.]}, \bibinfo{author}{Ketter\xfnm[ J.]}, \bibinfo{author}{Ramirez\xfnm[
  E.M.]}, \bibinfo{author}{Nesterenko\xfnm[ D.A.]},
  \bibinfo{author}{Novikov\xfnm[ Y.N.]}, \bibinfo{author}{Schweikhard\xfnm[
  L.]}.
\newblock \bibinfo{title}{{A phase-imaging technique for cyclotron-frequency
  measurements}}.
\newblock \emph{\bibinfo{journal}{Applied Physics B: Lasers and Optics}}
  \bibinfo{year}{2014};\bibinfo{volume}{114}(\bibinfo{number}{1-2}):\bibinfo{pages}{107--128}.
\newblock \URLprefix \url{http://dx.doi.org/10.1007/s00340-013-5621-0}.
  \DOIprefix\doi{10.1007/s00340-013-5621-0}.
%Type = Article
\bibitem[{Eliseev et~al.(2013)Eliseev, Blaum, Block, Droese, Goncharov, {Minaya
  Ramirez}, Nesterenko, Novikov and Schweikhard}]{Eliseev2013}
\bibinfo{author}{Eliseev\xfnm[ S.]}, \bibinfo{author}{Blaum\xfnm[ K.]},
  \bibinfo{author}{Block\xfnm[ M.]}, \bibinfo{author}{Droese\xfnm[ C.]},
  \bibinfo{author}{Goncharov\xfnm[ M.]}, \bibinfo{author}{{Minaya
  Ramirez}\xfnm[ E.]}, \bibinfo{author}{Nesterenko\xfnm[ D.A.]},
  \bibinfo{author}{Novikov\xfnm[ Y.N.]}, \bibinfo{author}{Schweikhard\xfnm[
  L.]}.
\newblock \bibinfo{title}{{Phase-imaging ion-cyclotron-resonance measurements
  for short-lived nuclides}}.
\newblock \emph{\bibinfo{journal}{Physical Review Letters}}
  \bibinfo{year}{2013};\bibinfo{volume}{110}(\bibinfo{number}{8}):\bibinfo{pages}{82501}.
\newblock \URLprefix
  \url{http://link.aps.org/doi/10.1103/PhysRevLett.110.082501}.
  \DOIprefix\doi{10.1103/PhysRevLett.110.082501}.
%Type = Misc
\bibitem[{Kramida et~al.(2020)Kramida, {Yu.~Ralchenko}, Reader and {and NIST
  ASD Team}}]{NIST_ASD}
\bibinfo{author}{Kramida\xfnm[ A.]}, \bibinfo{author}{{Yu.~Ralchenko}\xfnm[]},
  \bibinfo{author}{Reader\xfnm[ J.]}, \bibinfo{author}{{and NIST ASD
  Team}\xfnm[]}.
\newblock \bibinfo{howpublished}{{NIST Atomic Spectra Database (ver. 5.8),
  [Online]. Available: {\tt{https://physics.nist.gov/asd}} [2021, January 19].
  National Institute of Standards and Technology, Gaithersburg, MD.}};
  \bibinfo{year}{2020}.
%Type = Article
\bibitem[{Wang et~al.(2021)Wang, Huang, Kondev, Audi and Naimi}]{Wang2021}
\bibinfo{author}{Wang\xfnm[ M.]}, \bibinfo{author}{Huang\xfnm[ W.]},
  \bibinfo{author}{Kondev\xfnm[ F.]}, \bibinfo{author}{Audi\xfnm[ G.]},
  \bibinfo{author}{Naimi\xfnm[ S.]}.
\newblock \bibinfo{title}{The {AME} 2020 atomic mass evaluation ({II}). tables,
  graphs and references$^*$}.
\newblock \emph{\bibinfo{journal}{Chinese Physics C}}
  \bibinfo{year}{2021};\bibinfo{volume}{45}(\bibinfo{number}{3}):\bibinfo{pages}{030003}.
\newblock \URLprefix \url{https://doi.org/10.1088/1674-1137/abddaf}.
  \DOIprefix\doi{10.1088/1674-1137/abddaf}.
%Type = Misc
\bibitem[{NND(2021)}]{NNDC}
\bibinfo{title}{National nuclear data center}.
\newblock \bibinfo{howpublished}{Available at \url{https://www.nndc.bnl.gov/}
  (2020/4/7)}; \bibinfo{year}{2021}.
\newblock \URLprefix \url{https://www.nndc.bnl.gov/}.
%Type = Article
\bibitem[{Wiedeking et~al.(2016)Wiedeking, Krti\ifmmode~\check{c}\else
  \v{c}\fi{}ka, Bernstein, Allmond, Basunia, Bleuel, Harke, Daub, Fallon,
  Firestone, Goldblum, Hatarik, Lake, Larsen, Lee, Lesher, Paschalis, Petri,
  Phair, Scielzo and Volya}]{Wiedeking16}
\bibinfo{author}{Wiedeking\xfnm[ M.]},
  \bibinfo{author}{Krti\ifmmode~\check{c}\else \v{c}\fi{}ka\xfnm[ M.]},
  \bibinfo{author}{Bernstein\xfnm[ L.A.]}, \bibinfo{author}{Allmond\xfnm[
  J.M.]}, \bibinfo{author}{Basunia\xfnm[ M.S.]}, \bibinfo{author}{Bleuel\xfnm[
  D.L.]}, \bibinfo{author}{Harke\xfnm[ J.T.]}, \bibinfo{author}{Daub\xfnm[
  B.H.]}, \bibinfo{author}{Fallon\xfnm[ P.]}, \bibinfo{author}{Firestone\xfnm[
  R.B.]}, \bibinfo{author}{Goldblum\xfnm[ B.L.]},
  \bibinfo{author}{Hatarik\xfnm[ R.]}, \bibinfo{author}{Lake\xfnm[ P.T.]},
  \bibinfo{author}{Larsen\xfnm[ A.C.]}, \bibinfo{author}{Lee\xfnm[ I.Y.]},
  \bibinfo{author}{Lesher\xfnm[ S.R.]}, \bibinfo{author}{Paschalis\xfnm[ S.]},
  \bibinfo{author}{Petri\xfnm[ M.]}, \bibinfo{author}{Phair\xfnm[ L.]},
  \bibinfo{author}{Scielzo\xfnm[ N.D.]}, \bibinfo{author}{Volya\xfnm[ A.]}.
\newblock \bibinfo{title}{$\ensuremath{\gamma}$-ray decay from neutron-bound
  and unbound states in $^{95}\mathrm{Mo}$ and a novel technique for spin
  determination}.
\newblock \emph{\bibinfo{journal}{Phys Rev C}}
  \bibinfo{year}{2016};\bibinfo{volume}{93}:\bibinfo{pages}{024303}.
\newblock \URLprefix \url{https://link.aps.org/doi/10.1103/PhysRevC.93.024303}.
  \DOIprefix\doi{10.1103/PhysRevC.93.024303}.
%Type = Article
\bibitem[{Basu et~al.(2010)Basu, Mukherjee and Sonzogni}]{Basu2010}
\bibinfo{author}{Basu\xfnm[ S.K.]}, \bibinfo{author}{Mukherjee\xfnm[ G.]},
  \bibinfo{author}{Sonzogni\xfnm[ A.A.]}.
\newblock \bibinfo{title}{Nuclear data sheets for {A} = 95}.
\newblock \emph{\bibinfo{journal}{Nuclear Data Sheets}}
  \bibinfo{year}{2010};\bibinfo{volume}{111}(\bibinfo{number}{10-11}):\bibinfo{pages}{2555--2737}.
\newblock \DOIprefix\doi{10.1016/j.nds.2010.10.001}.
%Type = Book
\bibitem[{Thompson et~al.(2009)Thompson, Lindau, Attwood, Liu, Gullikson,
  Pianetta, Howells, Robinson, Kim, Scofield, Kirz, Underwood, Kortright,
  Wiliams and Winick}]{Thompson2009}
\bibinfo{author}{Thompson\xfnm[ A.]}, \bibinfo{author}{Lindau\xfnm[ I.]},
  \bibinfo{author}{Attwood\xfnm[ D.]}, \bibinfo{author}{Liu\xfnm[ Y.]},
  \bibinfo{author}{Gullikson\xfnm[ E.]}, \bibinfo{author}{Pianetta\xfnm[ P.]},
  \bibinfo{author}{Howells\xfnm[ M.]}, \bibinfo{author}{Robinson\xfnm[ A.]},
  \bibinfo{author}{Kim\xfnm[ K.]}, \bibinfo{author}{Scofield\xfnm[ J.]},
  \bibinfo{author}{Kirz\xfnm[ J.]}, \bibinfo{author}{Underwood\xfnm[ J.]},
  \bibinfo{author}{Kortright\xfnm[ J.]}, \bibinfo{author}{Wiliams\xfnm[ G.]},
  \bibinfo{author}{Winick\xfnm[ H.]}.
\newblock \bibinfo{title}{{X-RAY DATA BOOKLET}}.
\newblock \bibinfo{address}{Berkeley, California}: \bibinfo{publisher}{Lawrence
  Berkeley National Laboratory}; \bibinfo{year}{2009}.
%Type = Article
\bibitem[{Kellerbauer et~al.(2003)Kellerbauer, Blaum, Bollen, Herfurth, Kluge,
  Kuckein, Sauvan, Scheidenberger and Schweikhard}]{Kellerbauer2003}
\bibinfo{author}{Kellerbauer\xfnm[ A.]}, \bibinfo{author}{Blaum\xfnm[ K.]},
  \bibinfo{author}{Bollen\xfnm[ G.]}, \bibinfo{author}{Herfurth\xfnm[ F.]},
  \bibinfo{author}{Kluge\xfnm[ H.J.]}, \bibinfo{author}{Kuckein\xfnm[ M.]},
  \bibinfo{author}{Sauvan\xfnm[ E.]}, \bibinfo{author}{Scheidenberger\xfnm[
  C.]}, \bibinfo{author}{Schweikhard\xfnm[ L.]}.
\newblock \bibinfo{title}{{From direct to absolute mass measurements: A study
  of the accuracy of ISOLTRAP}}.
\newblock \emph{\bibinfo{journal}{European Physical Journal D}}
  \bibinfo{year}{2003};\bibinfo{volume}{22}(\bibinfo{number}{1}):\bibinfo{pages}{53--64}.
\newblock \URLprefix \url{http://dx.doi.org/10.1140/epjd/e2002-00222-0}.
  \DOIprefix\doi{10.1140/epjd/e2002-00222-0}.
%Type = Article
\bibitem[{Roux et~al.(2013)Roux, Blaum, Block, Droese, Eliseev, Goncharov,
  Herfurth, Ramirez, Nesterenko, Novikov and Schweikhard}]{Roux2013}
\bibinfo{author}{Roux\xfnm[ C.]}, \bibinfo{author}{Blaum\xfnm[ K.]},
  \bibinfo{author}{Block\xfnm[ M.]}, \bibinfo{author}{Droese\xfnm[ C.]},
  \bibinfo{author}{Eliseev\xfnm[ S.]}, \bibinfo{author}{Goncharov\xfnm[ M.]},
  \bibinfo{author}{Herfurth\xfnm[ F.]}, \bibinfo{author}{Ramirez\xfnm[ E.M.]},
  \bibinfo{author}{Nesterenko\xfnm[ D.A.]}, \bibinfo{author}{Novikov\xfnm[
  Y.N.]}, \bibinfo{author}{Schweikhard\xfnm[ L.]}.
\newblock \bibinfo{title}{{Data analysis of Q-value measurements for
  double-electron capture with SHIPTRAP}}.
\newblock \emph{\bibinfo{journal}{The European Physical Journal D}}
  \bibinfo{year}{2013};\bibinfo{volume}{67}(\bibinfo{number}{7}):\bibinfo{pages}{1--9}.
\newblock \URLprefix \url{http://dx.doi.org/10.1140/epjd/e2013-40110-x}.
  \DOIprefix\doi{10.1140/epjd/e2013-40110-x}.
%Type = Article
\bibitem[{Birge(1932)}]{Birge1932}
\bibinfo{author}{Birge\xfnm[ R.T.]}.
\newblock \bibinfo{title}{{The calculation of errors by the method of least
  squares}}.
\newblock \emph{\bibinfo{journal}{Physical Review}}
  \bibinfo{year}{1932};\bibinfo{volume}{40}(\bibinfo{number}{2}):\bibinfo{pages}{207--227}.
\newblock \URLprefix \url{http://link.aps.org/abstract/PR/v40/p207}.
  \DOIprefix\doi{10.1103/PhysRev.40.207}.
%Type = Article
\bibitem[{Huang et~al.(2021)Huang, Wang, Kondev, Audi and Naimi}]{Huang2021}
\bibinfo{author}{Huang\xfnm[ W.]}, \bibinfo{author}{Wang\xfnm[ M.]},
  \bibinfo{author}{Kondev\xfnm[ F.]}, \bibinfo{author}{Audi\xfnm[ G.]},
  \bibinfo{author}{Naimi\xfnm[ S.]}.
\newblock \bibinfo{title}{The {AME} 2020 atomic mass evaluation {(I)}.
  evaluation of input data, and adjustment procedures$^*$}.
\newblock \emph{\bibinfo{journal}{Chinese Physics C}}
  \bibinfo{year}{2021};\bibinfo{volume}{45}(\bibinfo{number}{3}):\bibinfo{pages}{030002}.
\newblock \URLprefix \url{https://doi.org/10.1088/1674-1137/abddb0}.
  \DOIprefix\doi{10.1088/1674-1137/abddb0}.
%Type = Article
\bibitem[{Langer and Wortman(1963)}]{Langer63}
\bibinfo{author}{Langer\xfnm[ L.M.]}, \bibinfo{author}{Wortman\xfnm[ D.E.]}.
\newblock \bibinfo{title}{Radioactive decay of ${\mathrm{nb}}^{95}$}.
\newblock \emph{\bibinfo{journal}{Phys Rev}}
  \bibinfo{year}{1963};\bibinfo{volume}{132}:\bibinfo{pages}{324--328}.
\newblock \URLprefix \url{https://link.aps.org/doi/10.1103/PhysRev.132.324}.
  \DOIprefix\doi{10.1103/PhysRev.132.324}.
%Type = Article
\bibitem[{Cretzu et~al.(1965)Cretzu, Hohmuth and Schintlmeister}]{CRETZU65}
\bibinfo{author}{Cretzu\xfnm[ T.]}, \bibinfo{author}{Hohmuth\xfnm[ K.]},
  \bibinfo{author}{Schintlmeister\xfnm[ J.]}.
\newblock \bibinfo{title}{Der zerfall von tc95m}.
\newblock \emph{\bibinfo{journal}{Nuclear Physics}}
  \bibinfo{year}{1965};\bibinfo{volume}{70}(\bibinfo{number}{1}):\bibinfo{pages}{129--140}.
\newblock \URLprefix
  \url{https://www.sciencedirect.com/science/article/pii/0029558265902294}.
  \DOIprefix\doi{https://doi.org/10.1016/0029-5582(65)90229-4}.
%Type = Article
\bibitem[{N.M.{Antoneva} et~al.(1974)N.M.{Antoneva}, A.V.{Barkov},
  A.V.{Zolotavin}, P.P.{Dmitriev}, S.V.{Kamynov}, G.S.{Katykhin},
  E.T.{Kondrat}, N.I.{Krasnov}, Y.N.{Podkopaev}, V.A.{Sergienko} and
  V.I.{Fominykh}}]{Anto74}
\bibinfo{author}{N.M.{Antoneva}\xfnm[]}, \bibinfo{author}{A.V.{Barkov}\xfnm[]},
  \bibinfo{author}{A.V.{Zolotavin}\xfnm[]},
  \bibinfo{author}{P.P.{Dmitriev}\xfnm[]},
  \bibinfo{author}{S.V.{Kamynov}\xfnm[]},
  \bibinfo{author}{G.S.{Katykhin}\xfnm[]},
  \bibinfo{author}{E.T.{Kondrat}\xfnm[]},
  \bibinfo{author}{N.I.{Krasnov}\xfnm[]},
  \bibinfo{author}{Y.N.{Podkopaev}\xfnm[]},
  \bibinfo{author}{V.A.{Sergienko}\xfnm[]},
  \bibinfo{author}{V.I.{Fominykh}\xfnm[]}.
\newblock \bibinfo{title}{The decay of $^{95}$*tc}.
\newblock \emph{\bibinfo{journal}{IzvAkadNauk}}
  \bibinfo{year}{1974};\bibinfo{volume}{SSSR}:\bibinfo{pages}{Ser.Fiz. 38, 48}.
%Type = Article
\bibitem[{J.A.{Pinston} et~al.(1968)J.A.{Pinston}, E.{Monnand} and
  A.{Moussa}}]{Pin68}
\bibinfo{author}{J.A.{Pinston}\xfnm[]}, \bibinfo{author}{E.{Monnand}\xfnm[]},
  \bibinfo{author}{A.{Moussa}\xfnm[]}.
\newblock \bibinfo{title}{Desintegration de $^{95}$ru}.
\newblock \emph{\bibinfo{journal}{JPhys(Paris)}}
  \bibinfo{year}{1968};\bibinfo{volume}{29}:\bibinfo{pages}{257}.
%Type = Article
\bibitem[{Sevestrean et~al.(2023)Sevestrean, Ni\ifmmode~\mbox{\c{t}}\else
  \c{t}\fi{}escu, Ghinescu and Stoica}]{SevestreanPRA2023}
\bibinfo{author}{Sevestrean\xfnm[ V.A.]},
  \bibinfo{author}{Ni\ifmmode~\mbox{\c{t}}\else \c{t}\fi{}escu\xfnm[ O.]},
  \bibinfo{author}{Ghinescu\xfnm[ S.]}, \bibinfo{author}{Stoica\xfnm[ S.]}.
\newblock \bibinfo{title}{Self-consistent calculations for atomic electron
  capture}.
\newblock \emph{\bibinfo{journal}{Phys Rev A}}
  \bibinfo{year}{2023};\bibinfo{volume}{108}:\bibinfo{pages}{012810}.
\newblock \URLprefix
  \url{https://link.aps.org/doi/10.1103/PhysRevA.108.012810}.
  \DOIprefix\doi{10.1103/PhysRevA.108.012810}.
%Type = Article
\bibitem[{Salvat and Fernández-Varea(2019)}]{SalvatCPC2019}
\bibinfo{author}{Salvat\xfnm[ F.]}, \bibinfo{author}{Fernández-Varea\xfnm[
  J.M.]}.
\newblock \bibinfo{title}{radial: A fortran subroutine package for the solution
  of the radial schrödinger and dirac wave equations}.
\newblock \emph{\bibinfo{journal}{Computer Physics Communications}}
  \bibinfo{year}{2019};\bibinfo{volume}{240}:\bibinfo{pages}{165--177}.
\newblock \URLprefix
  \url{https://www.sciencedirect.com/science/article/pii/S0010465519300633}.
  \DOIprefix\doi{https://doi.org/10.1016/j.cpc.2019.02.011}.
%Type = Article
\bibitem[{Hahn et~al.(1956)Hahn, Ravenhall and Hofstadter}]{HahnPR1956}
\bibinfo{author}{Hahn\xfnm[ B.]}, \bibinfo{author}{Ravenhall\xfnm[ D.G.]},
  \bibinfo{author}{Hofstadter\xfnm[ R.]}.
\newblock \bibinfo{title}{High-energy electron scattering and the charge
  distributions of selected nuclei}.
\newblock \emph{\bibinfo{journal}{Phys Rev}}
  \bibinfo{year}{1956};\bibinfo{volume}{101}:\bibinfo{pages}{1131--1142}.
\newblock \URLprefix \url{https://link.aps.org/doi/10.1103/PhysRev.101.1131}.
  \DOIprefix\doi{10.1103/PhysRev.101.1131}.
%Type = Article
\bibitem[{CAMPBELL and PAPP(2001)}]{CampbellADNDT2001}
\bibinfo{author}{CAMPBELL\xfnm[ J.]}, \bibinfo{author}{PAPP\xfnm[ T.]}.
\newblock \bibinfo{title}{Widths of the atomic k–n7 levels}.
\newblock \emph{\bibinfo{journal}{Atomic Data and Nuclear Data Tables}}
  \bibinfo{year}{2001};\bibinfo{volume}{77}(\bibinfo{number}{1}):\bibinfo{pages}{1--56}.
\newblock \URLprefix
  \url{https://www.sciencedirect.com/science/article/pii/S0092640X00908489}.
  \DOIprefix\doi{https://doi.org/10.1006/adnd.2000.0848}.
%Type = Book
\bibitem[{Behrens and B{\"{u}}hring(1982)}]{Behrens1982}
\bibinfo{author}{Behrens\xfnm[ H.]}, \bibinfo{author}{B{\"{u}}hring\xfnm[ W.]}.
\newblock \bibinfo{title}{{Electron Radial Wave Functions and Nuclear
  Beta-decay (International Series of Monographs on Physics)}}.
\newblock \bibinfo{address}{Oxford}: \bibinfo{publisher}{Clarendon press};
  \bibinfo{year}{1982}.
%Type = Book
\bibitem[{Suhonen(2007)}]{JSuhonen2007}
\bibinfo{author}{Suhonen\xfnm[ J.]}.
\newblock \bibinfo{title}{{From Nucleons to Nucleus}}.
\newblock \bibinfo{address}{Springer, Gaithersburg MD, 20899}:
  \bibinfo{publisher}{Springer--Verlag Berlin Heidelberg};
  \bibinfo{year}{2007}.
\newblock \URLprefix
  \url{https://www.springer.com/gp/book/9783540488590#aboutBook}.
  \DOIprefix\doi{10.1007/978-3-540-48861-3}.
%Type = Article
\bibitem[{Honma et~al.(2006)Honma, Otsuka, Mizusaki and
  Hjorth-Jensen}]{JPCSHonma2006}
\bibinfo{author}{Honma\xfnm[ M.]}, \bibinfo{author}{Otsuka\xfnm[ T.]},
  \bibinfo{author}{Mizusaki\xfnm[ T.]}, \bibinfo{author}{Hjorth-Jensen\xfnm[
  M.]}.
\newblock \bibinfo{title}{Effective interaction for f5pg9-shell nuclei and
  two-neutrino double beta-decay matrix elements}.
\newblock \emph{\bibinfo{journal}{Journal of Physics: Conference Series}}
  \bibinfo{year}{2006};\bibinfo{volume}{49}(\bibinfo{number}{1}):\bibinfo{pages}{45}.
\newblock \URLprefix \url{https://dx.doi.org/10.1088/1742-6596/49/1/011}.
  \DOIprefix\doi{10.1088/1742-6596/49/1/011}.
%Type = Article
\bibitem[{Barea et~al.(2013)Barea, Kotila and Iachello}]{PRCBarea2013}
\bibinfo{author}{Barea\xfnm[ J.]}, \bibinfo{author}{Kotila\xfnm[ J.]},
  \bibinfo{author}{Iachello\xfnm[ F.]}.
\newblock \bibinfo{title}{Nuclear matrix elements for
  double-$\ensuremath{\beta}$ decay}.
\newblock \emph{\bibinfo{journal}{Phys Rev C}}
  \bibinfo{year}{2013};\bibinfo{volume}{87}:\bibinfo{pages}{014315}.
\newblock \URLprefix \url{https://link.aps.org/doi/10.1103/PhysRevC.87.014315}.
  \DOIprefix\doi{10.1103/PhysRevC.87.014315}.
%Type = Article
\bibitem[{Suhonen(2017)}]{Suhonen2017}
\bibinfo{author}{Suhonen\xfnm[ J.T.]}.
\newblock \bibinfo{title}{Value of the axial-vector coupling strength in
  $\beta$ and $\beta \beta$ decays: A review}.
\newblock \emph{\bibinfo{journal}{Frontiers in Physics}}
  \bibinfo{year}{2017};\bibinfo{volume}{5}.
\newblock \URLprefix
  \url{https://www.frontiersin.org/articles/10.3389/fphy.2017.00055}.
  \DOIprefix\doi{10.3389/fphy.2017.00055}.
%Type = Article
\bibitem[{Mougeot(2018)}]{MougeotARI2018}
\bibinfo{author}{Mougeot\xfnm[ X.]}.
\newblock \bibinfo{title}{Improved calculations of electron capture transitions
  for decay data and radionuclide metrology}.
\newblock \emph{\bibinfo{journal}{Applied Radiation and Isotopes}}
  \bibinfo{year}{2018};\bibinfo{volume}{134}:\bibinfo{pages}{225--232}.
\newblock \URLprefix
  \url{https://www.sciencedirect.com/science/article/pii/S0969804317304372}.
  \DOIprefix\doi{https://doi.org/10.1016/j.apradiso.2017.07.027}.
%Type = Article
\bibitem[{Hinfurtner et~al.(1995)Hinfurtner, Hagn, Zech, Tröger and
  Butz}]{Hinfurtner1995}
\bibinfo{author}{Hinfurtner\xfnm[ B.]}, \bibinfo{author}{Hagn\xfnm[ E.]},
  \bibinfo{author}{Zech\xfnm[ E.]}, \bibinfo{author}{Tröger\xfnm[ W.]},
  \bibinfo{author}{Butz\xfnm[ T.]}.
\newblock \bibinfo{title}{Measurements of magnetic moments of tc isotopes and
  the hyperfine field of tc in fe and ni}.
\newblock \emph{\bibinfo{journal}{Zeitschrift für Physik A Hadrons and
  Nuclei}}
  \bibinfo{year}{1995};\bibinfo{volume}{350}(\bibinfo{number}{4}):\bibinfo{pages}{311--318}.
\newblock \DOIprefix\doi{10.1007/bf01291188}.
%Type = Article
\bibitem[{Proctor and Yu(1951)}]{Proctor1951}
\bibinfo{author}{Proctor\xfnm[ W.G.]}, \bibinfo{author}{Yu\xfnm[ F.C.]}.
\newblock \bibinfo{title}{On the nuclear magnetic moments of several stable
  isotopes}.
\newblock \emph{\bibinfo{journal}{Physical Review}}
  \bibinfo{year}{1951};\bibinfo{volume}{81}(\bibinfo{number}{1}):\bibinfo{pages}{20--30}.
\newblock \DOIprefix\doi{10.1103/physrev.81.20}.
%Type = Article
\bibitem[{Alzner et~al.(1984)Alzner, Bodenstedt, Gemünden, Herrmann, Münning,
  Reif, Rudolph, Vianden and Wrede}]{Alzner1984}
\bibinfo{author}{Alzner\xfnm[ A.]}, \bibinfo{author}{Bodenstedt\xfnm[ E.]},
  \bibinfo{author}{Gemünden\xfnm[ G.]}, \bibinfo{author}{Herrmann\xfnm[ C.]},
  \bibinfo{author}{Münning\xfnm[ H.]}, \bibinfo{author}{Reif\xfnm[ H.]},
  \bibinfo{author}{Rudolph\xfnm[ H.J.]}, \bibinfo{author}{Vianden\xfnm[ R.]},
  \bibinfo{author}{Wrede\xfnm[ U.]}.
\newblock \bibinfo{title}{Gyromagnetic ratios of the 3/2+ core vibration states
  of101ru and95mo}.
\newblock \emph{\bibinfo{journal}{Zeitschrift für Physik A Atoms and Nuclei}}
  \bibinfo{year}{1984};\bibinfo{volume}{317}(\bibinfo{number}{1}):\bibinfo{pages}{107--115}.
\newblock \DOIprefix\doi{10.1007/bf01420454}.
%Type = Book
\bibitem[{Stone(2021)}]{Stone2021}
\bibinfo{author}{Stone\xfnm[ N.]}.
\newblock \bibinfo{title}{Table of Nuclear Electric Quadrupole Moments}.
\newblock \bibinfo{year}{2021}.
\newblock \DOIprefix\doi{10.61092/iaea.a6te-dg7q}.
%Type = Article
\bibitem[{Brown and Rae(2014)}]{Brown2014}
\bibinfo{author}{Brown\xfnm[ B.]}, \bibinfo{author}{Rae\xfnm[ W.]}.
\newblock \bibinfo{title}{The shell-model code {NuShellX}@{MSU}}.
\newblock \emph{\bibinfo{journal}{Nuclear Data Sheets}}
  \bibinfo{year}{2014};\bibinfo{volume}{120}:\bibinfo{pages}{115--118}.
\newblock \URLprefix \url{https://doi.org/10.1016/j.nds.2014.07.022}.
  \DOIprefix\doi{10.1016/j.nds.2014.07.022}.
%Type = Article
\bibitem[{Machleidt(2001)}]{Machleidt2001}
\bibinfo{author}{Machleidt\xfnm[ R.]}.
\newblock \bibinfo{title}{High-precision, charge-dependent bonn nucleon-nucleon
  potential}.
\newblock \emph{\bibinfo{journal}{Physical Review C}}
  \bibinfo{year}{2001};\bibinfo{volume}{63}(\bibinfo{number}{2}):\bibinfo{pages}{024001}.
\newblock \DOIprefix\doi{10.1103/physrevc.63.024001}.
%Type = Article
\bibitem[{Vaquero et~al.(2020)Vaquero, Jungclaus, Aumann, Tscheuschner,
  Litvinova, Tostevin, Baba, Ahn, Avigo, Boretzky, Bracco, Caesar, Camera,
  Chen, Derya, Doornenbal, Endres, Fukuda, Garg, Giaz, Harakeh, Heil, Horvat,
  Ieki, Imai, Inabe, Kalantar-Nayestanaki, Kobayashi, Kondo, Koyama, Kubo,
  Martel, Matsushita, Million, Motobayashi, Nakamura, Nakatsuka, Nishimura,
  Nishimura, Ota, Otsu, Ozaki, Petri, Reifarth, Rodr\'{\i}guez-S\'anchez,
  Rossi, Saito, Sakurai, Savran, Scheit, Schindler, Schrock, Semmler, Shiga,
  Shikata, Shimizu, Simon, Steppenbeck, Suzuki, Sumikama, Symochko, Syndikus,
  Takeda, Takeuchi, Taniuchi, Togano, Tsubota, Wang, Wieland, Yoneda, Zenihiro
  and Zilges}]{Vaquero2020}
\bibinfo{author}{Vaquero\xfnm[ V.]}, \bibinfo{author}{Jungclaus\xfnm[ A.]},
  \bibinfo{author}{Aumann\xfnm[ T.]}, \bibinfo{author}{Tscheuschner\xfnm[ J.]},
  \bibinfo{author}{Litvinova\xfnm[ E.V.]}, \bibinfo{author}{Tostevin\xfnm[
  J.A.]}, \bibinfo{author}{Baba\xfnm[ H.]}, \bibinfo{author}{Ahn\xfnm[ D.S.]},
  \bibinfo{author}{Avigo\xfnm[ R.]}, \bibinfo{author}{Boretzky\xfnm[ K.]},
  \bibinfo{author}{Bracco\xfnm[ A.]}, \bibinfo{author}{Caesar\xfnm[ C.]},
  \bibinfo{author}{Camera\xfnm[ F.]}, \bibinfo{author}{Chen\xfnm[ S.]},
  \bibinfo{author}{Derya\xfnm[ V.]}, \bibinfo{author}{Doornenbal\xfnm[ P.]},
  \bibinfo{author}{Endres\xfnm[ J.]}, \bibinfo{author}{Fukuda\xfnm[ N.]},
  \bibinfo{author}{Garg\xfnm[ U.]}, \bibinfo{author}{Giaz\xfnm[ A.]},
  \bibinfo{author}{Harakeh\xfnm[ M.N.]}, \bibinfo{author}{Heil\xfnm[ M.]},
  \bibinfo{author}{Horvat\xfnm[ A.]}, \bibinfo{author}{Ieki\xfnm[ K.]},
  \bibinfo{author}{Imai\xfnm[ N.]}, \bibinfo{author}{Inabe\xfnm[ N.]},
  \bibinfo{author}{Kalantar-Nayestanaki\xfnm[ N.]},
  \bibinfo{author}{Kobayashi\xfnm[ N.]}, \bibinfo{author}{Kondo\xfnm[ Y.]},
  \bibinfo{author}{Koyama\xfnm[ S.]}, \bibinfo{author}{Kubo\xfnm[ T.]},
  \bibinfo{author}{Martel\xfnm[ I.]}, \bibinfo{author}{Matsushita\xfnm[ M.]},
  \bibinfo{author}{Million\xfnm[ B.]}, \bibinfo{author}{Motobayashi\xfnm[ T.]},
  \bibinfo{author}{Nakamura\xfnm[ T.]}, \bibinfo{author}{Nakatsuka\xfnm[ N.]},
  \bibinfo{author}{Nishimura\xfnm[ M.]}, \bibinfo{author}{Nishimura\xfnm[ S.]},
  \bibinfo{author}{Ota\xfnm[ S.]}, \bibinfo{author}{Otsu\xfnm[ H.]},
  \bibinfo{author}{Ozaki\xfnm[ T.]}, \bibinfo{author}{Petri\xfnm[ M.]},
  \bibinfo{author}{Reifarth\xfnm[ R.]},
  \bibinfo{author}{Rodr\'{\i}guez-S\'anchez\xfnm[ J.L.]},
  \bibinfo{author}{Rossi\xfnm[ D.]}, \bibinfo{author}{Saito\xfnm[ A.T.]},
  \bibinfo{author}{Sakurai\xfnm[ H.]}, \bibinfo{author}{Savran\xfnm[ D.]},
  \bibinfo{author}{Scheit\xfnm[ H.]}, \bibinfo{author}{Schindler\xfnm[ F.]},
  \bibinfo{author}{Schrock\xfnm[ P.]}, \bibinfo{author}{Semmler\xfnm[ D.]},
  \bibinfo{author}{Shiga\xfnm[ Y.]}, \bibinfo{author}{Shikata\xfnm[ M.]},
  \bibinfo{author}{Shimizu\xfnm[ Y.]}, \bibinfo{author}{Simon\xfnm[ H.]},
  \bibinfo{author}{Steppenbeck\xfnm[ D.]}, \bibinfo{author}{Suzuki\xfnm[ H.]},
  \bibinfo{author}{Sumikama\xfnm[ T.]}, \bibinfo{author}{Symochko\xfnm[ D.]},
  \bibinfo{author}{Syndikus\xfnm[ I.]}, \bibinfo{author}{Takeda\xfnm[ H.]},
  \bibinfo{author}{Takeuchi\xfnm[ S.]}, \bibinfo{author}{Taniuchi\xfnm[ R.]},
  \bibinfo{author}{Togano\xfnm[ Y.]}, \bibinfo{author}{Tsubota\xfnm[ J.]},
  \bibinfo{author}{Wang\xfnm[ H.]}, \bibinfo{author}{Wieland\xfnm[ O.]},
  \bibinfo{author}{Yoneda\xfnm[ K.]}, \bibinfo{author}{Zenihiro\xfnm[ J.]},
  \bibinfo{author}{Zilges\xfnm[ A.]}.
\newblock \bibinfo{title}{Fragmentation of single-particle strength around the
  doubly magic nucleus $^{132}\mathrm{Sn}$ and the position of the $0{f}_{5/2}$
  proton-hole state in $^{131}\mathrm{In}$}.
\newblock \emph{\bibinfo{journal}{Phys Rev Lett}}
  \bibinfo{year}{2020};\bibinfo{volume}{124}:\bibinfo{pages}{022501}.
\newblock \URLprefix
  \url{https://link.aps.org/doi/10.1103/PhysRevLett.124.022501}.
  \DOIprefix\doi{10.1103/PhysRevLett.124.022501}.
%Type = Article
\bibitem[{Lisetskiy et~al.(2004)Lisetskiy, Brown, Horoi and
  Grawe}]{Lisetskiy2004}
\bibinfo{author}{Lisetskiy\xfnm[ A.F.]}, \bibinfo{author}{Brown\xfnm[ B.A.]},
  \bibinfo{author}{Horoi\xfnm[ M.]}, \bibinfo{author}{Grawe\xfnm[ H.]}.
\newblock \bibinfo{title}{New $t=1$ effective interactions for the
  ${f}_{5/2}\phantom{\rule{0.3em}{0ex}}{p}_{3/2}\phantom{\rule{0.3em}{0ex}}{p}_{1/2}\phantom{\rule{0.3em}{0ex}}{g}_{9/2}$
  model space: Implications for valence-mirror symmetry and seniority isomers}.
\newblock \emph{\bibinfo{journal}{Phys Rev C}}
  \bibinfo{year}{2004};\bibinfo{volume}{70}:\bibinfo{pages}{044314}.
\newblock \URLprefix \url{https://link.aps.org/doi/10.1103/PhysRevC.70.044314}.
  \DOIprefix\doi{10.1103/PhysRevC.70.044314}.
%Type = Article
\bibitem[{Mach et~al.(1990)Mach, Warburton, Gill, Casten, Becker, Brown and
  Winger}]{Mach1990}
\bibinfo{author}{Mach\xfnm[ H.]}, \bibinfo{author}{Warburton\xfnm[ E.K.]},
  \bibinfo{author}{Gill\xfnm[ R.L.]}, \bibinfo{author}{Casten\xfnm[ R.F.]},
  \bibinfo{author}{Becker\xfnm[ J.A.]}, \bibinfo{author}{Brown\xfnm[ B.A.]},
  \bibinfo{author}{Winger\xfnm[ J.A.]}.
\newblock \bibinfo{title}{Meson-exchange enhancement of the first-forbidden
  $^{96}\mathrm{Y}^{\mathrm{g}}$(${0}^{\mathrm{\ensuremath{-}}}$)${\ensuremath{\rightarrow}}^{96}$${\mathrm{zr}}^{\mathit{g}}$
  (${0}^{+}$) \ensuremath{\beta} transition: \ensuremath{\beta} decay of the
  low-spin isomer of $^{96}\mathrm{Y}$}.
\newblock \emph{\bibinfo{journal}{Phys Rev C}}
  \bibinfo{year}{1990};\bibinfo{volume}{41}:\bibinfo{pages}{226--242}.
\newblock \URLprefix \url{https://link.aps.org/doi/10.1103/PhysRevC.41.226}.
  \DOIprefix\doi{10.1103/PhysRevC.41.226}.
%Type = Article
\bibitem[{Schweiger et~al.(2024)Schweiger, Braß, Debierre, Door, Dorrer,
  Düllmann, Enss, Filianin, Gastaldo, Harman, Haverkort, Herkenhoff,
  Indelicato, Keitel, Kromer, Lange, Novikov, Renisch, Rischka, Schüssler,
  Eliseev and Blaum}]{Schweiger2024}
\bibinfo{author}{Schweiger\xfnm[ C.]}, \bibinfo{author}{Braß\xfnm[ M.]},
  \bibinfo{author}{Debierre\xfnm[ V.]}, \bibinfo{author}{Door\xfnm[ M.]},
  \bibinfo{author}{Dorrer\xfnm[ H.]}, \bibinfo{author}{Düllmann\xfnm[ C.E.]},
  \bibinfo{author}{Enss\xfnm[ C.]}, \bibinfo{author}{Filianin\xfnm[ P.]},
  \bibinfo{author}{Gastaldo\xfnm[ L.]}, \bibinfo{author}{Harman\xfnm[ Z.]},
  \bibinfo{author}{Haverkort\xfnm[ M.W.]}, \bibinfo{author}{Herkenhoff\xfnm[
  J.]}, \bibinfo{author}{Indelicato\xfnm[ P.]}, \bibinfo{author}{Keitel\xfnm[
  C.H.]}, \bibinfo{author}{Kromer\xfnm[ K.]}, \bibinfo{author}{Lange\xfnm[
  D.]}, \bibinfo{author}{Novikov\xfnm[ Y.N.]}, \bibinfo{author}{Renisch\xfnm[
  D.]}, \bibinfo{author}{Rischka\xfnm[ A.]}, \bibinfo{author}{Schüssler\xfnm[
  R.X.]}, \bibinfo{author}{Eliseev\xfnm[ S.]}, \bibinfo{author}{Blaum\xfnm[
  K.]}.
\newblock \bibinfo{title}{Penning-trap measurement of the q value of electron
  capture in 163ho for the determination of the electron neutrino mass}.
\newblock \emph{\bibinfo{journal}{Nature Physics}}
  \bibinfo{year}{2024};\DOIprefix\doi{10.1038/s41567-024-02461-9}.

\end{thebibliography}
%\bibliography{my-final-bib-from-jabref_titles} 
%\bibliography{my-final-bib-from-jabref}
%\end{document}

\textbf{Declaration of competing interest}

The authors declare that there are no known competing financial interests or personal relationships that could have appeared to influence the work reported in this paper.

\textbf{Acknowledgements}

%\acknowledgments 

We acknowledge the staff of the Accelerator Laboratory of University of Jyv\"askyl\"a (JYFL-ACCLAB) for providing stable online beam. We thank the support by the Academy of Finland under the Finnish Centre of Excellence Programme 2012--2017 (Nuclear and Accelerator Based Physics Research at JYFL) and projects No. 306980, 312544, 275389, 284516, 295207, 314733,  315179, 327629, 320062, 354589, 345869 and 354968. The support by the EU Horizon 2020 research and innovation program under grant No. 771036 (ERC CoG MAIDEN) is acknowledged.  This project has received funding from the European Union’s Horizon 2020 research and innovation programme under grant agreement No. 861198–LISA–H2020-MSCA-ITN-2019. V.A.S., O.N., S.S., J.S., and J.K. acknowledge support from project PNRR-I8/C9-CF264, Contract No. 760100/23.05.2023 of the Romanian Ministry of Research, Innovation and Digitization. The work leading to this publication was supported by the Deutsche Forschungsgemeinschaft (DFG, German Research Foundation) - AY 155/2-1. The paper was supported by the DAAD Grant No. 57610603.

%We acknowledge the staff of the accelerator laboratory of University of Jyv\"askyl\"a (JYFL-ACCLAB) for providing stable online beam and J.~Jaatinen and R.~Sepp\"al\"a for preparing the production target. We thank the support by the Academy of Finland under the Finnish Centre of Excellence Programme 2012-2017 (Nuclear and Accelerator Based Physics Research at JYFL) and projects No. 306980, 312544, 275389, 284516, 295207, 314733, 318043, 327629, 320062 and 318043. The support by the EU Horizon 2020 research and innovation program under grant No. 771036 (ERC CoG MAIDEN) is acknowledged.


\bibliographystyle{model6-num-names}
%\bibliographystyle{cas-model2-names}

%\bibliographystyle{unsrt}
%model1-num-names.bst
%model1a-num-names.bst
%model2-names.bst
%model3-num-names.bst
%model4-names.bst
%model5-names.bst
%model6-num-names.bst

%\bibliography{}
%\bibliographystyle{elsarticle-num-name}
%\bibliography{cas-refs.bib}
%\bibliography{my-final-bib-from-jabref_titles}


\end{document}